\documentclass{article}
\usepackage{amstex}
\usepackage{graphicx,fullpage}

\begin{document}

\begin{titlepage}

\title{ {\bf
Conformal field theory approach to gapless
1D fermion systems and application to
the edge excitations of $\nu = 1/(2p+1)$ quantum Hall
sequences}}

\author{
P. Degiovanni\\
Laboratoire de Physique Th\'eorique  
 ENSLAPP\thanks{URS 1436 associ\'ee au CNRS, \`a l'\'Ecole normale sup\'erieure
de Lyon et \`a l'universit\'e de Savoie}\\   
ENS Lyon, 46 all\'ee d'Italie 69364 Lyon cedex 07, France\thanks{Work supported by TMR contract
FMRX - CT96 - 0012}\\
Email:~\texttt{Pascal.Degiovanni@@ens-lyon.fr}
\and
C.~Chaubet\\
GES, Universit\'e des Sciences et Techniques du Languedoc\\
Place Eug\`ene Bataillon, 34095 Montpellier cedex 05, France\\
Email:~{\ttfamily cris@@pollux.ges.fr}
\and
R.~M\'elin\\
Centre de Recherche sur les Tr\`es Basses 
Temp\'eratures\thanks{UPR 5001 du CNRS}\\
BP166, 25 avenue des martyrs, 38042 Grenoble Cedex 09, France\\
Email:~\texttt{melin@@crtbt.polycnrs-gre.fr}}

\maketitle
 
\begin{abstract} 
We present a comprehensive study of the 
effective Conformal Field Theory (CFT) describing 
the low energy excitations of 
a gas of spinless interacting fermions on 
a circle in the gapless regime (Luttinger liquid). 
Functional techniques and modular transformation properties are
used to compute all correlation functions in a finite size and at finite 
temperature. Forward scattering disorder is treated exactly. 
Laughlin experiments on charge transport in a Quantum Hall Fluid on a
cylinder are reviewed within this CFT framework. Edge excitations above
a given bulk excitation are described by a twisted version of the 
Luttinger effective theory. Luttinger CFTs
corresponding to the $\nu =1/(2p+1)$ filling fractions
appear to be rational CFTs (RCFT). Generators of 
the extended symmetry algebra are 
identified as edge fermions creators and annihilators, thus giving a physical 
meaning to the RCFT point of view on edge excitations of these sequences.
\end{abstract}
 
\vskip 1cm 
 
\rightline{{\small E}N{\large S}{\Large L}{\large A}P{\small P}-L-656/97}

\end{titlepage}

\newpage

\tableofcontents

\newpage

\section{Introduction}
\label{secIntroduction}

The many-body problem in condensed matter physics has
been a subject of intensive research for several decades.
The first of these theories dates back to 1956 and is the so-called
Landau Fermi liquid theory. It provides a powerful phenomenological
description of three dimensional normal systems of 
interacting fermions \cite{Landau:57-1,Landau:59-1}. One of the central concepts of
this theory is the notion of quasiparticle excitations that
are in a one-to-one correspondence with excitations of the non-interacting
gas,
via the adiabatic continuation principle \cite{AndersonBN}.
The Landau-Fermi liquid theory consists in a description of the
many-body state in terms of a few phenomenological parameters
(the effective mass and the so-called Landau parameters)
that can be extracted from experimental measurements such as
specific heat, compressibility, susceptibility and sound
velocity. For a detailed description of this
theory, we refer the reader to the many books
on this subject (see for instance
\cite{AndersonBN, AbrikosovGorkov}). The phenomenological
theory of Landau was put on a microscopic
basis by means of quantum field theory methods
\cite{AbrikosovGorkov,Nozieres:62-1,Nozieres:62-2}.
The adiabatic continuation of
\cite{AndersonBN} requires the inverse particle
life time $1/\tau({\bf k})$ to be much smaller than
the quasiparticle energy $\epsilon({\bf k})$ so that
the switching rate of interactions $R$ can be safely chosen such that
\begin{equation}
\frac{1}{\tau({\bf k})} \ll R \ll \epsilon({\bf k})
\label{branchement}.
\end{equation}

In three dimensions, the quasiparticle life time can
be evaluated from the ``golden rule'' \cite{AndersonBN}:
$\tau({\bf k}) \sim \epsilon^{-2}({\bf k})$. Henceforth
(\ref{branchement}) is valid at arbitrarily low
energies. It means that the quasiparticle should not disappear before the
end of the interaction switching and that the interaction switching rate should 
be compatible with Heisenberg uncertainty relations.
This condition breaks down in one dimension (1D),
indicating that the 1D interacting fermionic system
is not of the Fermi liquid type. 

\medskip

The failure of the Fermi liquid description for 1D 
interacting fermions dates back to
the pioneer work by Dzialoshinskii and Larkin \cite{Larkin:74-1}.
These authors showed that this 1D system has {\it no}
Fermi surface in the sense that the quasiparticle
residue vanishes and the fermionic occupation
numbers have a power law behavior without discontinuity.
Other striking features of 1D interacting fermionic
systems were derived latter. For instance, with local
interactions, the Green function vanishes \cite{Haldane:81-1}, 
in an orthogonality catastrophe-like scenario \cite{Anderson:67-1}.
Besides this point, the correlations decay algebraically and are 
governed by anomalous exponents
\cite{Haldane:81-1,Luther:74-1,Luther:74-2,Haldane:79-1}.
Also, the spin~1/2 liquid shows spin and charge separation
\cite{Voit:93-1}: an additional electron in this system
dynamically breaks into a spin packet and
a charge packet that propagate at different speeds. See \cite{Schulz:91-1}
and references therein for a review on the Fermi liquid/non-Fermi liquid
problematics. 

\medskip

The first steps in deriving exactly soluble 1D fermionic systems
were performed by Tomonaga \cite{Tomonaga:50-1}, Luttinger
\cite{Luttinger:63-1} and Mattis and Lieb \cite{Mattis:65-1}.
Technically, a linearization of the free electrons
dispersion relation around the two Fermi points provides
a mapping to a two dimensional Dirac theory with interactions (the massless 
Thirring model) which can then
be bosonized. It was pointed out by Haldane \cite{Haldane:81-1}
that these low energy properties are the ones of a wide
class of 1D systems, for instance fermionic 1D liquids
or some massless regimes of spin-1/2 chains.
In fact, it is possible to identify precisely the low energy
physics of some models with the Luttinger Liquid (LL).
For instance, this was done in \cite{Haldane:80-1}
for the spin~$1/2$ Heisenberg chain, with a mapping
to a fermionic liquid via a Jordan-Wigner transformation;
and in \cite{Schulz:93-2} for the 1D Hubbard model.
In this sense, the LL
is analogous to the Landau theory in 3D systems since
it provides a solvable universal effective low energy theory
with a few parameters.

\medskip

Two dimensional 
Conformal Field Theories (CFT) provide another unifying
framework for one dimensional quantum systems. Originally
studied in depth as a generic description of effective field theories for
two dimensional critical points \cite{BPZ}, they proved to be
useful for studying one dimensional {\em quantum} systems. This
is nothing but the equivalence between a $D$-dimensional quantum 
system at finite temperature and the appropriate 
classical system in dimension $D+1$ \cite{FeynmanHibbs}. Of course, since
CFTs essentially describe ``massless'' field theories, 
they are appropriate for describing one-dimensional {\em gapless}
systems. Then, one could expect, on these general grounds, that 
the LL admits such an effective low-energy 
description.

Many references now exist on CFTs. The interested reader is
of course referred to original papers (\cite{BPZ} for CFT on a plane, and 
\cite{Cardy} for CFT on the torus) but also to various reviews
on this vast subject: for example, Les~Houches 
lectures \cite{Cardy3} and \cite{Ginsp}, Itzykson and Drouffe book 
\cite[Chapter 9]{ItzDrou}, 
or more recently Ketov book \cite{Ketov:CFT}. Of special interest for
the present paper, let us quote \cite{DVVc1} which 
treats in great details $c=1$ 
CFT. Obviously, 
this small conformal bibliography is far from being exhaustive but should help
the non specialist of this subject!

\medskip

The aim of this article is to present a comprehensive study
of the relation between LLs and CFTs. More
precisely, we identify the spinless LL in a finite size and at finite 
temperature but
without {\em Umklapp} terms with the field theory of a compact boson
on a two-dimensional torus
with an appropriate topological weight. Let us recall
that, as far as 1D quantum liquids are concerned,
the most general model consists of the full $g-$ology.
It is shown in \cite{Solyom:79-1} that carrying out the
poor man scaling analysis (originally applied by Anderson
to the Kondo problem \cite{Anderson:70-1}), there exists a
region of interactions such that only the $(g_2,g_4)$
interactions survive at the fixed point. In this paper, we
shall limit ourselves to $g_2$ and $g_4$ interactions. These 
couplings
are marginal in the sense of the renormalization group,
just like the Landau parameters for the three dimensional
interacting Fermi system \cite{Shankar:94-1}. We shall see
precisely how the effective CFT of
the LL changes when these $g_{2,4}$ parameters are
varied.

\medskip

Identification between the LLs and the compactified boson
was already proposed in \cite{Houches:Affleck}
for quantum spin chains, and in the late 80s, string theorists also studied the
correspondence between Dirac theory and the compactified boson
but in modular invariant sectors \cite{Moore:86-1}.
By contrast, we shall deal with a specific sector of the fermionic 
LL theory which is obviously not modular invariant. 
Therefore, bosonization formulae need a careful treatment of
boundary conditions and this is why a topological term is needed.
For the sake of pedagogy,  
the appropriate formulae will be derived here in a elementary way. 
Nevertheless, they have already appeared sporadically in the literature
(see \cite{Moore:87-1,Saleur:87-1}) but have not been yet
exploited within the context of one-dimensional condensed matter systems. 
Nevertheless, we have found a few papers, unfortunately not very
well-known, that contain explicit references
to these subtle questions: in \cite{Wu:95-1}, Wu and Yu characterize the 
Luttinger CFT starting from the study of an ideal excluson gas, following
Haldane's ideas on generalized statistics \cite{Haldane:91-1}. In 
\cite{Klassen:92-1}, the Luttinger CFT is extracted from locality considerations
on the operator product algebra and its specific modular properties and
duality properties are noticed. This paper also contains a deep discussion
of the relation between the Thirring model and the Sine-Gordon theory and provides
a complementary view to the present article. 

Focusing back on more formal matters, a deep and quite complete
reference on bosonization is \cite{Gawedski:93-1}. See also the work by 
Jolicoeur and Le~Guillou \cite{Leguil3} who also use functional methods 
in the thermodynamic and zero temperature limits, that
is to say on the plane. And of course, classical works by Coleman
\cite{Coleman:75-1} and Mandlestam \cite{Mandelstam:75-1} provide operator based 
approaches to the bosonization of the massive Thirring model.

To conclude this small review on bosonization approaches to the Luttinger liquid, let
us quote \cite{Affleck:97-1} which 
contains a numerical study of an XXZ
spin $1/2$ 
quantum chain away from half-filling on a circle with periodic or open 
spin boundary conditions. According to the authors of this paper, 
CFT predictions for the spectrum agree with
numerical computations based on a density matrix renormalization group method.

\medskip

In the present article we shall explicitly
show that {\it all} physical quantities of the LL that can
be calculated within the framework of the bosonization used by Haldane in
\cite{Haldane:81-1} can also be explicitly calculated by 
functional integral techniques. 
The only
restriction is that the interactions should be local
so that conformal invariance holds. As shown in  
\cite{Melin:94-1}, non-local interactions induce a breakdown of the
Fermi liquid by loss of coherence for the quasiparticle 
but they do not preserve conformal 
invariance because of their length scale. 
However, at sufficiently
large distance all non-singular interactions can be
viewed as local. 

\medskip

One of the motivations for developing finite size functional
bosonization of the LL, besides clarifying the
literature on the subject, is to deal with 
an external magnetic field and
with forward-scattering disorder.
By forward-scattering disorder, we mean a random potential
coupled to long-wavelength density fluctuations. The problem
of a disordered LL was already considered
in \cite{Giamarchi:88-1}, within the framework of renormalization.
Compared to the present article, the model of \cite{Giamarchi:88-1}
involves backscattering and presents a localized phase. But then, 
it is not possible to perform a direct
functional integral computation of correlation functions. Forward 
scattering 
does not lead to a localized phase but, in this case, averages over
disorder of all products of correlation functions can be explicitly computed. 

\medskip

The CFT formulation of the LL surprisingly provides a
simpler point of view on the effective theory for edge excitations
of a Fractional Quantum Hall (FQH) fluid. Wen originally pointed out that
edge excitations of a FQH fluid were described by a LL
\cite{Wen:90-1}. The relation has been studied in depth for
edge states of a disk-shaped FQH fluid (in this case, a 
{\em chiral} LL arises at the boundary). In \cite{Haldane:94-1},
Haldane and Rezayi studied numerically
cylindrical geometries that may lead to a FQH fluid. Henceforth, it
is interesting to look for a precise identification of the field
theory of edge excitations on the cylinder.
Here, we argue that
the {\em non-chiral} LL studied here corresponds to
edge states of a FQH fluid on a cylinder (see Laughlin's
paper \cite{Laughlin:82-1}). More precisely, Luttinger CFTs
corresponding to the $\nu =1/(2p+1)$ filling fraction 
are Rational Conformal Field Theories (RCFTs) \cite{Friedan1}. Their partition
functions appear to coincide with the ones recently obtained
by Cappelli and Zemba \cite{Cappelli:96-1}. While they were obtained
from their modular invariant properties by these authors, here they
arise from a one dimensional effective theory of interacting fermions. 
Of course, in FQH fluids, these interactions
are induced by the strong correlations of electrons in the Hall fluid. We
have pursued this analysis by introducing an edge charge on the boundary
in order to describe the tower of edge excitations above a given gaped 
bulk excitation.
The partition function of this ``twisted'' Luttinger CFT should reflect
how states above a given bulk excitation are organized.
Following Fisher and Stone \cite{Stone:94-1}, it is also interesting
to identify edge excitations that correspond to introducing a physical
electron on one of the two edges of the cylinder. 
We show that the associated
conformal fields generate an extended maximal symmetry algebra for 
Luttinger CFTs, thus providing a simple view on edge excitations of the FQH
fluid. 

\medskip

The present article is organized as follows: section 
\ref{secCFTLuttinger} 
recalls the basics of Haldane work and connect them to the massless
Thirring model on the torus. Section \ref{secGauge} studies the coupling to
a gauge field, and gives the proof of our bosonization formulae by
functional techniques. In section \ref{secDisorder}, we explain how forward
scattering can explicitly be treated. Section \ref{secQJCorrelations} 
and \ref{secVertex} are
devoted to the computation of correlation functions: first, charge and 
current densities are computed. This provides us with the charge and 
current response to an external electric potential and to a magnetic
flux. Next, correlation functions of all vertex operators are 
computed. Among these are the Luttinger fermion operators.

Having completed the study of the generic Luttinger CFT, we shall show 
how some of
these CFTs provide a description for edge excitations of a FQH fluid 
on a cylinder. Section \ref{secLaughlin} explains how Laughlin {\em gedanken}
experiments on charge transport in a Hall sample can be understood
within the framework of Luttinger CFTs. This enables relating the
interaction parameter of the Luttinger CFT with the filling fraction
of the Hall fluid. In section \ref{secRational}, 
Luttinger CFTs corresponding to
the Laughlin series are considered as RCFTs. 
This analysis is extended in section \ref{secTwisted} 
where we introduce a fractional
charge on the edge in order to study edge excitations above bulk 
excitations of the Hall fluid. The resulting theory is again a 
RCFT and is a twisted version of the Luttinger CFT. The physical meaning of
the maximal symmetry algebra of these RCFTs is discussed in
the light of Stone and Fisher's identification of edge fermions in the
FQH effect. 

Last, most of the technical details and computations are in the appendices. In
particular, appendix \ref{secElliptic} contains
all the necessary definitions
and technicalities on elliptic and modular functions needed in this paper. 
Appendix \ref{secFermiSea} deals with the simple case of non-relativistic
free fermions, and shows that these results are compatible with the CFTs ones.
Appendix \ref{secOperators} contains operator computations
{\em \`a la} Haldane in the interacting theory.

\newpage

\newcommand{\NN}{\Bbb{N}}
\newcommand{\RR}{\Bbb{R}}
\newcommand{\ZZ}{\Bbb{Z}}
\newcommand{\QQ}{\Bbb{Q}}
\newcommand{\CC}{\Bbb{C}}
\section{CFT of the LL}
\label{secCFTLuttinger}

In this section, we recall how the long range physics of the
LL is described by an effective conformal field
theory, namely the theory of a free bosonic field compactified
on a circle. 

\subsection{The Luttinger model}

The LL is a theory of one dimensional 
interacting fermions on a circle
of perimeter $L$. Haldane has given a detailed discussion of the low
energy physics of the LLs \cite{Haldane:81-1}. 

\medskip

For non relativistic fermions,
the quadratic dispersion relation $\epsilon(k)=k^{2}/2m$
of the one dimensional Fermi gas is linearized in the vicinity
of the Fermi surface: $\epsilon_{lin}(k) = v_F(\alpha k - k_F)$,
where $\alpha=+1,-1=R,L$ denote the right and left
Fermi points, and $k_F$ is the Fermi wave vector. The Fermi
velocity is nothing but $v_F= {d\epsilon\over dk}(k_F)$.
This approximation is valid provided
the interactions are not too strong compared to the Fermi velocity. We shall
come back to this point in section \ref{secInstability}.
The linear dispersion relation is then extrapolated
to arbitrary energies: an infinite number of fermions (Dirac sea)
is then present in the ground state.
One ends up with two linear branches.
The system obtained in this way should capture all the long
range physics of the initial model. Results 
obtained through the Bethe Ansatz technique are
compatible with this effective theory \cite{Houches:Affleck}.

\subsubsection{Free theory}
\label{secFree}

In the infrared limit, the Fermi sea is replaced by a Dirac sea,
and we choose $k_F=0$ in what follows ($k_F \ne 0$
would lead to minor corrections).
The free Hamiltonian is then:
\begin{equation}
\label{eqHamiltDirac}
H^{(0)} = {2\pi v_F\over L}\sum_{n,\alpha} \alpha n :\;c_{n,\alpha}^{\dagger}
\,c_{n,\alpha}:
,
\end{equation}
where the $c$s are fermionic creation and destruction operator at
momenta $2\pi \alpha n /L$. These operators satisfy 
canonical anti-commutation relations:
\begin{eqnarray}
\{ c_{n,\alpha},c_{m,\alpha'} \} & = &
\{ c_{n,\alpha}^{\dagger},c_{m,\alpha'}^{\dagger} \}=0\\
\{ c_{n,\alpha}^{\dagger},c_{m,\alpha'} \} & = & \delta_{\alpha,\alpha'}
\delta_{n,m}\,\bold{1}.
\end{eqnarray}
Fermionic
normal ordering is defined by $:c_k^{\dagger} c_k:=c_k^{\dagger} c_k$
if $k>0$ and $:c_k^{\dagger} c_k:=- c_k^{\dagger} c_k$ if $k<0$. 
We shall use fermionic fields defined by
\begin{eqnarray}
\psi _R^{\dagger}(\sigma) & = & {1\over \sqrt{L}}\sum_n c^{\dagger}_{n,R}\, 
e^{-2\pi in\sigma/L}\\
\psi _L^{\dagger}(\sigma) & = & {1\over \sqrt{L}}\sum_n c^{\dagger}_{n,L}\, 
e^{2\pi in\sigma/L}.
\end{eqnarray}
As recalled in appendix \ref{secAppendRegis}, these fields are the spatially
slow-varying components of the initial Fermi fields, the fast
varying part originating from oscillations at the Fermi wave-vector.
The charge and current density operators are defined by:
\begin{eqnarray}
\label{eqDefQJ}
\rho (\sigma) & = &  :\psi _R^{\dagger}(\sigma)\psi_R(\sigma)
+ \psi _L^{\dagger}(\sigma)\psi_L(\sigma):\\
j(\sigma) & = & v_F\, :\psi _R^{\dagger}(\sigma)\psi_R(\sigma)
- \psi _L^{\dagger}(\sigma)\psi_L(\sigma):.
\end{eqnarray}
The interaction energy between the electrons and an external electrostatic or
vector potential is
linear:
\begin{equation}
\label{eqHamiltExtern}
H_{\mathrm{ext}} = \int_0^L d\sigma\, \left(V(\sigma) \rho(\sigma)
-A(\sigma)j(\sigma)\right).
\end{equation}
The central tool in the operator formulation of the model 
is the current algebra. Let us 
define the following operators ($n\in \Bbb{Z}$):
\begin{eqnarray}
\label{eqFourrier1}
{\cal J}_n & = & \int _0^L {\cal J}(\sigma)
\,e^{-2\pi i n\sigma /L} \,d\sigma\\
\label{eqFourrier2}
\overline{{\cal J}}_n & = & \int _0^L  \overline{\cal J}(\sigma)
\,e^{2\pi i n\sigma /L} \,d\sigma
,
\end{eqnarray}
with ${\cal J}(\sigma) = :\psi _R^{\dagger}(\sigma)\psi_R(\sigma):$
and $\overline{\cal J}(\sigma) = :\psi _L^{\dagger}(\sigma)\psi_L(\sigma):$
.
One can show that the modes (\ref{eqFourrier1}) and
(\ref{eqFourrier2}) satisfy the following commutation 
relations
\begin{equation}
\label{eqAffine}
[{\cal J}_n,{\cal J}_m]=n\,\delta _{n,-m}\,\bold{1}
,
\end{equation}
this anomalous commutator being central to mode bosonization. It can
be derived either from a point splitting regularization in real
space (see for instance \cite{Fradkin1}) 
or directly in Fourrier space (see for instance \cite{Haldane:81-1}).
In both cases, the anomaly originates from normal ordering of operators
with respect to the Dirac sea.
Moreover, the ${\cal J}_n$s commute with ${\cal \overline{J}}_m$s which 
also satisfy commutation relations (\ref{eqAffine}).
These relations define the infinite dimensional Heisenberg
algebra \cite{Kac}, also called the affine $U(1)$ algebra and denoted by 
$\widehat{U(1)}$. The symmetry of the model
is therefore given by two commuting copies of the Heisenberg algebra\footnote{This
factorization between a left and a right symmetry algebra is quite common in 
CFT.}.
The vacuum state of Dirac's 
theory $|0\rangle$ satisfies the highest weight conditions which implement 
Pauli's exclusion principle:
\begin{equation}
\label{eqHWcondition}
\forall n >0,\ {\cal J}_n|0\rangle = \overline{{\cal J}}_n|0\rangle =0.
\end{equation}
Finally, as shown by Haldane \cite{Haldane:81-1}, 
the free Hamiltonian (\ref{eqHamiltDirac})
can be expressed in terms of the currents:
\begin{equation}
\label{eqFreeHam}
H^{(0)}= {\pi v_F\over L}\, \sum _{n\in \Bbb{Z}}(
{\cal J}_n{\cal J}_{-n}+\overline{{\cal J}}_n\overline{{\cal J}}_{-n})
= \pi v_F \int_0^L \left( {\cal J}^2(\sigma) + \overline{\cal J}^2
(\sigma) \right)
.
\end{equation}
This bosonized form of the free Hamiltonian is the key point
of mode bosonization.

\subsubsection{Interacting theory}
\label{secInteractions}

In the following, local interactions between
fermions are assumed, although Haldane has given a solution for non local
interactions satisfying certain conditions. Short range 
interactions fall into this
class, and the local case should be considered as a limiting case of his
analysis. 
The local interaction term of the spinless LL is
\begin{equation}
H_{\mathrm{int}} = \frac{\pi}{L} \sum_{\alpha,\alpha'}
\sum_{k,k'} (g_{4} \delta_{\alpha,\alpha'}
+ g_{2} \delta_{\alpha,-\alpha'})\;
c_{k+q,\alpha}^{\dagger} c_{k,\alpha}
c_{k'-q,\alpha'}^{\dagger} c_{k',\alpha'}
.
\end{equation}
The total Hamiltonian is diagonalized via a bosonization
procedure \cite{Haldane:81-1}, 
using the current modes introduced in (\ref{eqFourrier1}) and 
(\ref{eqFourrier2}). The interactions are first
re-written in terms of the currents:
\begin{eqnarray}
\label{eqIntHam}
H_{\mathrm{int}} &=& {\pi \over L}\sum_{n\in \Bbb{Z}}\left(
2 g_2\, {\cal J}_n\overline{{\cal J}}_n + 
g_4({\cal J}_n{\cal J}_{-n}+
\overline{{\cal J}}_n\overline{{\cal J}}_{-n})
\right)\\
&=& 2 \pi g_2 \int_0^L  {\cal J}(\sigma) \overline{\cal J}(\sigma)
+ \pi g_4 \int_0^L \left( {\cal J}^2(\sigma)
+ \overline{\cal J}^2(\sigma) \right) d \sigma
.
\end{eqnarray}
The Bogoliubov transformation then consists in 
a redefinition of current algebra's generators. 
At the appropriate angle, it diagonalizes the Hamiltonian.
More precisely, let us define
\begin{eqnarray}
\label{Bogo-trans-I}
J_n & = & \cosh{(\varphi)}\, {\cal J}_n -\sinh{(\varphi)}\, 
\overline{{\cal J}}_{-n}\\
\overline{J}_{n}
& = & \cosh{(\varphi)}\, 
\overline{{\cal J}}_n -\sinh{(\varphi)}\, {\cal J}_{-n}
\label{Bogo-trans-II}
,
\end{eqnarray}
where
\begin{equation}
\label{eqDefPhi}
\tanh{(2\varphi)}=-{g_2\over v_F+g_4}
.
\end{equation}
In the interacting theory, these are the new symmetry generators.
They also
satisfy the commutation relations (\ref{eqAffine}). 
The ${\cal J}$s and $\overline{{\cal J}}$s should be understood as bare 
symmetry generators, whereas the $J$s and $\overline{J}$s are renormalized 
operators that implement the infinite dimensional affine symmetry in the
interacting theory. 
It is useful to Fourrier transform 
Laurent's modes $(J_n)_n$ and
$(\overline{J}_n)_n$ to
$J(\sigma)$ and $\overline{J}(\sigma)$. Then we have
\begin{equation}
\left(
\begin{array}{c}
J(\sigma )\\
\overline{J}(\sigma)
\end{array}
\right) = 
\left(
\begin{array}{cc}
\cosh{(\varphi)} & -\sinh{(\varphi)}\\
-\sinh{(\varphi)} & \cosh{(\varphi)}
\end{array}
\right)
\times
\left(
\begin{array}{c}
{\cal J}(\sigma)\\
\overline{\cal J}(\sigma)
\end{array}
\right)
.
\end{equation}
Using these notations, the free Hamiltonian plus
the interaction term is nothing but the one of a free bosonic
field theory:
\begin{equation}
\label{Hamil-J-inter}
H_{\mathrm{tot}} = \frac{\pi v_S}{L}
\sum _{n\in \Bbb{Z}}(
J_n J_{-n}+\overline{J}_n \overline{J}_{-n})
= \pi v_S \int_0^L \left( {J}^2(\sigma) + \overline{J}^2
(\sigma) \right)
.
\end{equation}
The density operator is {\em not} modified 
by switching on interactions since
it consists in counting particles:
$$\rho(\sigma)= :\,\psi^{\dagger}_R(\sigma)\psi_R(\sigma)\,:\, +
:\,\psi^{\dagger}_L(\sigma)\psi_L(\sigma)\,:\, = \alpha ^{-1/2}\,
(J(\sigma)+\overline{J}(\sigma)),$$
where we introduce the interaction parameter
$\alpha = e^{-2\varphi}$. To obtain the current, one should, as 
Haldane suggested in \cite{Haldane:81-1}, use charge conservation. Using the
total interacting Hamiltonian,
and the commutation relations of the $J(\sigma)$ and
$\overline{J}(\sigma)$ operators,
one easily finds the time evolution of these 
operators in the Heisenberg representation\footnote{The
the propagation of a wave packet created at time zero by 
$\psi _R^{\dagger}[\varphi(\sigma)](t=0)=
\int \varphi(\sigma)\psi_R^{\dagger}(\sigma,0) d\sigma $ is given by
$e^{-itH/\hbar }\,
\psi _R^{\dagger}[\varphi(\sigma)]\,e^{itH/\hbar}=
\psi _R^{\dagger}[f(\sigma-v_St)]$. As should be, a wave-packet
of right movers propagate to the right.}:
\begin{eqnarray}
J(\sigma,t) & = & {1\over L}\sum _{n\in \Bbb{Z}}
J_n\, e^{2\pi in(\sigma-v_St)/L}\\
\overline{J}(\sigma,t) & =  & {1\over L}\sum _{n\in \Bbb{Z}}
\overline{J}_n\, e^{-2\pi in(\sigma+v_St)/L}
.
\end{eqnarray}
Interactions are taken into account through the modification of the Fermi 
velocity $v_F\mapsto v_S$, with
\begin{eqnarray}
v_S &=& \sqrt{(v_F+g_4)^{2}- g_2^{2}}\label{eqVS}\\
\label{Charge_Velocity}
v_N &=& v_F + g_2 + g_4\\
v_J &=& v_F - g_2 + g_4. \label{eqVJ}
\end{eqnarray}
Then $\alpha=\sqrt{v_N/v_J}$.  To summarize, we have:
\begin{eqnarray}
\label{eqDensityCharge}
\rho(\sigma) & = & \alpha ^{-1/2}\,(J(\sigma)+\overline{J}(\sigma))\\
\label{eqDensityCurrent}
j(\sigma) & = & v_S\, \alpha ^{-1/2}\, (J(\sigma)-\overline{J}(\sigma))
.
\end{eqnarray}
Of course, these computations assume that the interactions
are local, {\it ie}, $\varphi$ does not depend on $n$. 
Since we are interested in the low energy properties of the model,
we may assume that this hypothesis holds, at least for all modes that
are likely to be excited 
within the appropriate temperature range. 

\medskip

The vacuum of the interacting theory
$|O_{\mathrm{Lutt}}\rangle$ also satisfies the highest weight 
conditions (\ref{eqHWcondition}) but for the 
$J_n$ and $\overline{J}_n$ operators 
(see \cite{Kac} for more information on
representation theory of infinite dimensional Lie algebras). 
As is well known, this state is orthogonal to the
original vacuum state. For non local interactions, 
the $\varphi$ angle depends on $n$. For the sake of regularization, let us
reestablish the
dependence over $n$ of $\varphi$\footnote{One should 
remember that there is always
some ultra violet cut-off that truncates the sum over $n$.}:
\begin{equation}
|O_{\mathrm{Lutt}}\rangle  =  {\cal Z}^{-1/2}\; \exp{\left(
- \sum _{n=1}^{+\infty} {\tanh{(\varphi_n)}\over n} {\cal J}_{-n}\,
\overline{{\cal J}}_{-n}\right)}\,|0\rangle,
\end{equation}
the prefactor ${\cal Z}$ being
\begin{equation}
{\cal Z} = \prod _{n=1}^{+\infty}\cosh^2{(\varphi_n)}
.
\end{equation}
The ${\cal Z}$ prefactor diverges in the thermodynamic limit where
more and more modes accumulate below some ultra violet cut-off.
It means that the vacuum of the interacting theory is 
orthogonal to the free theory's one. However, 
let us stress that the Hilbert space is still a representation of a
$\widehat{U(1)}_R\times \widehat{U(1)}_L$ algebra generated by the
``renormalized'' currents $J_n$ and $\overline{J}_n$. Turning on the 
interactions
preserves the symmetry of the Luttinger effective theory.

\medskip

The diagonalized Hamiltonian is then
\begin{equation}
\label{eqHamilt}
H_{\mathrm{Lutt}} = E_0 +
{2\pi v_S\over L} \sum_{n>0} n(N_{n,R}+N_{n,L}) +
\frac{\pi v_S}{2L} (\alpha N^{2} + \frac{1}{\alpha} J^{2}),
\end{equation}
where $E_0$ denotes the energy of the vacuum.
In (\ref{eqHamilt}), $N_{n,R}$ and $N_{n,L}$ denote
occupation numbers of bosonic modes:
\begin{equation}
\forall n>0,\ N_{R,n}={1\over n}\,J_{-n}J_n,
\quad N_{n,L}={1\over n}\,\overline{J}_{-n}\overline{J}_n
.
\end{equation}
The operators $N$ and $J$ are the
charge and current numbers in the $q=0$ mode: 
$N=J_0+\overline{J}_0$,
and $J=J_0-\overline{J}_0$.

\medskip

Let us notice that the physics of the LL
can be characterized through
two parameters: a renormalized Fermi velocity $v_S$ and a 
dimensionless interaction
parameter $\alpha$. At the Fermi liquid point, $v_S=v_F$ and $\alpha =1$.


\subsection{Field theoretical description}

The aim of this section is to introduce the field theoretical approach to
the LL. As explained before, all physical quantities of this 
quantum system at finite temperature can be derived from a Euclidian quantum
field theory in two dimensions, which will be determined in the following sections.

\subsubsection{CFT in two dimensions}

The basic idea consists in representing a one dimensional quantum system 
at finite temperature by a two dimensional statistical field theory. This
old idea \cite{FeynmanHibbs} 
proves to be useful in the present context since the LL
has gapless excitations. Therefore, one would naively expect
the corresponding two dimensional statistical theory to be scale 
invariant, and even
more, to be conformally invariant\footnote{It is important not to forget that these
arguments are quite heuristic. Their relevance in the present context
relies on the fact that we can recover Haldane's spectrum and other known low
energy properties of LLs using CFT.}. Before
studying in detail the effective CFT that describes the
low energy physics of LLs, let us recall a few 
basic facts about CFT \cite{BPZ,Ginsp,Cardy3,Ketov:CFT}.

\medskip

As usual in CFT,
the objects we shall be interested in are partition functions with 
a special twist: besides $\exp{(-\beta H)}$, a 
translation operator in the spatial 
direction $\exp{(i\theta P)}$ is introduced.
More precisely, we shall compute partition functions of the type
\begin{equation}
Z = \mathrm{Tr} \left( e^{- \beta H}
e^{i \theta P} \right).
\end{equation}
As explained by Cardy in \cite{Cardy}, this corresponds to computing functional
integrals on a torus. 
In our context, the relevant modular parameter $\tau $ of this torus is
\begin{equation}
\label{definitiondetau}
\tau = {\theta\over L} + i\, {\beta v_S\over L}
.
\end{equation}
As is usually done in CFT \cite{Cardy}, 
the Hamiltonian and momentum operators can be expressed in terms of Virasoro 
generators $L_0$ and $\overline{L}_0$ by
\begin{eqnarray}
H & = & {2\pi v_S\over L}\,\left(L_0+\overline{L}_0-{c\over 12}\right)\\
P & = & {2\pi v_S\over L}\,\left(L_0-\overline{L}_0\right)
.
\end{eqnarray}
The number $c$ (called the central charge) corresponds to
the ground state energy.
Results are usually expressed using the complex parameters
$q=\exp{(2 i \pi \tau)}$ 
and 
$\overline{q}=\exp{(-2 i \pi
\overline{\tau})}$. For example, the partition function
\begin{equation}
Z = \mathrm{Tr} \left( q^{L_0 - \frac{1}{24}}
\; \overline{q}^{\bar{L}_0 - \frac{1}{24}} \right)
\end{equation}
can be expressed as a double Puiseux expansion in $q$ and 
$\bar{q}$. Its coefficients are nothing but the degeneracy of states
at a given energy and momentum.

\subsubsection{Free fermions on a torus}

The Hamiltonian (\ref{eqHamiltDirac}) is nothing but the one arising from 
Dirac theory of free fermions. Let us introduce 
the Minkowskian gamma matrices: $\gamma _0= \sigma_x $ and
$\gamma _1=-i\,\sigma_y$. The Dirac Lagrangian density is nothing 
but 
\begin{equation}
{\cal L}= i\, \bar{\psi}\gamma ^{\mu}\partial _{\mu}\psi 
,
\end{equation}
where $\bar{\psi} = \psi ^{\dagger}\gamma ^0$.
We shall be interested in the thermodynamics of the LL.
Besides the operator approach in which
the density operator $\exp{(-\beta H)}$ plays a central role, we
shall use mainly 
the functional approach which involves functional Berezin 
integrals of the form: 
$$\int {\cal D}[\bar{\psi},\psi]\exp{\left(
-\int \bar{\psi}\gamma _E^{\mu}
\partial _{\mu}\psi \right)}, $$
where $(\gamma _E^{\mu})_{\mu}$ denote Euclidian gamma matrices
($\gamma _{E}^0=\sigma_x$ and $\gamma _E^1=\sigma_y$).

In the fermionic integral representing the partition 
function of our system, the Euclidean action is an integral 
over a torus $S_1\times S_1$. The spatial direction corresponds to
the first circle and the other one is associated with the imaginary time.
Obviously,
boundary conditions for fermionic fields on the torus play an 
important role which we shall discuss now.

\paragraph{Spatial boundary conditions}
In the space direction, fermions may be periodic ($n \in \ZZ$)
or anti-periodic ($n \in \ZZ+1/2$) \cite{sherk}. In the latter case, the
ground state is unique. In the former case, it has a four-fold
degeneracy since the occupation of the fermionic
levels located on the Fermi surface does not change the
total energy.
In \cite{Haldane:81-1},
the Fermi wave vector is an half-integer. Henceforth, 
we choose fermions in the anti-periodic sector.

Nevertheless, it is interesting to consider both cases. These
two sectors, called ``periodic'' and ``anti-periodic''
have their own Hilbert space denoted respectively by ${\cal H}_{P,A}$.
As we shall see later, other boundary conditions may be obtained
by tuning a magnetic flux through the circle.

\paragraph{Imaginary time boundary conditions}
In the functional integral formalism, it is well known 
(see \cite[chapter 9]{ItzZuber}) that 
computing the trace of an operator acting on a single
fermionic mode requires the evaluation of a Berezin integral for
its representing kernel with anti-periodic boundary conditions.
Remember also that
periodic boundary conditions amounts to 
introducing a $(-1)^F$ factor, with $F$ the fermion number:
\begin{eqnarray}
\mathrm{Tr}\,(A) & = & \int d\bar{\xi}d\xi\, e^{\xi\,\bar{\xi}}
A(-\bar{\xi},\xi)\\
\mathrm{Tr}\,((-1)^FA) & = & \int d\bar{\xi}d\xi\, e^{\xi\,\bar{\xi}}
A(\bar{\xi},\xi)
.
\end{eqnarray}
Therefore, in our setting, we end up with four different boundary
conditions for fermions on the torus: ``PP'', ``PA'', ``AP'' and ``AA''
where the first letter gives the periodicity in the spatial direction
and the second one in the imaginary time direction.
They are usually referred to as ``spin sectors''.
Finally we have:
\begin{eqnarray}
Z_{AP} &=&\mbox{Tr}_A \left( (-1)^{F} e^{-\beta H_A}\right) = \int _{AP}
 {\cal D}[\bar{\psi},\psi]\exp{\left(
-\int \bar{\psi}\gamma _E^{\mu}
\partial _{\mu}\psi \right)}
\label{ZAP}\\
Z_{PP} &=& \mbox{Tr}_P \left((-1)^{F} e^{-\beta H_P}\right)=\int _{PP}
 {\cal D}[\bar{\psi},\psi]\exp{\left(
-\int \bar{\psi}\gamma _E^{\mu}
\partial _{\mu}\psi \right)}
\label{ZPP}\\
Z_{PA} &=& \mbox{Tr}_P \left( e^{-\beta H_P}\right) = \int _{PA}
 {\cal D}[\bar{\psi},\psi]\exp{\left(
-\int \bar{\psi}\gamma _E^{\mu}
\partial _{\mu}\psi \right)}
\label{ZPA}\\
Z_{AA} &=& \mbox{Tr}_A \left( e^{-\beta H_A}\right) = \int _{AA}
 {\cal D}[\bar{\psi},\psi]\exp{\left(
-\int \bar{\psi}\gamma _E^{\mu}
\partial _{\mu}\psi \right)}
\label{ZAA}
.
\end{eqnarray}
We are indeed studying the two dimensional massless Dirac theory, which is
a CFT. But,  contrarily to the usual stringy point of view, we
consider here only one specific spin sector instead of summing over 
all of them. 

\subsubsection{The massless Thirring model}

The infrared behavior is described by an interacting theory as expressed
by the sum of Hamiltonians (\ref{eqHamiltDirac}) and (\ref{eqIntHam}). 
In Dirac's theory, the conserved current is given by
$J^{\mu}=\bar{\psi}\gamma^{\mu}\psi $. The interaction term 
of Haldane's Hamiltonian
can be expressed in terms of this current 
and therefore corresponds, in a Lagrangian formulation, to a quadratic 
current interaction of generic Thirring form:
\begin{equation}
\label{eqLagrangeJJ}
S_{\mathrm{int}}[\bar{\psi},\psi]=-{\kappa\over 2}\, \int  
(\bar{\psi}\gamma^{\mu}\psi)^2.
\end{equation}
This suggests that
the Euclidian quantum field theory to be considered in this paper therefore 
describes the massless Thirring model \cite{Thirring:58-1} in a specific spin sector,
which we call the {\em Luttinger CFT}. Its
Lagrangian is given by:
\begin{equation}
\label{eqLagrangeThirring}
{\cal L}= i\bar{\psi}\gamma ^{\mu}\partial _{\mu}\psi -{\kappa\over 2}\,  
(\bar{\psi}\gamma^{\mu}\psi)^2.
\end{equation}
Most of the analysis carried in sections 
\ref{secQJCorrelations} and \ref{secVertex} will consist in
performing explicit computations of correlation functions using 
a functional bosonization procedure. 
Let us point out that here, the relevant parameters are the Fermi velocity and 
the coupling constant $\kappa$. These are to be related with $v_S$ and 
$\alpha $ in Haldane's analysis. In order to obtain a perfect dictionary
between the Haldane's approach, the bosonization computation, and Thirring's 
model, one should of course be able to perform direct computations of
correlation functions in the massless Thirring model\footnote{This was
pointed out to us by R.~Stora.}. 
Reference papers on this particular point are \cite{Klaiber:67-1} and \cite{Glaser:58-1}.
The main point of these computations rely on the proper definition
for the current composite operators $\bar{\psi}\gamma^{\mu}\psi$ and 
$\bar{\psi}\gamma^{\mu}\gamma^5\psi$. However, since we know 
the spectra and the asymptotics of fermion correlators in Haldane and -- as we shall see later --
in the bosonic approach, all is needed here are relations (\ref{eqVS}) to (\ref{eqVJ}) and 
$\alpha$'s definition. It would however be interesting to clarify the relation between
these three approaches.

\subsubsection{Explicit computations in the operator formalism}

It is quite interesting to compute explicitly the partition functions
(\ref{ZAP}) to (\ref{ZPP}) 
of the Dirac theory, using the operator formalism.  Here, computations are given 
at zero magnetic field and chemical potential. But, anticipating
our needs, let us notice the interested reader that a more
complete discussion can be found in appendix \ref{secDirac}.

\paragraph{Definition of the operators}
In order to symmetrize both branches, we note
$b_{n,R}^{\dagger}=c_{n,R}^{\dagger}$ and 
$b_{n,L}^{\dagger}=c_{-n,L}^{\dagger}$,
and $N_{n,\alpha}=b_{n,\alpha}^{\dagger}b_{n,\alpha}$ if $n>0$,
$N_{n,\alpha}=b_{n,\alpha} b_{n,\alpha}^{\dagger}$ if $n<0$.
Then, using a zeta function renormalization to get rid of infinities, we
obtain:
\begin{equation}
H = {2\pi v_F\over L}\; \sum_n |n| \left( N_{n,R}+N_{n,L} \right)
-2\, \zeta _{a}(-1),
\end{equation}
where $a=0$ corresponds to the periodic sector and $a=-1/2$ to the 
anti-periodic one.
Here, 
$\zeta _{a}$ denotes the analytic continuation of
\begin{equation}
\label{zetadef}
\zeta _a(s)=\sum _{n=1}^{+\infty}{1\over (n+a)^s}.
\end{equation}
The right hand side of equation (\ref{zetadef}) 
is defined for $\Re{(s)}$ sufficiently large. 
We recall that \cite{Cartier:89-Houches}:
\begin{equation}
\label{zeta0}
\zeta_0 (-1) = -\frac{1}{12}\quad \mathrm{and} \quad
\zeta_{1/2} (-1) = 
\frac{1}{24}.
\end{equation}
The Hamiltonian is now completely determined in each sector:
\begin{eqnarray}
H_P & = & {2\pi v_F\over L}\; 
\sum_{n \in \Bbb{Z}}|n| \left( N_{n,R}+N_{n,L}
\right) + \frac{1}{6}\\
H_A & = & {2\pi v_F\over L}\; 
\sum_{n \in \Bbb{Z}+1/2} |n| \left( N_{n,R} + N_{n,L} \right)
- \frac{1}{12}
.
\end{eqnarray}
The momentum operator does not depend on the
regularization and is given by:
\begin{equation}
P = {2\pi v_F\over L}\; \sum_n |n| \left( N_{n,R} - N_{n,L} \right)
.
\end{equation}


\paragraph{Explicit expressions of the partition functions}

In the {\em free} case, all traces can be explicitly computed to obtain
\begin{eqnarray}
\label{ZAA3}
Z_{AA} & = & (q \overline{q})^{-1/24}
\left| \prod_{n=0}^{+\infty} (1 + q^{n+1/2})^{2} \right|^{2}\\
\label{ZPA3}
Z_{PA} & = & 4\, (q \overline{q})^{-1/24}
\left| q^{1/8} \prod_{n=1}^{+ \infty} (1+q^{n})^{2} \right|^{2}\\
\label{ZAP3}
Z_{AP} & = & (q \overline{q})^{-1/24}
\left| \prod_{n=0}^{+ \infty} (1-q^{n+1/2})^{2} \right|^{2}\\
Z_{PP} & = & 0
.
\label{ZPP2}
\end{eqnarray}
The vanishing of $Z_{PP}$ is due to the fact that, in the doubly
periodic sector, the occupation number of a state lying at the Fermi
surface does not change the energy nor the momentum (since $k_F=0$).
In the functional integral language, it
corresponds to the existence of fermionic zero modes in the 
``PP'' spin sector.

\medskip

Let us conclude that these partition functions can also be obtained by a
zeta regularization for the determinant of the Dirac operator on
the torus in each spin sector \cite{Ginsp}.

\paragraph{Partition functions of the LL}
Let us now infer from Haldane's spectrum the partition function
of the interacting Luttinger theory.
Using the $\zeta$ renormalization prescription, we obtain
\begin{equation}
E_0 = -{\pi v_S\over 6L} 
.
\end{equation}
Let us introduce the Dedekind function
\begin{equation}
\eta(q) = q^{\frac{1}{24}} \prod_{n=1}^{+ \infty}
(1-q^{n}).
\label{Dedekind}
\end{equation}
Then, the partition function of the LL is given by:
\begin{equation}
\label{eqZLutt}
Z_{AA} = \frac{1}{\left|\eta(q) \right|^{2}}
\left( \sum_{\shortstack{ \scriptsize $(n,m) \in \ZZ^{2}$\\
\scriptsize $m \equiv 0 (2)$}} +
\sum_{\shortstack{ \scriptsize $(n,m) \in (\ZZ+1/2) \times \ZZ$\\
\scriptsize $m \equiv 1 (2)$}} \right)
q^{\frac{1}{2}\left(n \sqrt{\alpha} + \frac{m}{2 \sqrt{\alpha}}
\right)^{2}}
\overline{q}^{\frac{1}{2}\left(n \sqrt{\alpha} - \frac{m}{2 \sqrt{\alpha}}
\right)^{2}}
,
\label{ZAA2}
\end{equation}
where we have simply replaced the sum over $J$ and $N$
of same parity by a sum over 
$(n,m)$ where $N=2n+m$ and $J=2n-m$. Our problem is now to find a
convenient way to describe the physics of the LL. More
precisely, we look for an effective field theory that would 
reproduce the partition function (\ref{ZAA2})
by functional techniques.


\subsection{Bosonic description of the Luttinger effective theory}

We explain here how to recover the partition function of the 
{\em interacting} Luttinger theory from a {\em free} bosonic
theory. All subtleties lye in the boundary conditions of the
bosonic field and we shall focus on this point. First
of all, we recall how to compute some partition functions
in the compactified bosonic theory, and then, the
identification of the ``AA'' fermionic partition function
will be performed. 
A standard argument of modular covariance enables to find
explicit expressions in two of the remaining sectors. 

\subsubsection{The compactified boson}

In this section, we give explicit expressions of partition functions
of the theory of a free boson, compactified on a circle of
radius $R$ \cite{DVVc1}. 
By this, we mean that
$\varphi$ and $\varphi+2\pi R$ are identified. We work on the
torus ${\bf T}_{\Gamma} = \CC/\Gamma$, where $\Gamma$ denotes the
lattice $\Gamma=\omega_1 \ZZ + \omega_2 \ZZ$. 
The modular parameter $\tau =\omega _2/\omega _1$ is assumed to have
a positive imaginary part.
The action
reads
\begin{equation}
\label{eqactionB}
S[\varphi] = \frac{g}{2\pi} \int_{{\bf T}_{\Gamma}}
|\nabla \varphi|^{2}\, .
\end{equation}
We consider the possible boundary conditions
$[\epsilon,\epsilon']$, where $\epsilon $  and
$\epsilon ' $ are given modulo $1$, defined by the monodromy conditions
\begin{equation}
\label{eqmonod}
\begin{cases}
\varphi(z+\omega_1)  \equiv  \varphi(z) + 2 \pi R \epsilon
\pmod{2 \pi R}\\
\varphi(z+\omega_2)  \equiv  \varphi(z) + 2 \pi R \epsilon'
\pmod{2 \pi R}
.
\end{cases}
\end{equation}
The partition function
\begin{equation}
\label{eqZtarget}
Z_{\epsilon,\epsilon'} = \int_{[\epsilon,\epsilon']}
{\cal D}[\varphi]\, e^{-S[\varphi]}
\end{equation}
is obviously Gaussian. It can be explicitly computed
using a quadratic expansion around classical solutions 
(also called instantons) 
compatible with 
the imposed boundary conditions. Such a computation is straightforward
and well known. Details are briefly recalled in appendices
\ref{secNormalisation} and \ref{secCalculs}.


\paragraph{Expression of the bosonic partition functions}
As shown in appendix \ref{secCalculs}, the partition
function of the compact boson is
\begin{equation}
Z_{[\epsilon,\epsilon']} (g R^{2})= \frac{1}{| \eta(q) |^{2}}
\sum_{{m \in \Bbb{Z} \atop n
\in \Bbb{Z}}}
e^{-2 i \pi m \epsilon'} q^{\frac{1}{2} p_{n+\epsilon,m}^{2}}
\overline{q}^{\frac{1}{2} \overline{p}_{n+\epsilon,m}^{2}},
\label{eqZepsilon}
\end{equation}
where we have introduced the momenta:
\begin{eqnarray}
\label{p}
p_{n,m} & = & n \sqrt{\alpha} + \frac{m}{2\sqrt{\alpha}}\\
\overline{p}_{n,m} & = & n\sqrt{\alpha} - \frac{m}{2\sqrt{\alpha}}.
\label{pbar}
\end{eqnarray}
Expression (\ref{eqZepsilon}) is
the building block of all the computations done in
this paper. As we shall see later, it will be used in the computation of 
correlation functions of the so-called vertex operators, among which renormalized 
fermions.


\subsubsection{Identification of the LL
to the compact boson}

String theorists are usually interested in the modular
invariant partition function \cite{Ginsp}
\begin{equation}
Z_{\mathrm{Dirac}} = \frac{1}{2} \left( Z_{AA}+Z_{AP}+Z_{PA}+Z_{PP} \right).
\end{equation}
This sum can be readily evaluated by the usual techniques \cite{Ginsp} 
and is equal to
\begin{equation}
Z_{\mathrm{Dirac}} = \frac{1}{\left| \eta(q) \right|^{2}}
\sum_{(m,n) \in \ZZ^{2}} q^{\frac{1}{8}(n+2m)^{2}}
\overline{q}^{\frac{1}{8}
(n-2m)^{2}}.
\end{equation}
Comparing to (\ref{eqZepsilon}) gives \cite{DVVc1}
\begin{equation}
Z_{\mathrm{Dirac}} = Z_{[0,0]}(1).
\end{equation}
Dirac theory thus coincides with the theory of a free compactified boson for 
$g R^{2}=1$. We now wish to express
each of the partition functions (\ref{ZAA3}), (\ref{ZPA3})
and (\ref{ZAP3}) in terms of bosonic functional integrals 
with $gR^{2}=1$. These formulae already
exist in the literature, at least in
\cite{Saleur:87-1} and \cite{Moore:87-1}. A very detailed study of bosonization in
a general charge and background metric may be found in \cite{Gawedski:93-1}.
Nevertheless these results
do not seem to be widely known and it is interesting to derive them
in an elementary way.

\paragraph{Bosonic expressions of the free partition functions}

Jacobi triple product identity reads \cite{whittaker-watson}:
\begin{equation}
\sum_{n=-\infty}^{+\infty} y^{n} q^{n^{2}/2} =
\prod_{n=1}^{+\infty}(1-q^{n}) \prod_{n=0}^{+\infty}
(1+y q^{n+1/2})(1+ y^{-1} q^{n+1/2}).
\end{equation}
Specializing to $y=1$,$y=q^{1/2}$ and $y=-1$ and combining with
(\ref{ZAA3}), (\ref{ZPA3}) and (\ref{ZAP3}) leads to
expressions that, 
as we shall explain, can be recovered from a bosonic theory:
\begin{eqnarray}
\label{Z_{AA}}
Z_{AA} &=& \frac{1}{|\eta(q)|^{2}} \left|
\sum_{n=-\infty}^{+ \infty} q^{\frac{1}
{2}n^{2}} \right|^{2}\\
\label{Z_{PA}}
Z_{PA} &=& \frac{1}{|\eta(q)|^{2}} \left|
\sum_{n=-\infty}^{+ \infty} q^{\frac{1}{2}
(n+\frac{1}{2})^{2}} \right|^{2}\\
\label{Z_{AP}}
Z_{AP} &=& \frac{1}{|\eta(q)|^{2}} \left|
\sum_{n=-\infty}^{+ \infty} (-1)^{n} q^{\frac{1}
{2}n^{2}} \right|^{2}.
\end{eqnarray}

\paragraph{Functional integral expressions}

Now, these bosonic partition functions will be identified with
bosonic functional integrals of the type (\ref{eqZepsilon}).
For this purpose, let us write explicitly (\ref{eqZepsilon}) in the
four sectors:
\begin{eqnarray}
Z_{[0,0]}(1) &=& \frac{1}{\left| \eta(q) \right|^{2}}
\sum_{(n,m) \in \ZZ^{2}} q^{\frac{1}{2}(n+\frac{m}{2})^{2}}
\overline{q}^{\frac{1}{2} (n-\frac{m}{2})^{2}}\\
Z_{[0,\frac{1}{2}]}(1) &=& \frac{1}{\left| \eta(q) \right|^{2}}
\sum_{(n,m) \in \ZZ^{2}} (-1)^{m}
q^{\frac{1}{2}(n+\frac{m}{2})^{2}}
\overline{q}^{\frac{1}{2} (n-\frac{m}{2})^{2}}\\
Z_{[\frac{1}{2},0]}(1) &=& \frac{1}{\left| \eta(q) \right|^{2}}
\sum_{(n,m) \in \ZZ^{2}} q^{\frac{1}{2}(n+\frac{1}{2}+\frac{m}{2})
^{2}}
\overline{q}^{\frac{1}{2} (n+\frac{1}{2}-\frac{m}{2})^{2}}\\
Z_{[\frac{1}{2},\frac{1}{2}]}(1) &=&
\frac{1}{\left| \eta(q) \right|^{2}}
\sum_{(n,m) \in \ZZ^{2}} (-1)^{m}
q^{\frac{1}{2}(n+\frac{1}{2}+\frac{m}{2})^{2}}
\overline{q}^{\frac{1}{2} (n+\frac{1}{2}-\frac{m}{2})^{2}}.
\end{eqnarray}
After some algebra, the ``AA''
partition function of the free Dirac theory can be expressed as:
\begin{equation}
Z_{AA} = \frac{1}{2} \left( Z_{[0,0]}(1) + Z_{[0,\frac{1}{2}]}(1)
+ Z_{[\frac{1}{2},0]}(1) - Z_{[\frac{1}{2},\frac{1}{2}]} (1)\right)
.
\label{idAA}
\end{equation}
Using the modular transformations $\tau \rightarrow \tau+1$
and $\tau \rightarrow -1/\tau$, equation (\ref{idAA}) leads to
\begin{eqnarray}
Z_{AP} & = & \frac{1}{2} \left( Z_{[0,0]}(1)
+ Z_{[0,\frac{1}{2}]}(1)
- Z_{[\frac{1}{2},0]}(1)
+ Z_{[\frac{1}{2},\frac{1}{2}]}(1)
 \right)\\
Z_{PA} & = & \frac{1}{2} \left( Z_{[0,0]}(1)
- Z_{[0,\frac{1}{2}]}(1)
+  Z_{[\frac{1}{2},0]}(1)
+  Z_{[\frac{1}{2},\frac{1}{2}]}(1) 
\right).
\end{eqnarray}
The key point is that any sector of the fermionic theory can be obtained 
from the ``AA'' sector using modular transformations. On the other hand,
the modular properties of the $Z_{[\epsilon,\epsilon']}$ bosonic
partition functions are obvious.

\paragraph{Interacting case}
As far as the interacting case in concerned, we use the form
(\ref{ZAA}) of the partition function of the LL to obtain
\begin{equation}
\label{eqBoundaryConditions1}
Z_{AA}^{(\mathrm{Lutt})} =\frac{1}{2}\left( Z_{[0,0]}(\sqrt{\alpha})+
Z_{[0,\frac{1}{2}]}(\sqrt{\alpha})+
Z_{[\frac{1}{2},\frac{1}{2}]}(\sqrt{\alpha})-
Z_{[\frac{1}{2},0]}(\sqrt{\alpha}) \right)
\label{idLutt}
.
\end{equation}
The LL with charge velocity $v_N$
and current velocity $v_J$ may be identified to the compact boson
with $gR^{2} = \sqrt{\alpha}$, where $\alpha=\sqrt{v_N/v_J}$.
The reader should keep in mind the very special boundary conditions
that are imposed to the bosonic field for this identification to hold.

Conformal spins appearing in (\ref{idLutt}) have the form
\begin{equation}
\Delta_{n,m} = \frac{1}{2} \left(n \sqrt{\alpha} +
\frac{m}{2 \sqrt{\alpha}} \right)^{2} -
\frac{1}{2} \left(n \sqrt{\alpha} - \frac{m}{2 \sqrt{\alpha}}
\right)^{2}
=n m.
\end{equation}
Therefore, no exotic statistics are involved in the interacting problem
in the following sense: all fields have integer of half-integer
conformal spin.

\medskip

Equation (\ref{idLutt}) describes the identification of spectra between the
LL studied by Haldane and the spectrum of a certain bosonic
CFT. It is based on an explicit comparison of
partition functions.
We will also need to couple the Luttinger system 
to an electromagnetic field in order
to understand the effects of a magnetic field 
and of an electric potential. Fermionic computations can be found in appendix
\ref{secDirac}. As we show there, the $AA$ sector does not lead to any
surprise. But, as explained in this appendix, the $PP$ sector deserves some
care. In order to go further in this analysis, the
next section will be devoted to the study of the 
bosonic Luttinger field theory coupled
to a gauge field.


\section{Coupling to a gauge field}
\label{secGauge}

The aim of this section is to give a bosonization prescription
of the LL coupled to a gauge field. We first
examine the case of the compact boson coupled to a gauge
field, and we end up with an identification of the
charge and current in terms of the bosonic field.
As an application, we can compute the partition function of the 
LL in the presence of a magnetic flux through the 
Luttinger ring. In section \ref{secQJCorrelations},
explicit expressions the generating 
functionals of charge and current density correlators will be obtained.

\subsection{Compact boson coupled to a constant gauge field}
\label{secBosonGauge}

In the bosonic theory, 
the gauge field is coupled via an exterior product, namely
the action is
\begin{equation}
\label{RegisSwedge}
S_{\wedge}[\varphi,A] = \frac{g}{2 \pi} \int (\nabla \varphi)^{2}
- {i\over \pi R}\,\int A \wedge d\varphi
.
\end{equation}
We first consider the case of a constant gauge potential with holonomies
\begin{equation}
\label{eqHolon}
\int _{(a)}A=2\pi \,a\ \mathrm{and}\quad 
\int _{(b)}A=2\pi \,b
.
\end{equation}
In a second step, non constant terms in the gauge potential
will be introduced.

\medskip

We are interested in a partition function with boundary conditions 
given by (\ref{eqmonod}). A computation detailed in appendix \ref{secCalculs}
gives the following result:
\begin{equation}
\label{def}
Z_{[\epsilon,\epsilon']}[A] 
=  \frac{1}{|\eta(q)|^2} \sum_{(m,n)\in \ZZ^2}
e^{2 i \pi m \epsilon'} e^{4i\pi (\epsilon+n)b}
q^{\frac{1}{2} p_{n+\epsilon,m+2a}^2}
\overline{q}^{\frac{1}{2} \overline{p}_{n+\epsilon,m+2a}^2},
\label{Z(A)}
\end{equation}
where $p_{n,m}$ and $\overline{p}_{n,m}$ are defined by
(\ref{p}) and (\ref{pbar}).


As a warm up exercise, let us focus on the non interacting case,
that is $\alpha=gR^{2}=1$
and let us compute the partition function
\begin{equation}
\label{eqBoundaryConditions2}
Z[A] = \frac{1}{2} \sum_{(\epsilon,\epsilon')
\in \{0,1/2 \}^{2}} (-1)^{4 \epsilon \epsilon'}
Z_{\epsilon,\epsilon'}[A].
\end{equation}
The signs appearing here correspond to the ones in 
identity (\ref{idLutt}). We notice that
\begin{equation}
\sum_{\epsilon'\in \{0,1/2\}} (-1)^{4 \epsilon
\epsilon'} e^{e i \pi m \epsilon'}
= \frac{1}{2}(1+(-1)^{2 \epsilon+m}).
\end{equation}
Therefore, $m \equiv 2 \epsilon \pmod{2}$.
Separating the two cases $\epsilon=0$
and $\epsilon=1/2$, and performing a change of indices leads to:
\begin{equation}
\label{ZAab}
Z[A]= \frac{1}{|\eta(q)|^{2}}
\sum_{(m,\overline{m}) \in \ZZ^{2}}
e^{2 i \pi b (m - \overline{m})}
q^{\frac{1}{2}(m-a)^{2}}
\overline{q}^{\frac{1}{2}(\overline{m}-a)^{2}}.
\end{equation}
Finally, expression (\ref{ZAab}) can be rewritten 
in terms of Riemann's theta function with characteristics
(see appendix \ref{secElliptic}), 
thus recovering the known results \cite{DVVc1,Moore:86-1,Sachs:96-1}:
\begin{equation}
Z[A_{a,b}]=\frac{1}{|\eta(q)|^{2}}
\vartheta \left[
{\begin{array}{c} a\\-b \end{array}}
\right] (0,\tau)\, \overline{
\vartheta \left[
{\begin{array}{c} a\\-\overline{b} \end{array}}
\right] (0,\tau)}
\end{equation}
Here, we have assumed that $b$ is complex and $a$ real, since the 
chemical potential corresponds to a purely imaginary $b$. This point is
important in the comparison of bosonic and fermionic partition functions
performed in appendix \ref{secDirac}.
As we shall see later, the interacting case is a straightforward
generalization of the non interacting case, provided the
theta function is replaced by its suitable generalization.


\subsection{Gauge transformations}

Strictly speaking, the previous computation has been done with a
constant gauge field. Knowing the transformation properties of
$Z_{[\epsilon,\epsilon']}[A]$ in normal
($A\mapsto A+d\chi $) and chiral 
($A\mapsto A+d^*\lambda$)
gauge transformations
provide us with $Z[A]$ for any gauge field $A$, since by the Hodge 
decomposition theorem, any 1-form $A$ can be decomposed in a unique way as:
\begin{equation}
\label{Hodge}
A=h+ d\chi + d^*\lambda
.
\end{equation}
Here, $h$ denotes an harmonic $1$-form, or equivalently 
on the torus, a constant connection.
We recall that  gauge parameters $\chi$ and $\lambda $ are well defined
functions on the torus (no monodromy).

\medskip

We first consider normal gauge transformations.
Using relation (\ref{eqFormuleIntegraleTopo}),
and the exactness of $d\chi$, we obtain:
\begin{equation}
S_{\wedge}[\varphi,A+d\chi] - S_{\wedge}[\varphi,A] = 
{i\over \pi R}\int d\chi \wedge d\varphi = 0.
\end{equation}
The partition function is exactly gauge invariant under
normal gauge transformations.

\medskip

We now turn to chiral gauge transformations:
\begin{equation}
A \rightarrow A + d^{*} \lambda,
\end{equation}
where $d^{*} \lambda = \epsilon^{\mu\nu}
(\partial_{\nu}\, \lambda) dx^{\mu}$.
Elementary manipulations lead to:
\begin{equation}
S_{\wedge}[\varphi,A+d^{*} \lambda] = S_{\wedge}[\varphi-\frac{i}
{gR} \lambda,A] + {1\over 2\pi} \int (d \lambda)^{2}+ {1\over \pi \alpha}
\int A\wedge d\lambda .
\end{equation}
To summarize, the partition function behaves as follows under
normal and chiral gauge transformations:
\begin{eqnarray}
\label{eqWID2}
Z[A+d\chi] & = & Z[A]\\
\label{eqWID1}
Z[A+d^*\lambda ] & = & Z[A]\, 
\exp{\left( -{1\over 2\pi \alpha}\int (d\lambda )^2 - {1\over \pi \alpha} 
\int A\wedge d\lambda \right)}
.
\end{eqnarray}
It is important to notice that the transformation properties of
bosonic functional integrals are completely independent of boundary
conditions in the bosonic functional integrals. Identities (\ref{eqWID2}) 
and  (\ref{eqWID1}) are
the Ward identities of the Luttinger CFT.


\subsection{Non chiral bosonization prescription}
\label{courant}

The aim of this section is to establish,
by means of functional integrals 
manipulations, the identification between fermionic functional integrals
of the massless 
Thirring model in the ``AA'' spin sector and an appropriate bosonic functional 
integral. This formula encodes the bosonization of the massless Thirring model
in a functional way.
This bosonization approach is different from the
usual one which relies on
chiral bosonization of left and right movers in the operator formalism
(see  \cite[Chapter 14]{Kac}). 
Here, both chiralities are treated at the same time. This is why
we call this procedure a {\it non-chiral bosonization}.

\subsubsection{Statement of the result}

Let us introduce the following notation:
\begin{equation}
\label{eqNotationBC}
\int_{\cal C} {\cal D}[\varphi] =
\frac{1}{2} \sum_{(\epsilon,\epsilon') \in \{0,1/2\}}
(-1)^{4 \epsilon \epsilon'} \int_{[\epsilon,\epsilon']} 
{\cal D}_R[\varphi],
\end{equation}
where the boundary conditions in the right hand side path
integral are
\begin{equation}
\label{eqBCexplicit2}
\begin{cases}
\varphi(z+1) \equiv \varphi(z) + 2 \pi \epsilon R
\pmod{2 \pi R}\\
\varphi(z+\tau) \equiv \varphi(z) + 2 \pi \epsilon' R
\pmod{2 \pi R}
,
\end{cases}
\end{equation}
and where ${\cal D}_R[\varphi]$ denotes the integration measure for 
a bosonic field compactified on a circle of radius $R$ (see appendix 
\ref{secNormalisation}).
We will show that the following equality between partition function
holds:
\begin{equation}
\label{eqbosonization}
\int _{AA}{\cal D}[\bar{\psi},\psi]\,
e^{-S_L[\bar{\psi},\psi] + i \int 
A_\mu\bar{\psi}\gamma_{E}^{\mu}\psi}=
\int _{\cal{C}}{\cal D}[\varphi]\,
e^{-S_{\wedge}[\varphi,A]}
,
\end{equation}
where $S_L$ denotes the action of the massless Thirring model,
defined by the Lagrangian (\ref{eqLagrangeThirring}),
and $S_{\wedge}$ is the action (\ref{RegisSwedge}) of the compact
bosonic field coupled to a gauge field via an exterior
product.
This formula expresses a non-chiral bosonization
of the Luttinger CFT
in the presence of a gauge background.
It also shows that the fermionic current can be expressed 
in terms of the bosonic field:
\begin{equation}
\label{eqcurrentexpr}
\overline{\psi}\gamma_E ^{\mu}\psi = {\epsilon^{\mu\nu}\partial _{\nu}
\varphi\over \pi R}
.
\end{equation}
This identity should be understood as an equality within correlation functions.

\subsubsection{The interacting case}
\label{secInstability}

We shall now prove (\ref{eqbosonization}) in full generality. The method
goes as follows: noticing that this equation is true for
a uniform gauge background, we first prove it for the Dirac theory
in a generic gauge background by using Ward identities. Then,
the interacting case will be analyzed with the help of a suitable Hubbard-Stratanovich transformation.
Let us now prove equation (\ref{eqbosonization}) 
for an arbitrary gauge field.
The proof goes as follows: we know the transformation law of
fermionic partition functions both under chiral gauge
transformations and normal gauge transformations. They are
invariant under the latter whereas the so-called chiral anomaly
term appears under chiral gauge transformations (see equations
(\ref{eqWID2}) and (\ref{eqWID1})).
Thus, the fermionic
partition functions and the bosonic functional integral
appearing in (\ref{eqbosonization}) have the same transformation
properties in normal and chiral 
gauge transformations and coincide for constant gauge fields.
Therefore, using the Hodge decomposition theorem for $1$-forms on the torus,
these functionals do indeed coincide.

\medskip

We now introduce the interaction term (\ref{eqLagrangeJJ}) of the
Thirring type. The four fermions interaction can be usefully decoupled
by introducing an auxiliary field $b_{\mu}$:
\begin{equation}
\exp{\left(-{\kappa \over 2}\int (\bar{\psi}\gamma _E^{\mu}\psi)^2
\right)} = \int {\cal D}[b_{\mu}(x)]\, 
\exp{ \left(
{1\over 2\kappa}\int b(x)^2 +
i\int b_{\mu}\bar{\psi}\gamma _E^{\mu}\psi\right)}
.
\end{equation}
Therefore, we are back to the previous case in presence of a fluctuating
vector potential:
\begin{eqnarray}
Z_{\mathrm{Lutt}}[A] & = &
\int_{AA} {\cal D}[\bar{\psi},\psi]\,
e^{-S_L[\bar{\psi},\psi]-{\kappa \over 2}\int
 (\bar{\psi}\gamma _E^{\mu}\psi)^2
+ i \int A_{\mu} \bar{\psi} \gamma^{\mu}_E \psi
}\nonumber \\
 & = &
 \int_{AA} {\cal D}[\bar{\psi},\psi]\,\int  {\cal D}[b_{\mu}(x)]\,
e^{-S_L[\bar{\psi},\psi]-{1\over 2\kappa}\int b(x)^2 +i
\int (b_{\mu}+A_{\mu})\bar{\psi}\gamma _E^{\mu}\psi}\nonumber
.
\end{eqnarray}
Let us now apply identity (\ref{eqbosonization})
which, up to now, has been proven to be valid  just
in the free theory and in the presence of a fluctuating
gauge potential.
We then obtain
\begin{equation}
Z_{\mathrm{Lutt}}[A] = 
\int _{{\cal C}}{\cal D}[\varphi]\int {\cal D}[b_{\mu}(x)]\,
e^{-{g\over 2\pi}\int (\partial \varphi)^2
+{i\over \pi R}\int (A+b)\wedge d\varphi-
{1\over 2\kappa }\int (b(x))^2}
.
\end{equation}
We now integrate again over the auxiliary field $b_{\mu}$ and
obtain:
\begin{equation}
Z_{\mathrm{Lutt}}[A]=
\int _{{\cal C}}{\cal D}[\varphi]\,
e^{-{g\over 2\pi}
(1+{\kappa \over \pi gR^2})
\int (\partial \varphi)^2
+{i \over \pi R}\int A\wedge d\varphi} 
.
\end{equation}
The effect of interactions as a change of the 
coupling constant (or equivalently of the compactification radius) is
summarized by:
\begin{equation}
\label{eqgchange}
{g'\over g}=1+{\kappa \over \pi \alpha}
.
\end{equation}

\paragraph{Discussion of instability}
Let us notice that the interacting theory is only defined for $\kappa >-\pi $. 
When $\kappa \leq -\pi $, then the bosonic action is no longer positive. 
Such a limitation already arises 
in the operator formalism. Formula (\ref{eqDefPhi})
shows that for $|g_2|\geq |v_F+g_4|$, the Bogoliubov transformation does
not exist anymore. More precisely, one easily 
sees that the spectrum of the bosonic 
Hamiltonian (\ref{eqIntHam}) is no longer bounded from below. This is the
sign of an instability, which we recover in the functional integral 
language since $\alpha^2 = (v_F+g_4-g_2)/(v_F+g_4+g_2)$. 

\medskip

It would be interesting to understand the origin of this instability 
directly in the
fermionic framework. Let us recall that
in the massive Thirring model, Korepin \cite{Korepin:80-1} 
has shown that for the so-called repulsive
regime ($-\pi < g < -\pi/2$), the spectrum 
includes bound states. Coming back to the massless case,
when $\kappa \rightarrow -\pi $,
the fermions tend to form a bound state condensate
and the model becomes unstable
at $\kappa =-\pi $. However, the precise relation
between this instability (in
the massive regime) and the one we find in the bosonic
theory is still an open question.

Finally, such instabilities are not 
surprising from the solid state physics point of
view. Indeed, they appear when the interactions,
measured in suitable units, are of 
the order of the Fermi velocity $v_F$. In this case, 
the electronic fluid is expected to develop instabilities (see 
for instance Anderson's book \cite{AndersonBN}).

\subsection{Evaluation of the effective action}

In this section, integration over matter
fields of the Luttinger model will be explicitly performed, 
thus providing an effective action for the
vector potential. From the field theoretical point of view, this
boils down to solving the Schwinger model \cite{Schwinger:62-1} on the 
torus in a specific spin sector.
This effective action can be
understood as a generating functional for charge and current densities 
correlation functions as we shall see in the next section.

\medskip

All computations will be done within the bosonic theory.
We shall first of all
isolate the contribution of the zero modes of the vector potential.
After this separation, the computation will be straightforward. 

\subsubsection{Separation of zero modes}

Let $A=A_{\mu}dx^{\mu}$ be a vector potential.
Hodge theory tells
us that it can be decomposed in a unique way as
$A=h_A+d\chi +d^*\lambda$, with
$h_A$ is a uniform vector potential. We then have:
\begin{equation}
Z[A]=Z[h_A]\, \exp{\left(-{1\over 2\pi \alpha}\int (d\lambda)^2\right)}
,
\end{equation}
where we have used the Ward identity (\ref{eqWID1}).

\subsubsection{Explicit evaluation}

We are now going to compute $\lambda$ in terms
of the vector potential $A$. Let
us introduce $B=A-h_A$, such that each component
of $B$ has zero average on the
torus. The one-form $d^*\lambda -B$ is exact and
therefore $d\, d^* \lambda =dB$. However, since
$d\, d^* \lambda = -(\Delta \lambda )^* $,
the scalar curvature $F_B$ of   $B$ is nothing but
\begin{equation}
\label{eqLapl}
F_B=\Delta \lambda
.
\end{equation}
The Laplacian is invertible on the kernel of
the normalized integral over the torus:
$$\int _N~: f\mapsto {1\over \cal{A}} \int f,$$
with $\cal{A}$ the total area of the torus.
Finding an inverse of the Laplacian involves solving a Green's 
equation:
\begin{equation}
\Delta_x G(x,y) = \delta (x-y) -{1\over \cal{A}}
.
\end{equation}
Different solutions to this equation differ by a constant
which is irrelevant since the Green's function is 
always applied to a zero average function.
Let $\Delta ^{-1}$ denote the Laplacian's inverse on this vector space.
Its explicit expression is obtained
in appendix \ref{secElliptic}.
Inverting equation (\ref{eqLapl}) and
introducing complex coordinates leads to
\begin{equation}
\label{eqEffectiveAction}
Z[A]  =  Z[h_A]\, \exp{(K[B_z,B_{\bar{z}}])}
,
\end{equation}
where
\begin{eqnarray}
\label{eqEffective}
K[B_z,B_{\bar{z}}] & = &
{2\over \pi \alpha }
\int d^2z\,d^2\xi \,(B_{z}B_{\xi}(
\partial_{\bar{z}}^2\Delta ^{-1})(z-\xi)
+B_{\bar{z}}B_{\bar{\xi}}(\partial_z^2\Delta ^{-1})(z-\xi))\nonumber \\
 & - & {1\over \pi \alpha }
\int B_z\,B_{\bar{z}}d^2z.
\end{eqnarray}
As recalled in appendix \ref{secElliptic}, both second derivatives
$\partial_z^2\Delta ^{-1}$ and 
$\partial_{\bar{z}}^2\Delta ^{-1}$ should be understood as derivatives of
distributions (see equation (\ref{eqDefDerDis}) of appendix 
\ref{secElliptic}). These distributions
are related to Weierstrass $\wp$
function by formula (\ref{eqDerGreenSecond}) 
and its complex conjugate. This subtlety 
is important when computing density-density correlation functions
in presence of an external potential
(see section \ref{secCFTQJCorr}).

\medskip

Other authors work on the plane, namely in the zero temperature
and thermodynamics limits. Under this circumstance, there are
two differences with the results we have obtained: first of all
the contribution
of $A$'s zero modes is not present. Next, the Green's function has a simpler
expression. Nevertheless, the idea is the same and explicit integration over
matter fields has long been a useful trick in these two-dimensional field
theories \cite{Schwinger:62-1}.


\section{LL with a forward scattering disorder}
\label{secDisorder}

The formalism developed in the previous sections shall now
be applied to the
study the LL
in presence of a magnetic flux and of a very special kind of disorder: 
a Gaussian external potential coupled to the fermionic
density $\rho(\sigma)$,
as defined in equation (\ref{eqHamiltExtern}). 
Let us stress that this disorder does not
include backscattering effects and therefore
localization is not
present in this toy-model. This is of course a severe
limitation of our study
but taking into account backscattering requires a detailed
study of Sine-Gordon like interactions, which goes beyond the scope
of the present work. When these interactions are relevant,
one is driven away from the conformal regime and other methods should be
used. A disorder more suited for systems such as
quasi-one-dimensional conductors would involve impurities along
the chains and therefore the existence of backscattering.
Several approaches, such as Berezinskii diagram techniques
(see for instance  \cite{Berezinskii:73-1,Gogolin:75-1,Gorkov:76-1,Ovchinnikov:77-1,Gogolin:77-1,Berezinskii:79-1,Antsygina:81-1,Gorkov:83-1})
were developed
in the seventies by the Russian school to treat such a kind
of disorder in the absence of electron-electron interactions. 
The conjugated effects of disorder and interactions
were analyzed in \cite{Giamarchi:88-1} by means of renormalization
group techniques. In these different works, the presence of
disorder easily gives way to massive phases,
or to massless phases with a singular low energy density of
states. In both cases, the resulting disordered system is
obviously not conformal and its treatment using methods
similar to the ones of the present article is an open
question.

\medskip

Nevertheless, in this framework, correlation functions and their
averages over
the aforementioned
toy-disorder can be computed without using the replica trick.
The method used here relies on transformation properties of the Lagrangian
under chiral gauge transformations and has been used by D.~Bernard 
\cite{Bernard:95-1,Bernard:95-2}
to solve the random vector potential model.
As we shall see in section \ref{secQJCorrelations},
transport properties are not affected by the toy-disorder,
as expected since no localization is involved.
We shall also compute the
specific heat of the LL in the regime where the
temperature is much larger than the inter-level spacing $2\pi v_S/L$,
showing that it is not altered by the simple toy-disorder
considered here.

\subsection{General setting and notations}
In this section, a random classical potential $V(\sigma)$
will be introduced. We also introduce a constant magnetic field
with a magnetic flux $\Phi$ 
through the ring, and 
$\chi=\Phi/\Phi_0$ denotes the number of flux quanta, with
$\Phi_0=2 \pi/e$ the flux quantum. The reason why we treat
simultaneously a coupling to an external potential and a coupling
to a magnetic field is that, as we shall see, both can
be naturally analyzed in the same framework. In fact, as we will
see later, they transform into one another by duality.

\medskip

The random potential distribution is taken to be Gaussian:
\begin{equation}
P[V(\sigma)] = \exp{\left( - \frac{1}{2 \gamma}
\int_0^{L} d \sigma\, V(\sigma)^2 \right)}.
\end{equation}
Thermal and quantum averages averages with
respect to the Luttinger system are denoted by 
$\langle {...} \rangle$:
\begin{equation}
\langle O[\overline{\psi},\psi]
\rangle_{[\chi,V(\sigma)]}
= \frac{\int {\cal D}[\overline{\psi},\psi]\,  e^{-S[
\overline{\psi},\psi,V,\chi]}\, O[\overline{\psi},
\psi] }
{ \int {\cal D}[\overline{\psi},\psi]\, e^{-S[\overline{\psi},\psi,V,\chi]}},
\end{equation}
where fermionic fields live in the ``AA''
sector. We denote by $\overline{X}$ the average of $X$
over the toy-disorder, and we are interested in correlation
functions of the type
\begin{equation}
\overline{\prod_{k=1}^{N} \langle O_k[\overline{\psi},
\psi] \rangle} = \int {\cal D}[V]\,
e^{-\frac{1}{2 \gamma} \int_0^{L} d\sigma\, V(\sigma)^{2}}
\prod_{k=1}^{N} \langle O_k[\overline{\psi},\psi]
\rangle_{[\chi,V(\sigma)]}.
\end{equation}
As we shall see in the following, these correlation functions
can be explicitly computed without using the replica trick. 
In order to achieve this goal, we shall first of all bosonize
the system as explained in the previous sections and then
use the quadraticity of the action both in the bosonic field and in
the random potential. At this point, the choice of a Gaussian
distributed random potential is crucial.

\paragraph{Feynman weight contribution of disorder and 
magnetic field}
The statistical weight associated with the disorder and
magnetic field is given by:
\begin{equation}
W[\overline{\psi},\psi] =
\exp{\left( i \int_0^{\beta} du \int_0^{L} d \sigma
\left[ \frac{e v_f \Phi}{L} \overline{\psi}
\gamma^{1} \psi + i e V(\sigma)
\overline{\psi} \gamma^{0} \psi \right] \right)},
.
\end{equation}
It may be usefully rewritten as $\exp{(i\int j\ldotp A)}$
where $j^{\mu}=\overline{\psi} \gamma^{\mu} \psi$,
and $A_0 = i e V(\sigma)$ and $A_1 = e v_f \Phi/L$.

\subsection{Correlation functions averages over disorder}

Let us now derive the general formula for computing averages
over disorder of any product of correlation functions.
We first split the potential $V(\sigma)$ into
its average part and a fluctuating part:
\begin{equation}
V(\sigma) = {\cal V}(\sigma) + \frac{Q}{L},
\end{equation}
and we introduce $\eta (\sigma)$ defined by
\begin{equation}
\eta(\sigma)=\int _0^{\sigma}{\cal V}(x)\, dx.
\end{equation}
The observable $O$ considered here is assumed to be expressed
as a functional of the bosonic field. In this case, 
we introduce the following notation:
\begin{equation}
Z_O[\chi,V(\sigma)] =
\int_{\cal C} {\cal D}[\varphi]\, e^{-S[\varphi,A]}
O[\varphi].
\end{equation}
The idea is now to incorporate the fluctuating part
$\eta (\sigma)$ of the potential in a redefinition of
$\varphi$. Readers familiar with CFTs can
recognize here the
abelian version of Polyakov-Wiegmann's identity \cite{KZ}.
After simple algebraic manipulations, we get
\begin{equation}
Z_O[\chi,V(\sigma)] = e^{\frac{e^{2} \beta}
{2 \pi g R^{2}} \int_0^{L} d \sigma \eta'(\sigma)^{2}} 
\,Z_{O[\varphi+ e \eta/gR]}[\chi,Q/L],
\label{eqZidentity}
.
\end{equation}
The field $\varphi$ has been shifted by $ e \eta/gR$.
This shift does {\em not} modify the boundary conditions
for our bosonic field since $\eta(0) = \eta(L) = 0$.
Since the exponential prefactor appearing
in the right hand side of
(\ref{eqZidentity}) is independent of the observable $O$,
it cancels when averages over $O$ are taken:
\begin{equation}
\langle O[\varphi] \rangle_{[\chi,V(\sigma)]}
= \langle O[\varphi+\frac{e}{gR} \eta]\rangle_{[\chi,
Q/L]}.
\label{averages}
\end{equation}
Using these equations, averages
over the toy-disorder can be computed:
\begin{equation}
\overline{\prod_{k=0}^{N} \langle O_k[\phi] \rangle}
 =   \int_{-\infty}^{+\infty} \frac{dQ}{\sqrt{2 \pi
\gamma L}}\, e^{-{Q^2\over 2\gamma L}}  \int {\cal D}[\eta(\sigma)]
\delta(\eta(0))
\,  e^{-\frac{1}{2 \gamma}
\int_0^{L} d \sigma \eta'(\sigma)^{2}}
\prod_{k=0}^{N} \langle O_k[\phi+\frac{e}{gR} \eta]
\rangle_{[\chi,Q/L]}.
\end{equation}
We have therefore achieved our goal: averages over
the toy-disorder
can be computed without the replica trick. 
In the forthcoming sections, we shall
give explicit examples of correlation functions, averaged over
the toy-disorder.
But before doing that, we turn now to the problem of computing the
average free energy over the toy-disorder.


\subsection{Average free energy, thermal capacity}
We first rewrite the partition function of the Luttinger CFT
(\ref{eqZLutt}) as
\begin{equation}
Z_{\mathrm{Lutt}}=
\frac{1}{|\eta(q)|^{2}} \sum_{(m,\overline{m})\in \ZZ^{2}}
q^{\frac{1}{2} \left( \frac{\sqrt{\alpha}+\sqrt{\alpha}^{-1}}
{2} m + \frac{\sqrt{\alpha}-\sqrt{\alpha}^{-1}}{2}
\overline{m} \right)^{2}}
\overline{q}^{\frac{1}{2} \left(
\frac{\sqrt{\alpha}-\sqrt{\alpha}^{-1}}
{2} m + \frac{\sqrt{\alpha}+\sqrt{\alpha}^{-1}}{2}
\overline{m} \right)^{2}}.
\end{equation}
We are interested in the behavior of the partition function
in the limit $\tau=i \beta v_S/L \rightarrow 0^{+}$.
Physically, it means that the temperature is much larger than 
the inter-level spacing. Henceforth, a high number
of energy levels are excited. We recall that a real system may not 
be described at all energies by the effective Luttinger theory considered
here. If $E^*$ denotes the energy scale where the 
effective description by an interacting massless Thirring model 
breaks down, we assume that $E^*>>2\pi v_S/L$.
The energy scale $E^*$ may originate from band curvature,
non local interactions,....
Our analysis will be
valid in a temperature regime such that
\begin{equation}
\label{eqcriteretemp}
{2\pi v_S\over L}<<\, k_BT\,<< E^*
.
\end{equation}
In terms of $x=2\pi v_S\beta /L$, the relevant limit is 
$x\rightarrow 0^+$.
Without any twist in the spatial direction (in other words,
$\theta=0$ in (\ref{definitiondetau})),
we have $q=\overline{q}=e^{-x}$.
In terms of $x$, the partition function is
\begin{equation}
Z_{\mathrm{Lutt}}(\alpha,[A]) = \frac{1}{\eta (ix/2 \pi)^2}
\sum_{(n,\overline{n})\in \ZZ^{2}}
\exp { \left( - \frac{x}{4} \left( \alpha
(n+\overline{n})^{2} + \alpha^{-1} (n-\overline{n}
-2 a)^{2} \right)\right)}
\exp{\left( 2 i \pi b(n+\overline{n}) \right)}.
\end{equation}
Its limiting behavior for
$x \rightarrow 0^{+}$ is easily obtained
with the help of
Dedekind's function asymptotics in the limit $\tau \rightarrow i\,0^{+}$.
Substituting $b=i \beta v_S Q/2 \pi L$,
we finally obtain:
\begin{equation}
\ln{(Z_{\mathrm{Lutt}}}(\alpha,[A])) \sim
\frac{\pi L}{6 \beta v_S} + \frac{\beta v_S Q^{2}}
{2 \pi \alpha L}.
\end{equation}
Before averaging over the quenched toy-disorder, let us notice that
the effect of the magnetic field and of the electric potential
boils down to a free energy decrease of $v_f Q^{2}/2 \pi \alpha L$,
independent on the temperature and on the
magnetic field. In particular, the specific heat
is not affected neither by the
electric potential nor by the magnetic field.
Of course, the same conclusions are valid after averaging over the
disorder.

\section{Charge-charge and current-current correlations}
\label{secQJCorrelations}

As an application of previous formalism, correlations
of charge and current densities in Luttinger CFT will be computed. 
Our strategy will rely on the evaluation 
of the generating functional for charge density correlators and
current density correlators.
We are interested in correlation
functions in a regime of a fixed total charge in the system.
This total charge is denoted here by $q$.
Averaging over $q$ amounts to
considering a set of many isolated Luttinger rings,
some of which contain an odd number of charge carriers, and others
carry an even number of carriers. Again, our aim is not to
describe in a ``realistic'' way the physics of permanent
currents in mesoscopic rings, would involve
multichannel effects, impurities, spin effects,..., and goes much
beyond the scope of the present formalism.
We rather want to demonstrate that the
formalism developed in the previous sections is an operational
tool from which physical quantities such as permanent currents
in a Luttinger ring can be calculated. In order to be more
general, we will first calculate the generating functional
of the $n$ points density and current correlations.

\subsection{Charge and current density correlators}
\label{secCFTQJCorr}

\paragraph{Statement of the results}

The generating functional for charge density correlators is defined by
\begin{equation}
\label{d-gene-func}
W^{(0)}_{[V(\sigma),\chi,q]}[b(z)]=\langle 
\exp{\left( \int d\sigma\, dt\, b(\sigma,t)\rho (\sigma ,t)\right)}
\rangle
.
\end{equation}
An explicit expression can be derived from the effective
action (\ref{eqEffectiveAction}). It has the form
\begin{equation}
\label{eqCorrCharge}
W^{(0)}[V(\sigma),\chi,q]=
\exp{\left(L_0[V(\sigma),b(z)]+F_0[b(z)]\right)}
,
\end{equation}
involving a linear contribution:
\begin{equation}
\label{eqCorrQA}
L_0[V(\sigma),b(z)]=
\int d\sigma \, du \, \left({q\over L}-
{1\over \pi \alpha}{\cal V}(\sigma)\right)b(\sigma,u)
,
\end{equation}
and a quadratic contribution
\begin{equation}
\label{eqCorrQB}
F_0[b(z)]  =
\exp{\left( {1\over 4\pi \alpha}\int d^2z\,d^2\,\xi\; b(z)b(\xi)G_0(z-\xi)
\right)}
,
\end{equation}
the kernel $G_0$ being
\begin{equation}
\label{eqCorrQC}
G_0(z-\xi)  =  
-\Re{\left({\wp (z-\xi)\over \pi}\right)}+\delta (z-\xi)
.
\end{equation}
Of course, the Weierstrass function
contribution should be understood as 
a regularized distribution, just like
in formula (\ref{eqDerGreenSecond}) in
appendix \ref{secElliptic}. Details of the computation will be given in the
next subsection.

\medskip

The generating functional for current correlators is defined by
\begin{equation}
\label{eqCorrJA}
W^{(1)}_{[V(\sigma),\chi,q]}[c(z)]=\langle 
\exp{\left( \int d\sigma\, dt\, c(\sigma,t)\, j(\sigma ,t)\right)}
\rangle
.
\end{equation}
It can be explicitly computed and we obtain
\begin{equation}
\label{eqCorrJ}
W^{(1)}_{[V(\sigma),\chi,q]}[c(z)]=
{Z_q[0,\chi +{1\over 2\pi \beta}\int c(z)d^2z]\over 
Z_q[0,\chi]}\times \exp{(F_1[\tilde{c}(z)])}
,
\end{equation}
where
$\tilde{c}(z)=c(z)-{1\over {\cal A}}\int c(\xi)d^2\xi$.
The quadratic contribution $F_1$ is
\begin{equation}
\label{eqCorrJF}
F_1[\tilde{c}(z)]=-{1\over 4\pi \alpha}
\int d^2z\, d^2\xi \, \tilde{c}(z)\tilde{c}(\xi)\,
\left(\Re\left(
{\wp(z-\xi)\over \pi}\right)+\delta (z-\xi)\right)
.
\end{equation}

\paragraph{Details of computation}

Let us now show how to obtain formulae
(\ref{eqCorrCharge}) to (\ref{eqCorrQC}). 
We shall start from the following potential: $A_{\sigma}=0$ and 
$A_u=i({\cal V}(\sigma)+b(\sigma,u))$. This implies $B_z=-B_{\bar{z}}=
({\cal V}-\tilde{b})/2$, with
$\tilde{b}(z)=b(z)-{1\over {\cal A}}\int b(\xi)d^2\xi$.
Plugging this in formula (\ref{eqEffective}) leads
to three distinct contributions:
\begin{equation}
\begin{cases}
{1\over 2\pi \alpha}\int \left((\partial ^2+\overline{\partial}^2)\Delta^{-1}+\delta/2
\right)(z-\xi)\, \tilde{b}(z)\tilde{b}(\xi)\, d^2z\,d^2\xi\\
{1\over 2\pi \alpha} \int \left((\partial ^2+\overline{\partial}^2)\Delta^{-1}+\delta/2
\right)(z-\xi)\, {\cal V}(\sigma_z){\cal V}(\sigma_{\xi})\, d^2z\,d^2\xi\\
{1\over \pi \alpha} \int \left((\partial ^2+\overline{\partial}^2)\Delta^{-1}+\delta/2
\right)(z-\xi)\, \tilde{b}(z){\cal V}(\sigma_{\xi})\, d^2z\,d^2\xi
.
\end{cases}
\end{equation}
The first one directly 
gives $F_0[b(z)]$ and the second one compensates while dividing
by the partition function. Only the third one deserves attention. The main
point is to remember that derivatives of $\Delta^{-1}$ should be understood as
derivatives of a distribution as explained in appendix \ref{secLaplacienInv}.
For simplicity, computations use the $\Bbb{Z}\oplus\tau\Bbb{Z}$ lattice.
Let $f$ be a periodic function, of period 1, of a {\em real} variable. 
We define
a $\Bbb{Z}\oplus \tau\Bbb{Z}$ periodic function $f_1$ by $f_1(z)=f(\Re{(z)})$. 
The problem is now
to evaluate the distribution $T$ defined by 
\begin{equation}
\label{eqDefT}
T\ldotp f = \partial_z^2[\Delta^{-1}]\ldotp f_1
.
\end{equation}
Noticing that $f_1$ only depends on $\Re{(z)}$, we thus get
$\partial _zf_1=(\partial _z+\partial _{\bar{z}})f_1$
and therefore
$
\partial_z^2[\Delta^{-1}]\ldotp f_1 = 
\partial _{\bar{z}}\partial _z [\Delta ^{-1}]\ldotp f_1$.
It is now possible to apply formula (\ref{eqEllipticDelta}) to 
get
\begin{equation}
\label{eqEvalT}
T\ldotp f = \partial_z^2[\Delta^{-1}]\ldotp f_1 = 
{1\over 4}\,\left(\delta -1\right)\ldotp f
,
\end{equation}
where the distributions are now acting on $f$, which depends on a real 
variable. In other words, for a zero-average test function $f$:
$$\int f(\sigma_z)g(\xi){\wp(z-\xi)\over \pi}\, d^2zd^2\xi =
\int  f(\sigma)g(\sigma,u)\;d\sigma\, du.$$
This shows that the ${\cal V}\times 
b$ contribution is given by the ${\cal V}$-dependent part
of $L_0[V,b]$ and that the second contribution is nothing but 
$$\exp{\left({1\over 2\pi \alpha}\int d^2z\,{\cal V}(\sigma_z)^2 \right)},$$
a result that can easily be deduced from the Ward identity (\ref{eqWID1}). 

\medskip

This computation deserves some comments: it shows that properly considering
correlation functions of a CFT as distributions
enables to extract some ultra violet properties of the model from
its infrared effective description.
More explicitly, the coupling between $b$ and 
the potential, which indeed contains the charge response to an external potential, 
is {\em local} in the CFT approximation. Nevertheless, it {\em can} 
be extracted from
a long distance effective theory since it relies on the symmetry properties of
the system, that is to say, its transformation properties under chiral gauge 
transformations, which we expect to be independent of the description of 
the system.

\medskip

The proof of equations (\ref{eqCorrJ}) and (\ref{eqCorrJF}) goes along the same
lines. We have to compute $K[C+D,C-D]$ where $C=\tilde{c}(z)/2$ and 
$D={\cal V}(\sigma)/2$. This gives two terms, the first one being
$$F_1[c(z)]+{1\over 2\pi \alpha}\int d^2z {\cal V}(\sigma_z)^2,$$
and the second one contains a crossed term of the type
$\tilde{c}\times {\cal V}$:
$$
{i\over 2\pi \alpha}\lim_{\varepsilon \rightarrow 0^+}
\int_{|z-\xi|>\varepsilon} 
d^2zd^2\xi\; \tilde{c}(z){\cal V}(\sigma_{\xi})\;
\Im{\left({\wp(z-\xi)\over \pi}\right)}
.
$$
The point is then to notice that, taking the imaginary part of 
$\wp$ in equation (\ref{eqEvalT}) gives zero.
Therefore, the crossed term vanishes.

\subsection{Density response to an external potential}

\paragraph{Density modifications induced by the external potential}
Differentiating once with respect to $b(\sigma,t)$ provides
us with the average density in the presence of the potential
$V(\sigma)$. 
We obtain:
\begin{equation}
\label{eqChargeResp}
\rho _{\mathrm{av}}(\sigma)={q\over L}-{1\over \pi v_N}\left(V(\sigma)-
{1\over L}\int _0^LV(\sigma)\,d\sigma \right)
.
\end{equation}
The first term in (\ref{eqChargeResp})
ensures charge conservation, with a total charge of $q$.
The response function to an external potential
is thus purely local.
This result is expected
from the density response of non interacting, non relativistic
fermions in one dimension (see appendix \ref{secFermiSea} where
the calculations in this limit are recalled).
Turning on local interactions
in the LL only amounts to replacing the Fermi velocity
$v_F$ by the charge velocity $v_N$ defined in equation
(\ref{Charge_Velocity}) in terms of the interaction parameters.
This result could maybe have been anticipated from the beginning
on physical grounds. 

\paragraph{Correlation functions}
The two point connected  correlator of the densities has a singularity at 
equal times due to the delta function in time {\em and} space. It
is interesting to understand the origin of this short distance 
singularity. For this reason, we have recalled the corresponding
computation for non relativistic free fermions on a circle in
appendix \ref{secFermiSea}. This computation has been
performed at zero temperature for simplicity but
may easily be extended to a finite temperature.
In the non relativistic model, the Fermi sea has a
finite depth $k_F$ 
and the density-density correlations show a peak spreading
over a  distance $1/k_F$.
The two points correlations also show some
short distance oscillations
of the form $\sin^2(k_F\sigma)$ which are {\em not} present in our
CFT computation. These two apparent discrepancies with
conformal results have the following interpretation:
first of all,
observed at distances small compared to the size $L$ of the circle
but large compared to the ``microscopic'' length scale $1/k_F$,
the short distance
peak becomes a delta function, the coefficient of which is 
proportional to $k_F$ (and therefore diverges in the infrared limit). 
This corresponds to the double delta function in
the conformal result.
Next, the average of 
$\sin^2(k_F\sigma)$ over distance much greater than $k_F^{-1}$ is
$1/2$ and we recover the conformal result.

\medskip

Finally, no orthogonality-like scenario is operational in
the density-density correlations, unlike the case of fermion
correlations, as we shall see later. The reason for this
is that, even though density operators are defined in
terms of chiral fermions, density operators are {\em
composite} operators. It is also {a-priori} obvious that
no orthogonality-like behavior is expected since
integrating the $n$-point correlations
with respect to all the coordinates leads to a
non vanishing term even in the thermodynamic limit,
thus excluding the possibility of exponentially small
prefactors.

\medskip

In a sector of fixed charge, the generating functional of
density correlations is {\em Gaussian} (see equation
(\ref{eqCorrCharge})).
The same is true for the current density fluctuations. Apart
from the term associated with the permanent current, the functional 
(\ref{eqCorrJ}) is Gaussian. 
The reason for that lies in the fact that the fermionic density can
be viewed as a ``coordinate'' for this quantum field theory. The current
density is then conjugated to it. In term of these 
coordinates, the 
Luttinger CFT is essentially free, therefore explaining the
Gaussian nature of these functionals.

\subsection{Permanent currents in the Luttinger ring}
\label{secCurrents}

As emphasized by many authors, and as observed experimentally, 
mesoscopic rings exhibit permanent currents of quantum
origin in the presence of a magnetic flux (see for instance
\cite{Levy:90-1,Mailly:93-1}).
In the LL, these currents can be exactly computed at
a finite temperature. 
They have a ``universal'' expression in the sense that
all the interaction dependence
is encoded in the $\alpha $ and $v_S$ Luttinger parameters, 
as expected on general grounds exposed in the introduction.

\medskip

\begin{figure}[h]
\includegraphics[angle=270]{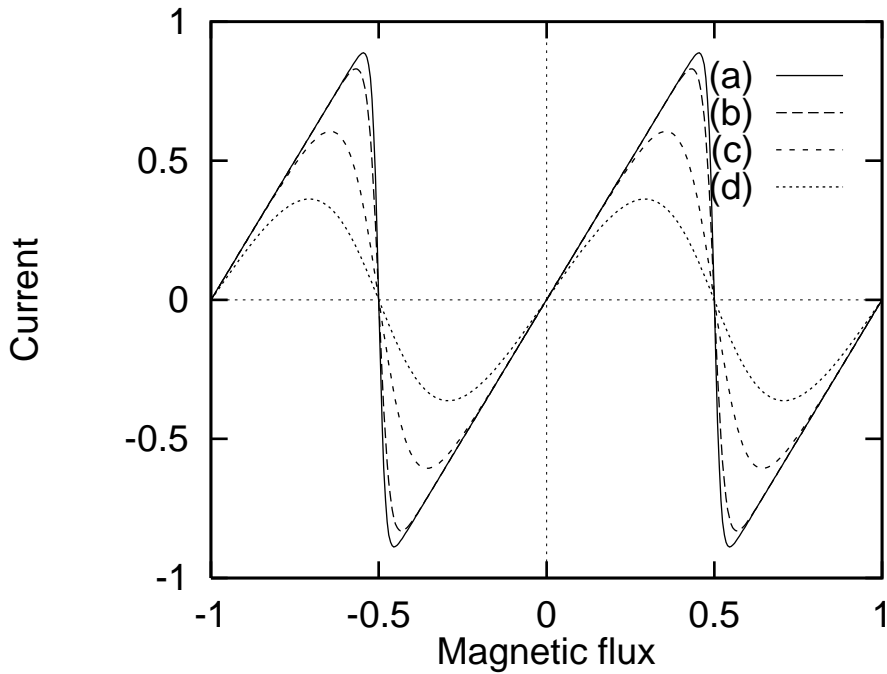}
\caption{\label{fig1}
Variations of the permanents currents (in units of $2 \pi e v_F/ \alpha L$) 
around the Luttinger ring with $q=0$
as a function of the flux inside the ring (in units of
the flux quantum).
The different curves correspond to
a ratio $x/\alpha$ of (a):~50, (b):~30, (c):~10, and (d):~5.}
\end{figure}

The permanent current in an isolated system of charge $q$
is proportional to the derivative of the free 
energy with respect to the magnetic flux:
\begin{equation}
I[V(\sigma),\chi,q]=\left(
{\partial F\over \partial \Phi}\right)=
{e\over h}\,\left(
{\partial F\over \partial \chi}\right)
.
\end{equation}
A straightforward computation leads to
\begin{equation}
I[V(\sigma),\chi,q]=-{e\over \beta h}{\partial \over \partial \chi}
\left(
\log{\left(
\sum _{n\in \Bbb{Z}}
\exp{\left(-{x\over \alpha }\left({q\over 2}+\chi +n\right)^2
\right)}\right)}\right)
.
\end{equation}
In the zero temperature limit, this obviously reduces
to the famous ``saw-tooth''
curve. The maximal value is given by $\pi e v_J/L$.
The current variations as a function of $\chi$ are shown
on figure \ref{fig1} as a function for different
values of the prefactor $x/\alpha$.
As this ratio decreases the ``saw-tooth'' curve is more
and more smoothened.
The currents are
periodic in the magnetic flux, with a unit flux quantum periodicity,
as can also be seen from figure \ref{fig1}.
Even though we stress that the LL is far from a
proper modeling of mesoscopic rings, this unit quantum flux
periodicity was observed experimentally in a single ring in \cite{Mailly:93-1}.
The currents also show a periodic behavior in the total charge
$q$ but with a period of two unit charges, as expected.
Of course, if an average
over the parity of the charges is taken, the current shows
a half flux quantum periodicity. This situation is the analogous
of the experimental realization of a large ensemble of mesoscopic
ring, where half-flux-quantum periodic currents were observed
\cite{Levy:90-1}.
Finally, the current 
decays exponentially as soon as the temperature exceeds the level
spacing $2\pi v_S/L$, as expected on physical grounds.

\section{Correlation functions of vertex operators}
\label{secVertex}

The aim of this section is to compute correlation functions  of vertex
operators. Let us recall that in a free bosonic CFT,
$\widehat{U(1)}$ primary fields are in a one to one correspondence
with chiral operators. They constitute the building blocks of the
Luttinger CFT. As we shall see in section \ref{secFermions}, the renormalized
fermion in the interacting theory correspond to one of these vertex 
operators. 

After having identified the effective theory for edge
excitations of a FQH fluid
on a cylinder as some of the Luttinger CFTs (see section \ref{secLaughlin}), 
we shall see that physical 
fermions localized on one edge of the Hall sample are created and destroyed by 
specific vertex operators. Therefore computations performed in this section
give access to physically important correlation functions.

\medskip

We start by recalling a definition of vertex operators well-suited to
functional integral computations. Then, we derive correlation functions
on the sphere for warming up and on the torus. Finally, the identification of
Luttinger fermion and the effect of the toy-disordered are discussed.

\subsection{Vertex operators}
\label{secVertex1}

Instead of studying vertex operators themselves, we shall be interested in
specific products of them. More precisely, we shall compute objects such as
averages of
\begin{equation}
\label{Vertexgeneral}
\prod_{k=1}^n \exp{\left({i\,e_k\over R}\int _{C_k}d\varphi +{e_k^*\over R}
\int _{C_k}d^*\varphi \right)}
.
\end{equation}
The parameters $(e_k,e^*_k)_k$ are fixed real numbers and $C_k$ are some open oriented
curves starting at $a_k$ and ending at $b_k$. Of course,
the object (\ref{Vertexgeneral}) should
be renormalized precisely, and this will be explained in 
the forthcoming subsections. 
Nevertheless, it is first interesting to understand its meaning.

\medskip

The first integral in (\ref{Vertexgeneral}) defines a product of 
usual vertex operators since one can easily recognize
$$\exp{(i\,e_k\, (\varphi (b_k)-\varphi (a_k))/R)}.$$ 
To understand the physical meaning of the second contribution
in (\ref{Vertexgeneral}), 
let us choose locally some coordinates near
a point $x$ on the curve $C_k$. The spatial coordinate $l_x$ will be chosen locally
along the curve $C_k$ and the imaginary time 
coordinate $u_x$ is chosen normally to
$C_k$ at point $x$. Then, the local contribution to the second integral is
$$-(\partial _{u_x}\varphi )(l_x,u_x)\, dl_x.$$
But the imaginary time derivative is proportional to the moment
associated with 
coordinate $\varphi $. More precisely, we know that 
the conjugate moment of $\varphi$ is nothing but
$$\Pi_{\varphi} ={g\over \pi}\, \partial _t\varphi.$$
Therefore, if we denote by $\Pi _{C_k}$ the moment on the
curve $C_k$, we  have 
\begin{equation}
\exp{\left({e_k^*\over R}
\int _{C_k}d^*\varphi \right)}
=\exp{\left( -i{\pi \, e_k^*\over gR}\, \Pi _{C_k}\right)}
.
\end{equation}
The second integral thus corresponds to introducing a defect along
curve $C_k$. It is nothing but a {\em disorder operator}. 

\medskip

With this interpretation, we immediately see that both $e_k$ and $e_k^*$ are
quantized. Taking into account $\varphi $'s compactification at radius $R$
immediately gives $e_k\in \Bbb{Z}$. The disorder part introduces a cut of
$\pi e_k^*/gR$ for the field $\varphi$ along curve $C_k$. Then, imposing 
that this shift be a multiple of $2\pi R$ gives the second quantization condition:
$e^*_k\in 2\alpha \Bbb{Z}$. In fact these are the quantization conditions
of
the usual bosonic theory, in the modular invariant sector. In order to 
understand this, and as a warm up exercise, let us
first compute correlation functions on the sphere. 

\subsection{Correlation functions on the sphere}
\label{secCorrS2}

Let us compute the two points function on the Riemann sphere:
\begin{equation}
\label{eqCorrFct}
{1\over Z_{\Bbb{P}_1(\Bbb{C})}}\,\int {\cal D}[\varphi]
\exp{\left({i\,e\over R}\int _{C}d\varphi +
{e^*\over R}
\int _{C}d^*\varphi -{g\over 2\pi}\int (d\varphi )^2\right)}
,
\end{equation}
where $C$ starts at point $a$ and ends at point $b$.
This is a Gaussian integral to be computed using a saddle point
method. In the following, $\delta _C$ will denote 
the delta distribution located
on the oriented curve $C$ and $\partial _{n}\delta _C$ will denote its derivative
with respect to the normal coordinate on the curve. The saddle point equation is
then:
\begin{equation}
\Delta \varphi = {i\pi e\over gR}\, (\delta _a - \delta _b)
-{\pi \,e^*\over gR}\partial _{n}\delta _C
.
\end{equation}
As expected, when crossing the curve $C$ in the direct sense with respect to $C$'s 
orientation, $\varphi $ drops by $\pi \,e^*/gR$. Up to an additive constant, 
the solution to this equation can be expressed in terms
of a determination of the
complex logarithm. Let us define $h_{a,b}$ by:
\begin{equation}
h_{a,b}(z)={z-a\over z-b}
.
\end{equation}
If $\log_C$ denotes the determination of the logarithm such that 
$\log_C(h_{a,b}(z))$ has a cut along $C$, we have:
\begin{equation}
\varphi_0(z)={i\over 4g\,R}\left((e-e^*)\,\log_C{(h_{a,b}(z))}
+(e+e^*)\,\overline{\log_C{(h_{a,b}(z))}}\right)
.
\end{equation}
The Gaussian integral to be computed here is of the form
$$\int dX\; \exp{\left(-{1\over 2}\;{}^tX\ldotp A\ldotp X +{}^tX\ldotp J\right)}.$$
The saddle point evaluation of the exponential gives
\begin{equation}
\exp{\left({1\over 2}\;{}^tJ\ldotp X_0\right)}
\quad \mathrm{with}\ 
 A\ldotp X_0 =J
.
\end{equation}
As we shall see, using this Ansatz requires some care because of divergences.

\medskip

Let us compute the first contribution to the saddle point evaluation: 
we have to renormalize
\begin{equation}
\exp{\left(
{i\, e\over 2R}\,(\varphi_0(b)-\varphi_0 (a))\right)}
.
\end{equation}
The prescription chosen here is to replace $\varphi_0(a)$ and 
$\varphi_0(b)$ by their averages of
circles of {\it radii} $\varepsilon $, centered on $a$ and $b$ respectively.
This will provide us with regularized quantities depending on 
$\varepsilon$. Then, we shall use
a minimal renormalization scheme, that is
we will forget about divergences.
Let us denote by $\Gamma^{(1)} _{\varepsilon}(a,b,e,e^*)$
the regularized quantity defined by 
\begin{equation}
\langle e^{
{i\, e\over 2R}(\varphi_0(b)-\varphi_0 (a))}\rangle_{\varepsilon}
=e^{-\Gamma^{(1)} _{\varepsilon}(a,b,e,e^*)}
.
\end{equation}
We have, at dominant orders:
\begin{equation}
\Gamma^{(1)} _{\varepsilon}(a,b,e,e^*)\simeq -{e^2\over 2\alpha}\, \log{(\varepsilon)}
+{e^2\over 2\alpha} \log{(|h_{a,b}(z)|)}-
{e\,e^*\over 2\alpha} \log{\left({h_{a,b}(z)\over \overline{h_{a,b}(z)}}\right)}
.
\end{equation}
In order to obtain the other contribution, we simply notice that
\begin{eqnarray}
d\varphi & = & \partial _z\varphi \; dz +  \partial _{\bar{z}}\varphi \; d\bar{z}\\
d^*\varphi  & = &  i\, (\partial _{\bar{z}}\varphi \; d\bar{z}-\partial _z\varphi \; dz)
,
\end{eqnarray} 
and therefore, the second contribution is obtained from the first by
exchanging $e$ and $e^*$. In the end:
\begin{equation}
\Gamma^{(2)}_{\varepsilon}(a,b,e,e^*)
\simeq -{(e^*)^2\over 2\alpha}\, \log{(\varepsilon)}
+{(e^*)^2\over 2\alpha} \log{(|h_{a,b}(z)|)}-
{e\,e^*\over 2\alpha} \log{\left({h_{a,b}(z)\over \overline{h_{a,b}(z)}}\right)}
.
\end{equation}
Taking into account the two contributions and eliminating the overall divergence
$\varepsilon ^{-(e^2+(e^*)^2)/2\alpha}$
leads to the required correlation function since
the partition function on the sphere cancels between 
the numerator and the denominator.
We finally get:
\begin{equation}
\label{eqVertexCorr}
\langle \;  \ e^{{i\,e\over R}\int _{C}d\varphi + {e^*\over R}
\int _{C}d^*\varphi } \;
\rangle_{\Bbb{P}_1(\Bbb{C})}=
(b-a)^{-(e-e^*)^2/4\alpha}\;
(\overline{b-a})^{-(e+e^*)^2/4\alpha}
.
\end{equation}
Remembering the quantization conditions obtained in section
\ref{secVertex1} and 
the usual form of two point correlation functions in 2D CFT \cite{BPZ}, 
we recover the spectrum of the 
free compactified boson in the modular invariant sector, as expected:
$e^*=2\alpha n$ and $e=-m$. In the following, $\phi _{n,m}(z,\bar{z})$ will
denote the $(p_{n,m},\overline{p}_{n,m})$ $\widehat{U(1)}$ primary field. Equation 
(\ref{eqVertexCorr}) simply 
gives $\langle \phi _{n,m}^{\dagger}(b)\phi_{n,m}(a)\rangle $.

\subsection{Correlation functions on the torus}

In order to perform this computation on the torus, one has to proceed in 
three steps. The idea is to split the bosonic field into several parts:

\begin{itemize}

\item The usual instanton contribution, depending on winding numbers along
the two homology cycles of the torus. 

\item A saddle point contribution with vanishing monodromy along
homology cycles of the torus but which takes into account vertex operator
insertions.

\item The fluctuation contribution.

\end{itemize}

Renormalization issues only appear in the 
computation of the fluctuation contribution.
As for the computation of partition 
functions, a Poisson resummation has to be performed.
Here, only the final results will be given and commented.
Some details may be found in appendix \ref{secCalculs}.

\medskip

Let us introduce 
\begin{equation}
Z(a,b)=\omega _1 {\vartheta _1\left({b-a\over \omega _1},\tau\right)
\over \vartheta '_1(0,\tau)}
,
\end{equation}
and 
\begin{equation}
{\cal F}_{[(e_k)_k]}(z_1,\ldots ,z_n)=
\prod _{k\neq l}Z(z_k,z_l)^{e_k\,e_l/ 8\alpha}
.
\end{equation}
The following divisors \cite{Bost:89-Houches} on the torus 
($e_k^{\pm}=e_k\pm e_k^*$): 
\begin{equation}
\label{eqDiviseurs}
D_- = \sum _{k=1}^n e^-_k\, {z_k\over \omega _1}\ \mathrm{and}\quad
\overline{D}_+ =  \sum _{k=1}^n 
e^+_k\, {\overline{z_k}\over \overline{\omega_1}}
\end{equation}
are needed 
in order to define a Theta function contribution:
\begin{equation}
\label{eqCorrTheta}
\Theta _{\Gamma_L}(\tau,\overline{\tau},D_-,\overline{D}_+)=
\sum _{(p,\bar{p})\in \Gamma _L}
q^{{1\over 2}\,p^2}\, \overline{q}^{{1\over 2}\, \bar{p}^2}\;
\exp{\left(i{\pi\over \sqrt{\alpha}} (p\,D_-+\bar{p}\,
\overline{D}_+)\right)}
.
\end{equation}
Then the correlation function is given by the rather complicated expression:
\begin{equation}
\langle \prod _{k=1}^nV_{(e_k,e^*_k)}(z_k.\overline{z_k})
\rangle _{\Bbb{T}_{\tau}}=
{\Theta _{\Gamma_L}(\tau,\overline{\tau},D_-,\overline{D}_+)\over 
\Theta _{\Gamma_L}(\tau,\overline{\tau},0,0)}\times 
{\cal F}_{[(e^-_k)_k]}(z_1,\ldots ,z_n)\;
\overline{{\cal F}_{[(e^+_k)_k]}(z_1,\ldots ,z_n)}
.
\end{equation}
It captures all finite size effects in correlation functions. Let us 
determine
the asymptotics of this correlation function
in the limit of coinciding points, that is when 
$$\forall k\neq l,\quad |z_k-z_l|<<\mathrm{min}(|\omega _1|,|\omega_2|).$$
In this limit, $D_-$ and $\overline{D}_+$ vanish,
and $Z(a,b)\simeq b-a$. Therefore, 
we recover the expression obtained before for correlation
functions on the Riemann sphere.

\medskip

Let us notice the familiar CFT structure \cite{Friedan1} 
of these correlation functions
which appear as sesquilinear combinations of conformal blocks:
\begin{equation}
\sum _{I,\bar{I}}{\cal F}_I(x)\, \overline{{\cal F}_{\bar{I}}}(\overline{x})
,
\end{equation}
where $x$ denotes some complex coordinates over the relevant Teichm\H{u}ller
space. Here this space is nothing but the one associated with complex tori
with $n+1$ marked points\footnote{The $n+1$-th point is the reference point $P_0$.
It does not appear in the expression.} and the appropriate spin
structure on the 
elliptic curve.
The expression just obtained assumes that this correlation has been computed
using some special region in Teichm\H{u}ller space: namely all points $z_k$ 
and $z_0$ belong to the fundamental cell of the period lattice of the torus.
Locally, $\widehat{U(1)}$ conformal blocks are given by:
\begin{equation}
{\cal F}_{[(e_k)_k,p]}=
{q^{{1\over 2}\, p^2}\over \eta(\tau)}\,
e^{i{\pi\over \sqrt{\alpha}}\, p\,D_-}\,
{\cal F}_{[(e^-_k)_k]}(z_1,\ldots ,z_n)
,
\end{equation}
where $p$ is of the form $p_{l/2,m}$ with $l\equiv m\pmod{2}$. 

\medskip

During the eighties, very 
similar expressions appeared in the context of string theory and in
CFT \cite{DVVc1}. 
However, exactly as for partition functions, people
were interested in the modular invariant sector. More precisely, 
expression obtained for correlators were exactly the same, except for
the sum over all the moments $(p,\bar{p})$.
In the modular invariant sector, one 
has to use the lattice 
$\{ (p_{n,m},\overline{p}_{n,m})/\ (n,m)\in \Bbb{Z}^2\}$ which corresponds to 
the moments appearing $Z_{0,0}$.
In the Luttinger case, these moments belong
to the lattice 
$$\Gamma _{\mathrm{Lutt}}=
\{ (p_{l/2,m},\overline{p}_{l/2,m})/\ (l,m)\in \Bbb{Z}^2\quad l\equiv m
\pmod{2}\}.$$

\subsection{Free fermions}
\label{secFermions}

It is interesting to specialize these expressions to the case of free
fermions ($\alpha =1$), in order to compute the correlation functions
of fermionic operators. 
Right fermions have conformal spin $1/2$. 
They are $\widehat{U(1)}$ primary fields, 
{\it id est} chiral vertex operators with some specific choice of $(e,e^*)$. 
The following choice has conformal spin $1/2$: $e=1$ and
$e^*=-\alpha $ ($n=-1/2$ and $m=1$).
Let us now compute the two point
function on the torus:
\begin{equation}
G_{\alpha=1}(a,b)=
\langle 
\exp{\left({i\over R}\int_{C_{ab}} d\varphi -{\alpha \over R}
\int _{C_{ab}}d^*\varphi \right)}\ 
\rangle _{\omega _1\Bbb{Z}\oplus \omega _2\Bbb{Z}}
.
\end{equation}
We specialize to $\alpha =1$. In this case $e_-=2$ and $e_+=0$. The fluctuation 
contribution only contains a holomorphic contribution
$$\omega _1^{-1}\, { \vartheta '_1(0,\tau)\over \vartheta _1(z,\tau)},$$
where $z=(b-a)/\omega _1$.
The theta function contribution is a quotient of two theta functions. 
For $\alpha =1$, expression 
(\ref{eqCorrTheta}) specializes to:
\begin{equation}
\sum _{(p,\bar{p})\in \Gamma _L}
q^{{1\over 2}\,p^2}\, \overline{q}^{{1\over 2}\, \bar{p}^2}\;
e^{2i\pi p\,z}
.
\end{equation}
In the free case, the lattice $\Gamma _{\mathrm{Lutt}}$ 
coincides with $\Bbb{Z}^2$ and therefore
this theta function factorizes between holomorphic and anti-holomorphic
parts. 

\medskip

Putting all parts together we get:
\begin{equation}
\label{eqFermionCorr}
G_{\alpha=1}(a,b)=\omega _1^{-1} { \vartheta '_1(0,\tau)\over \vartheta _1(z,\tau)}
{\vartheta_3(z,\tau)\over \vartheta _3(0,\tau)}
,
\end{equation}
which may be rewritten as 
\begin{equation}
\label{eqFermionCorr2}
G_{\alpha=1}(a,b)={\pi \over \omega _1} N(\tau)\, H(z,\tau) 
,
\end{equation}
where ($y=e^{2\pi iz}$)
\begin{eqnarray}
N(\tau) & = & \prod _{n=1}^{+\infty}\left ({1-q^n\over 1+q^{n-1/2}}\right)^2\\
H(z,\tau) & = & {1\over \sin{(\pi z)}}\times
\prod _{n=1}^{+\infty}
{
(1+yq^{n-1/2})(1+y^{-1}q^{n-1/2})
\over 
(1-yq^n)(1-y^{-1}q^n)
}
.
\end{eqnarray}
This is nothing but the two point correlator of the free Dirac fermions 
on the torus.
This may be understood by showing that expression (\ref{eqFermionCorr}) 
satisfies the Dyson-Schwinger equations of Dirac theory.
A direct fermionic computation also gives the same result.

\medskip

Of course, when $\alpha \neq 1$, these correlation functions mix holomorphic and
anti-holomorphic parts. Moreover,
vertex operators considered here
($n=\pm 1/2$, $m=\pm 1$) are renormalized fermions in the interacting theory. More 
precisely, if $\psi _{0,R}^{\dagger}$ creates a bare right fermion, 
we have $\phi _{1/2,1}=Z^{-1/2}\, \psi _{0,R}^{\dagger}$ 
where $Z$ is a multiplicative renormalization constant. $Z$ vanishes
for local interactions, in an orthogonality catastrophe-like
scenario \cite{Anderson:67-1}.
Its physical
meaning is quite clear: $\phi _{1/2,1}$ creates a normed state
whereas $Z$ is the square norm of the state obtained by adding a bare electron to
the system. $Z=0$ simply means that adding a bare electron to the system drives 
it in a state orthogonal to all eigenstates of the interacting Hamiltonian. 
Saying this another way round: the eigenstates of the
interacting Hamiltonian have nothing
to do with the original fermions. The renormalisation constant $Z$ is also 
related to the discontinuity of the momentum distribution function at $k_F$ 
through Migdal's theorem (see \cite[Page 42 and 63]{AbrikosovGorkov}). Therefore, the
vanishing of $Z$ means that there is no discontinuity in the electron's momentum 
distribution at the Fermi surface.

\subsection{Inclusion of the toy-disorder}

Thanks to the results obtained in section \ref{secDisorder}, 
the effect of the toy-disorder 
on vertex operators correlators can be explicitly studied. 
Let us consider the product of vertex operators (\ref{eqCorrFct}).
We have
\begin{equation}
{\cal O}[\varphi +{\eta \over gR}] = {\cal O}[\varphi] \times
\exp{\left(\sum _{k=1}^n {i\,e_k\over \alpha}(\eta (\sigma_k)-\eta (\sigma_0))
+{e_k^*\over \alpha}\int _{C_{0k}}d^*\eta
\right)}
.
\end{equation}
Of course $d\eta = \eta'(\sigma)\,d\sigma$ and $d^*\eta =
\eta'(\sigma)\,dt$. The functional integral
\begin{equation}
F_1(\sigma_1,\ldots,\sigma_n)=
\int {\cal D}[\eta (\sigma)]
\,\delta(\eta(0))\; 
e^{-{1\over 2\gamma}\int _0^L(\eta '(\sigma))^2\,d\sigma }
\, e^{i\sum _k{e_k\over \alpha }(\eta (\sigma_k)-\eta (\sigma_0))}
\end{equation}
is easy to compute by the saddle point method. The result is expressed in
terms of the $\sigma_{kl}$s which are the representatives of 
$\sigma _k-\sigma _l\in \Bbb{R}/L\Bbb{Z}$ that belong to $[0,L[$.
\begin{equation}
F_1(\sigma_1,\ldots,\sigma_n)=
\exp{\left(-{\gamma \over 4\alpha ^2}\sum _{{k\neq l\atop (k,l)\in <1,n>^2}}
e_ke_l\, {\sigma_{kl}(L-\sigma_{kl})\over L}\right)}
.
\end{equation}
The other contribution is 
\begin{equation}
F_2(z_1,\ldots,z_n)=
\int {\cal D}[\eta (\sigma)]
\,\delta(\eta(0))\; 
e^{-{1\over 2\gamma}\int _0^L(\eta '(\sigma))^2\,d\sigma }
\, e^{\sum _k{e_k^*\over \alpha }\int _{C_{0k}}\eta '(\sigma)dt }
.
\end{equation}
For the sake of simplicity, let us give the result for a two point function.
The curve $C_{ab}$ we have used in our computations is nothing but the 
line connecting $z_1$ to $z_2$.  
We have
$$\int_{C_{12}} d^*\eta = {\Im{(z_{12})}\over \Re{(z_{12})}}\, \left(
\eta (\Re{(z_{2})})-\eta(\Re{(z_{2})})\right),$$
and using this expression we obtain:
\begin{equation}
F_2(z_1,z_2)=
\exp{\left({\gamma (e^*)^2\over 2\alpha ^2}
{u_{12}^2(L-\sigma_{12})\over L\,u_{12}}
\right)}
.
\end{equation}

These results deserve some comments. First of all, an exponential decrease of 
correlations in the spatial direction is observed. This is the net effect
of the toy-disorder.
Let us stress that the exponents of algebraic decays are not modified.
Restoring a finite wave-vector $k_F$ would show that
this exponential decrease
is in fact a $2 k_F$ contribution and
is {\em not} a localization effect.

\section{Laughlin's experiment and LLs}
\label{secLaughlin}

The aim of this section is to discuss the relation between
some incompressible Hall fluids
and the LL. The basic idea is that, in a certain limit, the Hall
fluid in the annular geometry is equivalent to a LL with
suitable parameters. 
In particular, we discuss the behavior 
of FQH fluids within the framework of the Luttinger CFT.

\medskip

We begin by recalling how to compute the physical charge and
current densities in the LL from the conserved $U(1)$ currents
of the underlying CFT. Then, revisiting Laughlin
{\it gedanken } experiment \cite{Laughlin:82-1}, 
we relate the filling fraction $\nu=1/(2p+1)$ of the 
QHE to the interaction parameter of the LL.
We are then able to express the correspondence between quantum Hall fluids
and LLs. In the forthcoming sections,
we discuss rational LLs. Some of these RCFTs
describe Laughlin Hall states (at filling fraction 
$\nu =1/(2p+1)$). After identifying these RCFTs,
we introduce a fractional charge on the edge
in order to describe the tower of excited states above
a given bulk excitation,
and discuss in a precise manner the resulting twisted theories. 

\subsection{Physical charge and current densities of the LL}
As we have already explained in previous sections, 
the LL admits an infrared
description by an effective CFT. This effective theory is
nothing but the theory of a free massless compactified boson, with special
boundary conditions that take into account the boundary conditions of
the underlying bare fermionic fields. As a free bosonic
CFT, it has a 
$\widehat{U(1)}\times \widehat{U(1)}$ symmetry generated by a spin one current.

\medskip

As explained in section \ref{secInteractions}, 
the physical electric charge and current 
densities are directly 
related to the CFT $U(1)$ currents of the Luttinger CFT.
A careful restitution of the speed and charge factors
into (\ref{eqDensityCharge}) and
(\ref{eqDensityCurrent}) yields
\begin{eqnarray}
\rho & = & {e\over \sqrt{\alpha}}\, 
\left(J(\sigma,t) + \overline{J}(\sigma,t)\right)   \\
j & = & {v_S\, e\over \sqrt{\alpha}}\, 
\left(J(\sigma,t) - \overline{J}(\sigma,t)\right) 
,
\end{eqnarray}
$e$ being the unit charge.
These densities can be expressed in terms of the Laurent modes 
$(J_n)_{n\in \Bbb{Z}}$ and 
$(\overline{J}_n)_{n\in \Bbb{Z}}$
of currents on the 
Riemann sphere \cite{BPZ}:
\begin{eqnarray}
\rho(\sigma,t) & = & {e\over L\sqrt{\alpha}}\, \sum _{n\in \Bbb{Z}}\left(
J_n\, e^{2\pi i\, {n\over L}\,(\sigma-v_St)}
+\overline{J}_n\, e^{-2\pi i\, {n\over L}\,(\sigma+v_St)}
\right)\\
j(\sigma,t) & = & 
{ev_S\over L\sqrt{\alpha}}\, \sum _{n\in \Bbb{Z}}\left(
J_n\, e^{2\pi i\, {n\over L}\,(\sigma-v_St)}
-\overline{J}_n\, e^{-2\pi i\, {n\over L}\,(\sigma+v_St)}\right)
.
\end{eqnarray}
Therefore, the total charge and average electric current around the circle
are given in terms of Fourrier modes of the currents in the
underlying CFT by
\begin{eqnarray}
q & = & {e\over \sqrt{\alpha}} \, (J_0+\overline{J}_0)\\
I & = & {ev_S\over L\sqrt{\alpha}}\,
(J_0-\overline{J}_0)
.
\end{eqnarray}
It is now convenient to introduce some charge densities on each branch 
$\rho _R$ and $\rho _L$ such
that
\begin{equation}
\label{eqCurrentA}
\rho = \rho _L+\rho _R,
\quad
j =v_S\,(\rho _R-\rho _L)+ 2e\, {v_S\over L\alpha}\, \chi
.
\end{equation}
The second relation takes into account the effect of the potential 
vector on the fermion momenta and therefore on the electric current.
Let us recall the usual non relativistic 
expression of the current:
\begin{equation}
J_{\mu} = {e\,\hbar \over 2m\,i}\left(\psi \partial _{\mu}\psi^*-
\psi ^*\partial _{\mu}\psi\right) -{e^2\over 2m}|\psi|^2\,A_{\mu}
.
\end{equation}
The second term, proportional to the vector potential, is imposed
by gauge invariance of the current. Formula (\ref{eqCurrentA})
may be guessed from the study of a free fermion gas in two dimensions, 
confined on a cylinder. 
Chiral charges are expressed in terms of 
CFT operators:
\begin{equation}
q_R={e\over \sqrt{\alpha}}\,J_0-e\,{\chi \over \alpha}\quad
q_L={e\over \sqrt{\alpha}}\,\overline{J}_0 +e\,{\chi \over \alpha}
.
\end{equation}
These expressions will now be used for understanding the
relation between the LL physics and the QHE. 

\subsection{Charge transport between the chiralities in a LL}

Laughlin \cite{Laughlin:82-1} 
and Halperin \cite{Halperin:82-1} considered a 
quantum Hall fluid localized on a ring
or on a cylinder
threaded by a
magnetic flux $\Phi$. In fact, even though the cylinder
geometry cannot be realized experimentally, it is
more convenient for our purposes to consider
the quantum Hall effect on the cylinder.
These authors study charge
transport between the two edges 
when this flux is adiabatically switched on.
In the integer QHE,
when the flux is increased by one flux quantum $\Phi _0$, 
the system returns in the same
state except that some electrons have moved from one edge
to the other. 

In the FQHE, a fractional charge is 
transferred \cite{Tao:84-1}, and we have to shift
the magnetic flux by several flux quanta before the
system gets back on a state with an integer number of
transferred electrons. More precisely,
for a filling fraction $\nu $, the charge transferred during an increase 
$\Phi \mapsto \Phi + \Phi_0$ is equal to $\nu e$. For $\nu =q/p$, after 
$\Phi \mapsto \Phi +p\phi_0$, $q$ electrons are transferred.
This may be understood as
the effect of the electric field induced by the flux variation, in presence
of a Hall conductivity $\sigma_H=\nu e^2/h$. 

\medskip

Let us now consider the LL with an interaction parameter $\alpha$.
The partition function with a magnetic flux $\chi \Phi_0$ is
equal to:
\begin{equation}
\label{eqRatio1}
Z_{\mathrm{Lutt}}(\chi)= {1\over |\eta (\tau)|^2}\,
\sum _{{(n,m)\in (\Bbb{Z}/2)\times \Bbb{Z}\atop
2n\equiv m\pmod{2}}}
q^{{1\over 2}p_{n,m+2\chi}^2}
\overline{q}^{{1\over 2}\overline{p}_{n,m+2\chi}^2}
,
\end{equation}
where 
\begin{equation}
\label{eqRatio2}
p_{n,m}=n\sqrt{\alpha}+{m\over 2\sqrt{\alpha}}\quad
\overline{p}_{n,m}=n\sqrt{\alpha}-{m\over 2\sqrt{\alpha}}
.
\end{equation}
States of interest will be $\widehat{U(1)}\times \widehat{U(1)}$ 
highest weight states. Let us introduce the following notation:
\begin{equation}
|n,m\rangle_{\chi}=|p_{n,m+2\chi},\bar{p}_{n,m+2\chi}\rangle 
.
\end{equation}
The $|n,m\rangle_{\chi}$ can be viewed as the 
$|p_{n,m},\bar{p}_{n,m}\rangle $ state 
continuously deformed by the external magnetic flux. 

\medskip

The partition function (\ref{eqRatio1}) gives
the spectrum of charges on both
edges. For the $|n,m\rangle_{\chi}$ state, we have:
\begin{equation}
\label{eqSpecCharge}
\begin{cases}
q_R = e\,\left(n+{m\over 2\alpha}\right)\\
q_L =  e\,\left(n-{m\over 2\alpha}\right)
.
\end{cases}
\end{equation}
The total charge is therefore $2n\, e$ which is {\em always} an integer. But
charges on each edge can be non integer. 
Note that charge density fluctuations, which corresponds to 
$\widehat{U(1)}$ descendants, are {\em globally neutral}: they don't change
left and right total charges.

\medskip

We now study in more details charge transport between chiralities in
the Luttinger CFT. This will enable us to guess the correspondence 
between edges excitations of the FQH fluid at $\nu =1/q$ 
and an appropriate Luttinger CFT.

\subsection{Laughlin experiment revisited}

Let us now start from a Luttinger CFT with an interaction parameter 
$\alpha $. 
At zero temperature and zero magnetic flux, the system is in the 
$|p_{0,0},\bar{p}_{0,0}\rangle=|0,0\rangle_{\chi=0}$ state.
The chiral charges $q_R$ and $q_L$ both vanish in this state.
If we increase adiabatically the magnetic flux from $\chi =0$ to 
a finite value $\chi \in [0,1/2[$,
the system goes adiabatically to the state 
$|0,0\rangle_{\chi}$. 
In this intermediate
state, we still have
$q_L=q_R=0$
but there is a non zero current:
$I={2e\,v_S\over L\alpha}\, \chi.$.
When $\chi $ reaches $1/2$, the lowest energy state is no longer
$|0,0\rangle_{\chi}$ but for $1/2<\chi <3/2$, it is
$|0,-2\rangle_{\chi}$.  
But in this new state, a charge has been transferred from one chirality to 
the other:
$$
\begin{cases}
q_R = e/\alpha \\
q_L = -e/\alpha
.
\end{cases}
$$
We immediately notice that this behavior is very similar 
to the one of a FQH fluid
as described by Laughlin \cite{Laughlin:82-1} 
and by Tao and Wu \cite{Tao:84-1}.
If we now identify the left and right branches of the LL
with the two different edges of the cylindrical sample, 
the charge transferred from one {\em edge} to the other is equal to $e/\alpha$
as Laughlin considered. 
Therefore, the
filling fraction $\nu$ should be related to the interaction parameter by
$\alpha \,\nu =1$.
Assuming this relation, 
it is straightforward to compute the Hall conductivity from Luttinger CFTs:
let us assume that the two edges are not connected to any charge reservoir and 
assume that we increase the magnetic flux $\chi $. 
While increasing the flux, the system
goes through various states: 
\begin{equation}
|0,0\rangle _{\chi} \longrightarrow |0,-2\rangle _{\chi}
\longrightarrow |0,-4\rangle _{\chi} \longrightarrow \ldots
\end{equation}
The charge on the edge is therefore a step function of the magnetic 
flux. More precisely, for
$k+1/2< \chi < k+3/2$, we have $q_R=-q_L=e\nu\, k $. When $k$ reaches
$p-1/2$, we have $q_R=-q_L=e\nu p = qe$, as expected from Tao and Wu.
As we shall
see in section \ref{secFusion}, the state $|0,-2q\rangle_{\chi}$ is obtained by 
destroying electrons on the left edge and re-creating them on the right edge.
Let the magnetic
flux be linearly increased with time, then by
induction, an orthoradial electric field appears. As we have
just seen, charges are then transferred from one edge to the other. 
The voltage between the two edges is given by
$$U= {h\over e}\, {d\chi (t)\over dt},$$
and charges are transferred at the following rate:
$$I_{\perp}= e\nu \, {d\chi (t)\over dt}.$$
Therefore, we find:
\begin{equation}
\label{eqClassicHall}
I_{\perp} = {e^2\over h}\,\nu \, U
,
\end{equation}
which is the correct form of Hall conductivity. As we have already seen, 
the longitudinal intensity is a periodic function of the flux, the 
average of which 
is zero. Therefore, the longitudinal conductivity is zero.

\subsection{Another transport experiment}

Another transport experiment consists in bringing the two edges of the sample
at different potentials. In this case, a current should appear along the 
Hall ring. We shall see how to recover this result
from the effective Luttinger CFT, thus supporting our identification of 
the effective theory of edge excitations by an appropriate Luttinger CFT.

\medskip

Let us couple
the two edges of our cylinder to different electric potentials:
\begin{equation}
H_{elec}=-\int _0^L(V_R(\sigma)\,\rho_R(\sigma)
+V_L(\sigma)\,\rho_L(\sigma))
\,d\sigma
.
\end{equation}
Assuming $\chi=0$, this can be rewritten as
\begin{equation}
H_{elec}=-\int _0^LV(\sigma)\rho(\sigma)\,d\sigma
-\int_0^L
{V_R(\sigma)-V_L(\sigma)\over2v_S}\,j(\sigma)\,d\sigma
,
\end{equation}
where $V(\sigma)=(V_R(\sigma)+V_L(\sigma))/2$ and
$U_{\perp}=V_R-V_L$.
Therefore,  
the electric potential between the two edges can be viewed
as a vector potential $A_{\mathrm{eff}}=U_{\perp}/2v_S$ in the effective 
1D LL.
The effective magnetic flux is given by
\begin{equation}
\chi_{\mathrm{eff}}= {eL\over 2hv_S}\, U_{\perp}
.
\end{equation}

\medskip

We may apply our previous results about the electric current induced by
a magnetic flux in a LL. 
However, 
as we computed in section \ref{secCurrents}, 
the current in our mesoscopic cylinder should
be periodic in the magnetic 
flux. But that is {\em not} the expected behavior in a Hall 
experiment. The solution to this apparent paradox lies in the fact that, 
in a Hall experiment, 
edges are connected to
an electric generator which, besides imposing $U_{\perp}$, 
can bring charges into the system. Let us discuss this
point more precisely, assuming $\alpha =q\in 2\Bbb{N}+3$. 

When the
flux reaches $1/2$, a charge $\pm e/q$ appears at the edge. The external generator
does not bring any charge in to neutralize it, since only integer charges 
can be brought by the external generator. Hence, nothing special happens
compared to the discussion of an isolated Luttinger ring.
Current periodicity breaks down when the total charge becomes $\pm e$ on each
edge. Then, the generator may neutralize the charge on each edge. For
$\chi $ slightly less than $q-1/2$, the system is in the state
$$|0,2(1-q)\rangle_{\chi}$$
Then when $\chi $ becomes larger than $q-1/2$, 
instead of going back in the state $$|p_{0,0},\bar{p}_{0,0}\rangle
= |0,-2q\rangle_{\chi},$$
with left and right charges $q_L=-q_R=1$, 
the system jumps to the state
$|0,0\rangle_{\chi}$
In this state, we again have $q_{L,R}=0$ but the current is non zero. 
Notice that the energy varies due to edge effects.
The energy change is brought in by the external electric generator. 
The electric Hall conductivity can easily 
be computed.
In terms of the effective magnetic flux,
the ``classical'' Hall current is given by
$$I={2v_Se\over L\alpha}\, \chi_{\mathrm{eff}}.$$
Therefore, reintroducing the potential difference
between the two edges of the system, we obtain
\begin{equation}
I = {e^2\over h}\, \alpha ^{-1}\, U_{\perp}= {e^2\over h}\, \nu\, U_{\perp}
.
\end{equation}
Of course, the result obtained above crucially relies
on the charge transfer 
mechanism
between the electric generator and the Hall sample. A correct 
treatment of this situation should
use an explicit description of the connection of the two edges
to charge reservoirs at different chemical potentials and of course a 
quantum description of these two reservoirs. For completeness, let us recall
that tunneling between chiralities of an Hall bar has been widely studied
these last years: see \cite{Kane:94-1} and also section V.A of reference
\cite{Fisher:96-1}. Recent experimental work \cite{Etienne:97-1} 
has been performed on tunneling
experiments in the $\nu =1/3$ case. This experiment, based on a measurement of
the current noise due to charge tunneling between two edges,
seem to establish the existence of fractionally charged
edge excitations.

\medskip
To summarize, we have shown that transport
properties of the quantum Hall fluid can be understood
within the framework of the Luttinger CFT.
It is important to notice that not all LLs
can be obtained from quantum Hall fluids: the interaction parameter $\alpha $
must be rational. Of course, our approach does not describe the transition
between Hall plateaux. We only provide an effective theory for edge
excitations within a given plateau.

It is interesting to understand more precisely
why these rational values are special. This is our main motivation for studying
Luttinger CFTs from the point of view of RCFTs. 

\section{Rational Luttinger CFTs}
\label{secRational}

RCFTs arose originally from Friedan and Shenker
work in the late 80s \cite{Friedan1}. 
Roughly speaking, these are CFTs with
a very large symmetry algebra; so large that the Hilbert space of states
is a direct sum of a {\em finite} number of irreducible representations of
this symmetry algebra (see \cite{MSRev} for a review). 
By analogy, a
LL is called rational if and only if the effective
theory is a RCFT.

\medskip

We begin by deriving a rationality condition for the 
LL. Generalized characters with respect to the extended 
symmetry algebra will be computed.
This will enable us to re-express the partition function
in terms of the extended characters. As we shall see, partition functions
recently discovered by Cappelli and collaborators
\cite{Cappelli:96-1} belong to our list.

\subsection{Rationality criterion for the Luttinger CFT}

The partition function for the Luttinger CFT is given
by equations (\ref{eqRatio1}) and (\ref{eqRatio2}). 
If the effective theory is a RCFT, then there should exist some 
conformal primary fields of dimensions $(h,0)$ which extend the 
$\widehat{U(1)}$ algebra. This implies a condition similar to the one
found for the usual free modular invariant bosonic theory: 
$\alpha /2\in \Bbb{Q}$. In the present situation, we find 
that $\alpha $ is rational. Let us write $\alpha =q/p$ where
$q$ and $p$ are coprime positive integers. 

\medskip

The conformal dimensions $(h,\bar{h})$ of $\phi _{l/2,m}$ are given by:
\begin{equation}
\label{eqRatio3}
\left(
{(lq+mp)^2\over 8pq},{(lq-mp)^2\over 8pq}
\right)
.
\end{equation}
Here $l$ and $m$ are integers of same parity (congruence condition).
From Gauss lemma \cite{Serre}, 
we deduce that if $\phi _{l/2,m}$ has a vanishing $\bar{h}$
dimension, then there exists $k\in \Bbb{Z}$ such that $l=kp$ and
$m=kq$. 
In this case, we have 
$$h_{kp/2,kq}={pq\over 2}k^2.$$
Nevertheless, in the Luttinger model, one has the congruence condition 
$l\equiv m\pmod{2}$ and therefore, two cases will be discussed separately:

\paragraph{ Case $p\equiv q\pmod{2}$} Since $p$ and $q$ are coprime,
both of them are odd.
The congruence condition is satisfied for all
$k\in \Bbb{Z}$. 
Holomorphic $\widehat{U(1)}$ primaries are 
$\phi _{kp/2,kq}$ for any $k\in \Bbb{Z}$. For $k$ even, the $\bar{h}$ dimension
is an integer but for $k$ odd, it is half integer. 
Let us denote by $\cal{A}$ the chiral algebra generated by the $U(1)$
current $J(z)$ and all fields $\phi _{kp/2,ku}$ for $k\in \Bbb{Z}$. 
This operator algebra 
${\cal A}$ is clearly $\Bbb{Z}_2$
graduated by $2h \pmod{2}$ and we write ${\cal A}={\cal A}_+\oplus 
{\cal A}_-$. 
Fields 
$\phi _{kp/2,ku}$ for $k\in 2\Bbb{Z}+1$ belong to ${\cal A}_-$ and fields
$\phi _{kp,2ku}$ belong to ${\cal A}_+$.

\paragraph{Case $p\equiv q+1\pmod{2}$} One of these two integers is even. 
Therefore, the
congruence condition is satisfied only when $k$ is even.
The holomorphic $\widehat{U(1)}$ primaries are 
$\phi _{kp,2kq}$ for all $k\in \Bbb{Z}$.
The chiral algebra only consists of integer spin fields.

\subsection{Extended characters}

Extended characters are defined in CFT as follows: let us consider ${\cal V}$ an
irreducible representation of a chiral algebra\footnote{Which of course 
contains the 
Virasoro algebra.}, then
\begin{equation}
\chi _{{\cal V}}(\tau )=\mathrm{Tr}_{{\cal V}}(q^{L_0-c/24})
,
\end{equation}
where $q=\exp{(2\pi i\tau)}$. As shown by Cardy \cite{Cardy}, 
these characters are building
blocks for the partition function of CFTs on the torus. 

\medskip

Here, the extended algebra is formed by the integer spin holomorphic fields of
the Luttinger CFT. It is therefore quite obvious to obtain extended characters.
Fields $\phi _{kp,2kq}$ and their $\widehat{U(1)}$ (right) descendants generate
the chiral algebra ${\cal A}_c$. In terms of $J_0$'s eigenvalue $p$, they
correspond to $p\in 2\sqrt{pq}\Bbb{Z}$.
An irreducible 
representation of ${\cal A}_c$ 
is a direct sum of $\widehat{U(1)}$ irreducible representations, and
may be characterized by a set of $J_0$
eigenvalues of the form $p_0+2\sqrt{pq}\,\Bbb{Z}$. 
Let us now determine what are the possible values of $p_0$.
For this, we impose that the Laurent modes of all chiral 
vertex operators generating ${\cal A}_c$ and of the chiral vertex
operator corresponding to $p_0$ satisfy {\em commutation relations}. 
The operator product expansion of {\em chiral} vertex
operators
\begin{equation}
\label{eqChiralOPE}
V_{p_1}(z)\,V_{p_2}(\xi)=(z-\xi)^{p_1p_2}\; V_{p_1+p_2}(\xi)
\end{equation}
provides us with such a condition. The Laurent modes of
$V_{p_1}$ and $V_{p_2}$ satisfy commutation relations if and only if 
$p_1p_2\in \Bbb{Z}$. Therefore, we
have $p_0\in (2\sqrt{pq})^{-1}\Bbb{Z}$. 
The extended characters are therefore given by
\begin{equation}
\label{eqRatio5}
\chi_{\lambda }(\tau)={1\over \eta(\tau)}\sum _{n\in \Bbb{Z}}
q^{(\lambda +4npq)^2/8pq}
,
\end{equation}
where $\lambda \in {\Bbb{Z}/4pq\Bbb{Z}}$.
In terms of theta functions with characteristic, we have:
\begin{equation}
\label{eqRatio6}
\chi_{\lambda }(\tau)={\vartheta \left[
\begin{array}{c}
\lambda /4pq\\
0
\end{array}\right]
(0,4pq \tau)\over \eta(\tau)}
.
\end{equation}
When $p\equiv q\pmod{2}$, it is convenient to introduce extended
characters\footnote{Analogous to the super-characters of super-conformal field 
theories.} defined by:
\begin{equation}
\label{eqRation7}
\varphi ^{(\pm)}_{\lambda} = \chi _{\lambda}\pm \chi _{\lambda +2pq}
,
\end{equation}
where $\lambda \in \Bbb{Z}/2pq\Bbb{Z}$.
Their modular transformation properties can easily be computed. A
subset of them, namely the $\varphi _{2\lambda}^{(+)}$ where 
$\lambda \in \Bbb{Z}/pq\Bbb{Z}$, forms a unitary representation of
the modular group $\Gamma (S,T^2)$ \cite{Koblitz1}. 
The corresponding $S$ matrix \cite{Cardy} is given by: 
\begin{equation}
\label{eqRatio8}
\varphi _{2\lambda}^{(+)}(-1/\tau)=
{1\over \sqrt{pq}}\sum _{\lambda '\in \Bbb{Z}/pq\Bbb{Z}}
e^{ -2\pi i{\lambda \lambda '\over pq}}\,
\varphi _{2\lambda}^{(+)}(\tau ).
\end{equation}
We also have:
\begin{equation}
\label{eqRatio9}
\varphi _{2\lambda}^{(+)}(\tau +2)=
e^{2\pi i ({\lambda ^2\over 4pq} -{1\over 12})}\,
\varphi _{2\lambda}^{(+)}(\tau)
.
\end{equation}
In conclusion, we notice that the 
underlying CFT is nothing but a
$\Bbb{Z}/N\Bbb{Z}$ CFT in the sense of \cite{Degio2} where
$N=4pq$. However, since the boundary conditions to be used in the Luttinger
model are not modular invariant with respect to the full modular group
$SL(2,\Bbb{Z})$, the partition function of a Luttinger CFT does not belong to 
the list obtained in
\cite{Degio2}. Instead of giving a complete classification of 
possible partition functions (see \cite{Gannon:96-1} for classification 
results), we shall express the partition function of the
rational Luttinger model in terms of these extended characters.

\subsection{Partition functions of rational Luttinger CFTs}
\label{secRationalZ}

We now express the partition function 
of the rational Luttinger models in terms of
the extended characters. These formulae will give us more insight into
the structure of these theories. Of course, two cases will be analyzed separately
according to $p$ and $q$'s relative parity.

\subsubsection{Case $p$ and $q$ odd}

We can decompose the Luttinger partition function into two terms:
$$Z_1={1\over |\eta (\tau)|^2}\sum _{(r,s)\in \Bbb{Z}^2}
q^{{(rq+sp)^2\over 2}}
\overline{q}^{{(rq-sp)^2\over 2}},$$
and
$$Z_2={1\over |\eta (\tau)|^2}\sum _{(r,s)\in \Bbb{Z}^2}
q^{{(rq+sp+pq)^2\over 2}}
\overline{q}^{{(rq-sp)^2\over 2}}.$$
Since $p$ and $q$ are coprime, there exists $(u,v)\in \Bbb{Z}^2$ such that
$qu-pv=1$. These integers are not unique and indeed $qu+pv$ is only 
defined modulo $2pq$. Let
us denote its class modulo $2pq$ by $\omega\in \Bbb{Z}/2pq\Bbb{Z}$. 
Let us now introduce the following subsets of $\Bbb{Z}^2$: 
\begin{eqnarray}
C_{\lambda } & = & \left((\{\lambda\}+2pq\Bbb{Z})\times 
(\{\overline{\,\lambda}\,\}+2pq\Bbb{Z})\right)\\
R_1 & = & \{ (rq+sp,rq-sp)\ ;\quad (r,s)\in \Bbb{Z}^2\}\\
R_2 & = & \{ (rq+sp,rq-sp+pq)\ ;\quad (r,s)\in \Bbb{Z}^2\}
,
\end{eqnarray}
where $(\lambda,\overline{\lambda}) \in (\Bbb{Z}/2pq\Bbb{Z})^2$.
It is obvious to prove that 
$(C_{\lambda,\omega\lambda})_{\lambda \in \Bbb{Z}/2pq\Bbb{Z}}$ is a partition
of $R_1$ and that 
$(C_{\lambda,\omega\lambda+pq})_{\lambda \in \Bbb{Z}/2pq\Bbb{Z}}$ is a partition
of $R_2$.
Therefore, we get:
\begin{eqnarray}
\label{eqRatio22}
Z_1 & = & \sum _{\lambda \in \Bbb{Z}/2pq\Bbb{Z}}
\chi _{2\lambda }(\tau)
\,\overline{\chi _{2\omega\lambda }(\tau)}\\
Z_2 & = & \sum _{\lambda \in \Bbb{Z}/2pq\Bbb{Z}}
\chi _{2\lambda +2pq }(\tau)
\,\overline{\chi _{2\omega\lambda }(\tau)}
.
\label{eqRatio23}
\end{eqnarray}
Summing these two contributions gives
\begin{equation}
\label{eqRatioZ}
Z_{\mathrm{Lutt}}(\tau,\bar{\tau})=\sum_{\lambda \in \Bbb{Z}/pq\Bbb{Z}}
\varphi ^{(+)}_{2\lambda }(\tau)
\,\overline{\varphi ^{(+)}_{2\omega\lambda }(\tau)}
,
\end{equation}
which is exactly the partition function found by Cappelli and 
his collaborators in \cite{Cappelli:96-1}.

\subsubsection{Case $p$ or $q$ even}

The same kind of computation can be carried 
out in this case.
Again, let us decompose the partition function  into two terms:
$$Z_1={1\over |\eta (\tau)|^2}\sum _{(r,s)\in \Bbb{Z}^2}
q^{{(rq+sp)^2\over 2}}
\overline{q}^{{(rq-sp)^2\over 2}}
,
$$
and
$$Z_2={1\over |\eta (\tau)|^2}\sum _{(r,s)\in \Bbb{Z}^2}
q^{{(2(rq+sp)+p+q)^2\over 2}}
\overline{q}^{{(2(rq-sp)+q-p)^2\over 2}}.
$$
Exactly as in the previous case, let us apply Bezout theorem to find 
$(u,v)\in\Bbb{Z}^2$ such that $qu-pv=1$. These two numbers are not unique:
$(u+kp,v+kq)$ where $k\in \Bbb{Z}$ satisfies the same property. If we
define $\omega=pu+qv$, we have $\omega\, (p+q)=q-p+2pq(u+v)$. Since $p$ and
$q$ do not have the same parity, it is possible to choose $(u,v)$ such
that $u+v$ is even. With such a choice, $\omega\, (p+q)\equiv q-p\pmod{4pq}$. 
But now, the freedom on $u$ and $v$ is restricted and therefore
$\omega \in \Bbb{Z}/4pq\Bbb{Z}$. We also have $\omega ^2=1$ in 
$\Bbb{Z}/4pq\Bbb{Z}$.
Finally
we get:
\begin{eqnarray}
\label{eqRatio27}
Z_1 & = & \sum _{\lambda \in \Bbb{Z}/2pq\Bbb{Z}}
\chi _{2\lambda }(\tau)
\,\overline{\chi _{2\omega\lambda }(\tau)}\\
Z_2 & = & \sum _{\lambda \in \Bbb{Z}/2pq\Bbb{Z}}
\chi _{2\lambda +1 }(\tau)
\,\overline{\chi _{\omega(2\lambda+1) }(\tau)}
\label{eqRatio28}
,
\end{eqnarray}
and therefore
\begin{equation}
\label{eqRatioZ2}
Z_{\mathrm{Lutt}}(\tau,\bar{\tau})  =  \sum _{\lambda \in \Bbb{Z}/4pq\Bbb{Z}}
\chi _{\lambda }(\tau)
\,\overline{\chi _{\omega\lambda}(\tau)}
.
\end{equation}
We have thus obtained the field contents of all rational 
Luttinger CFTs. 

\section{Introduction of an edge charge on the boundary}
\label{secTwisted}

The purpose of this section is to present a CFT
description of
edge excitations above a given bulk excitation in a quantum Hall
fluid. We begin by explaining the motivations of our computations.
They are based on the introduction of a fractional edge charge in the
Luttinger CFT. Then, we shall compute the partition functions
of these ``twisted'' Luttinger CFTs with edge charges and express them in terms of
generalized characters in the rational cases. Then, it will be appropriate to
provide a physical interpretation of the chiral algebra of the theory. Following
Fisher and Stone \cite{Stone:94-1}, fermions localized on one edge of the 
sample will enter the game. Duality properties of the Luttinger CFT will also
be discussed. Finally, 
examples such as the Fermi
liquid and Laughlin Hall fluids will be considered. 

\subsection{Why introducing an edge charge?}

It has been known since Laughlin that a quantum Hall fluid
at filling fraction $1/(2p+1)$ 
is incompressible \cite{Laughlin1}.
This means that all bulk excitations have a gap in energy: it is therefore 
difficult to create them contrarily to the case of the Fermi liquid for example,
which has a finite compressibility \cite{AbrikosovGorkov} since
low lying quasi-particle quasi-hole excitations can be easily excited.
As explained by Laughlin, bulk 
excitations can be described and correspond to objects with fractional charge
and anyonic statistics \cite{WilczekAnyons}. 

\medskip

However, some low energy excitations do
exist in a quantum Hall fluid: these are the edge excitations. They are present
above any bulk excitation. We have argued in the previous sections that
edge excitations above the bulk ground state were described by the
Luttinger CFT
at $\alpha =\nu^{-1}$. Here, we would like to give 
a description of edge excitations
above the excited bulk states.

\medskip

Starting from a quantum Hall fluid on a cylinder and creating a bulk excitation
implies creating a fractional charge excess or 
defect in the bulk \cite{Laughlin1}. 
But by
charge conservation, the opposite charge should appear on the edge. Therefore
a fractional charge appears on the edge. But then, creating edge 
excitations should not change the charge on the edge (except by 
integers). This leads us to considering some twisted sectors of the Luttinger
CFT. These sectors are defined by a congruence condition on
the total charge: 
\begin{equation}
\label{eqExc1}
{q_R+q_L\over e}\equiv  r \pmod{1}
.
\end{equation}
The aim of this section is to study these twisted sectors.

\subsection{Partition functions with an edge charge}

We now compute the partition function of the Luttinger model
in presence of an edge charge satisfying the condition (\ref{eqExc1}). 
We will first of all compute it for generic $\alpha$. Then we specialize
$\alpha $ to some rational value and express the partition function in terms of
generalized characters.

\subsubsection{Computation of the partition function}

Because of the identification of the physical charge in the 
Luttinger model, we modify our bosonic field boundary conditions
to impose a given charge modulo $1$ on the boundary. More precisely, we
impose
$$\int _{(a)}d\varphi \equiv \pi R\,r \pmod{\pi R}.$$
In the previously considered cases, $r=0$. The
modified functional integral will then be
\begin{equation}
\label{eqBoundaryTwisted}
\int _{{\cal C}(r)} {\cal D}[\varphi]\, W[\varphi]
= {1\over 2}\sum _{(\varepsilon ,\varepsilon ')\in \{0,1/2\}^2}
(-1)^{4\varepsilon\varepsilon '}
\int _{
{\Delta _{\sigma}\varphi \equiv 2\pi R(r/2+\varepsilon) \pmod{2\pi R}
\atop {\Delta _{u}\varphi \equiv 2\pi R\varepsilon' \pmod{2\pi R}}}}
{\cal D}_R[\varphi]\, W[\varphi]
.
\end{equation}
All computations can be carried out exactly as in the usual case and
we obtain, in the presence of a magnetic flux $\chi\, \Phi_0$ through the
ring:
\begin{equation}
Z_{\mathrm{Lutt}}^{(r)}(\tau,\bar{\tau},\chi)=
{1\over |\eta (\tau)|^2}
\sum _{{(n,m)\in (\Bbb{Z}/2)\times \Bbb{Z}\atop
2n\equiv m\pmod{2}}}
q^{{1\over 2}p_{n+r/2,m+2\chi}^2}\,\bar{q}^{{1\over 2}\bar{p}_{n+r/2,m+2\chi}^2}
.
\end{equation}
The $(n,m)$ term in this sum corresponds to a total charge of
$(2n+r)\, e$ as expected.

\subsubsection{Expression in terms of generalized characters}

Exactly as before, it is quite interesting to express these partition functions
for $\alpha $ rational and
in a zero magnetic field in terms of generalized characters. Of course, although
the vacuum structure depends on the sector, fields of the chiral algebra
can still be used for going from one representation of $\widehat{U(1)}$
to the other. 
We choose $\alpha =q/p$. Again,
two cases must be treated separately according to $p+q$'s parity.

\paragraph{$p$ and $q$ odd}
Conformal dimensions of $\widehat{U(1)}$ primaries are given by
$${((r+l)q\pm (m+2\chi ))^2\over 8pq},$$
where $(l,m)\in \Bbb{Z}^2$ have the same parity. The interesting case
arises when $r=a/q$ with $a\in \Bbb{Z}$. 
Then for $\chi =0$, the spectrum of conformal dimensions becomes
$${(a+lq\pm m)^2\over 8pq}.$$
When $a$ is even, we note that $lq\pm mp +q\equiv 0\pmod{2}$ and therefore, 
exactly as in section \ref{secRationalZ}, we expect the partition function
to be a sesquilinear combination of the generalized characters 
$\varphi ^{(+)}_{2\lambda}$. We can easily prove that in general:
\begin{equation}
Z_{\mathrm{Lutt}}^{({a\over q})}(\tau,\bar{\tau})=
\sum_{\lambda \in \Bbb{Z}/2pq\Bbb{Z}}
\left(
\chi _{a+2\lambda}(\tau)+\chi_{a+2\lambda+2pq}(\tau)
\right)\,
\overline{\chi_{a+2\omega\lambda}(\tau)}
.
\end{equation}
This partition function can be rewritten in terms of generalized characters
\begin{equation}
Z_{\mathrm{Lutt}}^{({a\over q})}(\tau,\bar{\tau})=
\sum_{\lambda \in \Bbb{Z}/pq\Bbb{Z}}
\varphi^{(+)}_{a+2\lambda}(\tau)\,
\overline{\varphi^{(+)}_{a+2\omega\lambda}(\tau)}
.
\end{equation}

\paragraph{$p$ or $q$ even}
The same kind of analysis yields the following result:
\begin{equation}
Z_{\mathrm{Lutt}}^{({a\over q})}(\tau,\bar{\tau})=
\sum_{\lambda \in \Bbb{Z}/4pq\Bbb{Z}}
\chi _{a+\lambda}(\tau)
\,
\overline{\chi_{a+\omega\lambda}(\tau)}
,
\end{equation}
which reduces to (\ref{eqRatioZ2}) when $a=0$.
We have therefore obtained the operator content of the twisted sectors of
Luttinger rational models in terms of the
characters relative to the maximal 
chiral algebra. The study of this partition function tells us how
gapless are excitations are organized above bulk excitations.

\subsection{Fusion rules in the rational Luttinger models}
\label{secFusion}

Let us now  discuss the fusion rules \cite{MSRev} of these models. 
The partition 
functions are of the following form: 
$$\sum _i \chi _i (\tau) \, \overline{\chi _{\sigma(i)}(\tau)},$$
where $i$ runs over a finite set of indices and $\sigma $ 
is a bijection of this set. Of course, this decomposition is relative
to the maximal symmetry algebra of the model ${\cal A}\otimes {\cal A}$
\cite{MS2}.
In this case, the generalized primary field can be indexed by a 
chiral index $i$.
We would like to find the selection rules for the
operator product algebra \cite{BPZ} of the Luttinger CFT. 

Invariance under $\tau \mapsto -1/\tau$ shows, thanks to Verlinde
formula \cite{Ver}, that $i\mapsto \sigma (i)$ defines an automorphism of
the fusion algebra \cite{DV}. Therefore, the selection rules of the operator
algebra of our Luttinger CFT are given by the fusion
rules for the chiral fields, exactly as in usual CFTs.

\paragraph{$p$ and $q$ odd}
Here we may consider the theory with respect to ${\cal A}_+$ or to 
the extended algebra ${\cal A}={\cal A}_+\oplus {\cal A}_-$. For the sake
of simplicity, we consider here the
maximal algebra ${\cal A}$.

\medskip

Let us introduce $\Phi _{\lambda }(a)$ corresponding 
to the extended character $\varphi _{a+2\lambda}^{(+)}$. It is 
then clear that the following fusion rules hold:
\begin{equation}
\Phi _{\lambda}(a)\,\Phi_{\lambda'}(a')=\Phi _{\lambda+\lambda'}(a+a')
.
\end{equation}
There are $\Bbb{Z}/pq\Bbb{Z}$ fusion rules in the $a=0$ sector as noticed 
by Cappelli and his collaborators \cite{Cappelli:96-1}. 
However let us point out that the chiral algebra
${\cal A}$ contains fields with half integer spins. Taking a chiral algebra
containing only integer spin fields gives us $\Bbb{Z}/4pq\Bbb{Z}$ 
fusion rules. 

\paragraph{$p$ or $q$ even}
Let us denote by $\Phi _{\lambda }(a)$ the primary field of indices 
$(a+\lambda,a+\omega\lambda)$. It lives in the sector of charge $a/q$. Using
the fusion rules for the $\widehat{U(1)}$ chiral algebra, we obtain:
\begin{equation}
\Phi _{\lambda }(a)\Phi _{\lambda' }(a')=\Phi  _{\lambda+\lambda' }(a+a')
,
\end{equation}
which expresses both the conservation of the boundary charge $a/q$ 
and the fact that we
have a $\Bbb{Z}/4pq\Bbb{Z}$ CFT. Starting from 
$\Phi _{\lambda }(a)$, other fields $\Phi _{\lambda' }(a)$ are obtained by fusion with $\Phi _{\mu }(0)$.

\paragraph{Physical interpretation}
The field $\Phi _{\lambda }(a)$ not only creates a boundary excitation but also
changes the sector of our Luttinger CFT. In some sense
these fields create bulk excitations. All descendants of the state
created by $\Phi _{\lambda}(a)$ satisfy $q_L+q_R\equiv a/q\pmod{1}$. We can
shift $\lambda $ at a fixed $a$ with an operator
of the form $\Phi _{\mu }(0)$ that does not change the sector. 

Finally, let us stress that even in the twisted sectors, the Hilbert space of
states is still a representation of the same extended chiral algebra ${\cal A}$
as in the bulk vacuum sector. In this sense, this chiral algebra may be
viewed as the symmetry algebra for single branch FQH fluid edge excitations.
This automatically raises the question of the physical meaning of the fields
that generate the extended symmetry algebra. It turns out that this is related to the question of fermions localized on one edge.

\medskip

In the cylindrical geometry, it should be physically possible to introduce
electrons on one of the edges of the system (through a tunnel junction
connected to one of the edges). 
Fields corresponding to the creation or destruction
of these edge fermions should therefore be present in all twisted 
Luttinger CFTs corresponding to the $\nu=1/(2p+1)$ sequence.
Moreover, they
should carry an electric charge localized only on one edge. Because of the
relation between the conformal dimensions and the edge charge of 
$\widehat{U(1)}$ primary fields, fields that carry a zero charge on the 
left (respectively right) 
edge correspond to holomorphic (respectively anti-holomorphic) 
primary fields, and therefore to
the chiral algebra of the Luttinger CFT.

Let us recall that, for $\alpha=q \in 2\Bbb{N}+3$ 
(corresponding to Laughlin fluids),
the chiral algebra is generated by the
two fields $\phi _{1/2,q}$ and 
$\phi _{1/2,-q}$. The following table gives the $(n,m)$ parameters,
the left and right charges and the conformal dimensions of these fields 
and their hermitian conjugates\footnote{The $e$ subscript recalls 
that we are dealing 
with edge fermions, not to be confused with the Luttinger fermions.}:
\begin{center}
\begin{tabular}{|l|r|r|rr|cc|}\hline
Field
& $n$ & $m$ & $q_R$ & $q_L$ & $h$ & $\bar{h}$ \\ \hline \hline
$\psi_{R,e}^{\dagger}$ & $1/2$ & $q$ & $1$ & $0$ & $q /2$ & 0 \\ \hline
$\psi_{R,e}$ & $-1/2$ & $-q$ & $-1$ & $0$ & $q /2$ & $0$ \\ \hline
$\psi_{L,e}^{\dagger}$ & $1/2$ & $-q$  & $0$ & $1$ & $0$ & $q /2$\\ \hline
$\psi_{L,e}$ & $-1/2$ & $q$ & $0$ & $-1$ & $0$ & $q /2$ \\ \hline
\end{tabular}
\end{center}

As expected, they carry a unit charge localized on one of the two edges,
which is a strong support for their identification with 
edge fermions of the FQH fluid. The first column of the table
details the action of these fields.
They were identified with edge fermions of the FQH fluid by
Stone and Fisher \cite{Stone:94-1}. Their argumentation was based on
the problem of establishing a relation between correlation functions for the 
Luttinger CFT and Laughlin wave function. 

\medskip

Finally, we recall that the $U(1)$ currents, which belong
to the chiral algebra of 
the Luttinger CFT for {\em all} values of $\alpha$,
generate neutral excitations
which are incompressible deformations of the FQH fluid.
Therefore, the structure of edge excitations appears to be remarquably simple:
they are all obtained from a finite set of elementary excitations 
corresponding to primary fields $\Phi _{\lambda}(a)$ 
with respect to the
extended algebra ${\cal A}$ by
adding or subtracting fermions localized on the edges, and deforming
the edges boundary through the use of $U(1)$ currents. Luttinger fermions
generate all the excitations in the untwisted ($a=0$) Luttinger CFT. Only for
$\alpha =1$ do they coincide with edge fermions. The same table as above reads for
Luttinger fermions:

\begin{center}
\begin{tabular}{|l|r|r|rr|cc|}\hline
Field
& $n$ & $m$ & $q_R$ & $q_L$ & $h$ & $\bar{h}$ \\ \hline \hline
$\psi_{R,\mathrm{Lutt}}^{\dagger}$ & $1/2$ & $1$ & $(1+q^{-1})/2$ & $(1-q^{-1})/2$ 
& $(q^{1/2}+q^{-1/2})^2 /8$ & $(q^{1/2}-q^{-1/2})^2 /8$  \\ \hline
$\psi_{R,\mathrm{Lutt}}$ & $-1/2$ & $-1$ & $-(1+q^{-1})/2$ & $-(1-q^{-1})/2$ & 
$(q^{1/2}+q^{-1/2})^2 /8$ & $(q^{1/2}-q^{-1/2})^2 /8$ \\ \hline
$\psi_{L,\mathrm{Lutt}}^{\dagger}$ & $1/2$ & $-1$  & $(1-q^{-1})/2$ & $(1+q^{-1})/2$ 
& $(q^{1/2}-q^{-1/2})^2 /8$ & $(q^{1/2}+q^{-1/2})^2 /8$\\ \hline
$\psi_{L,\mathrm{Lutt}}$ & $-1/2$ & $1$ & $-(1-q^{-1})/2$ & $-(1+q^{-1})/2$ & 
$(q^{1/2}-q^{-1/2})^2 /8$ & $(q^{1/2}+q^{-1/2})^2 /8$ \\ \hline
\end{tabular}
\end{center}

\subsection{Duality for the Luttinger CFT}

The free bosonic modular invariant theory possesses an
electro-magnetic duality which plays an important role in the 
discussion of $c=1$ CFTs. It is therefore interesting to 
determine if such a duality is present in the Luttinger CFT.
Not surprisingly, the
answer is yes but there are a few differences with the usual modular 
invariant theory.

\medskip

In order to understand duality properties of the usual free boson 
and Luttinger CFTs coupled to a gauge field, we derive them using 
functional methods. This involves performing a Hubbard-Stratanovich 
transformation 
which introduces a $1$-form $b=b_{\mu}dx^{\mu}$. 
Integration over the original bosonic field
constrains the harmonic, exact and co-exact parts of $b$ which are used to 
build the dual field. For the sake of clarity, all necessary
technical details are given in appendix \ref{secDuality}. 

\medskip

The main point is that duality acts differently on coupling constants of the
usual bosonic theory and of the Luttinger theory.
In the usual, modular invariant
bosonic theory, $\alpha $ is sent on $1/4\alpha$ whereas in the Luttinger CFT
$\alpha $ is sent on $1/\alpha$. Dirac theory now appears as the fixed
points for duality transformations acting on Luttinger CFTs instead
of $\alpha =1/2$ in the modular invariant sector\footnote{Which is 
known to be the $SU(2)$ Wess-Zumino-Witten model at level one.}. Of course, in 
both cases, duality exchanges the roles of electric and magnetic
potentials. For the Luttinger CFT, we have:
\begin{equation}
\label{eqLuttingerDualOrig}
Z^{(r)}_{\mathrm{Lutt}}[\alpha,A] = 
Z^{(0)}_{\mathrm{Lutt}}[\alpha ^{-1},{iA^*\over \alpha}+b_r]\times 
\exp{\left({1\over 2\pi \alpha }\int A^2\right)}
,
\end{equation}
where $b_r$ corresponds to a magnetic flux of $r\,\Phi _0/2$. This shows that
a given edge charge in the original theory becomes a magnetic flux in the dual
theory. 

\medskip

In an operator approach to the Luttinger CFT, the original field derivative
with respect to the spatial coordinate
appears to be the charge density. Therefore, the bosonic field may be interpreted
as a spatial deformation. A strong coupling regime is a regime of low fluctuations
of the charge density.
On the other hand, in the dual theory, the dual field's 
derivative coincides with the electric current. The strong coupling regime of
the original theory corresponds to a large fluctuation regime for the dual 
theory.

\medskip

Finally, we may ask for the structure of the moduli space of Luttinger CFTs. 
For modular invariant $c=1$ CFTs, this moduli space has been described
in the late eighties (see \cite{Ginsp} for a review). Here, the moduli space
is simpler.

In a zero magnetic field, 
we have a line of effective theories indexed by $\alpha >0$. As usual, we may look
for marginal integrable operators. Such an operator gives rise to the line of
orbifolds in the modular invariant case \cite{DVVc1}. But it is not present in
the Luttinger case: looking for operators of conformal dimensions $(1,1)$
gives $nm = 0$ and  $n^2\alpha + {m^2\over 4\alpha }=2$.
Besides that, the integrability condition, which only relies on fusion rules of
the bosonic model, forces us to limit to the $n=0$, $m=1$ operator 
which is {\em not} present
in the spectrum of a Luttinger CFT. Taking into account the magnetic field
would also eliminate this possibility since there is no operator that has
conformal dimensions $(1,1)$ for {\em all} values of $\chi $. 

To conclude, the only deformation 
parameters of Luttinger CFT are the interaction strength $\alpha$, the magnetic
flux $\chi $, and the external potential. Of course, there exists 
{\em relevant} perturbations which drive us away from CFT. As
an example, $\psi _R^{\dagger}\psi_L+\psi_L^{\dagger}\psi_R$ corresponds
to $\cos{(2R^{-1}\varphi)}$ and becomes relevant for $\alpha >1/2$ in zero 
magnetic field. This 
operator corresponds to a mass perturbation of the Thirring
model \cite{Thirring:58-1}. One could also get
interested in other operators, corresponding to the
{\em Umklapp} processes, such as 
$(\psi _L^{\dagger}\psi _R)^2$ which 
correspond to a momentum transfer of $-4k_F$. 
This operator appears in a fermion lattice
model as explained in appendix \ref{secAppendRegis}. It is responsible 
for the phase transition 
of the XXZ antiferromagnetic 
spin~$1/2$ chain from the massless phase to the Ising 
massive phase as the anisotropy increases. It becomes relevant for $\alpha >2$. 

Analogous criteria are used in the
study of the resonant tunneling 
of a LL through an impurity. In this
problem, the
{\em total} density (including $2pk_F$ Fourier components) is coupled
to a local potential $U(\sigma)$ localized near a particular value of
$\sigma$:
$H_{\mathrm{imp}}= \int U(\sigma)\rho _{\mathrm{total}}(\sigma)\, d\sigma$
(see \cite{Kane92-1} and section IV.E of \cite{Fisher:96-1}).

\subsection{Examples}

\subsubsection{The Fermi liquid}

This basic example corresponds to $\alpha =1$, that is to say $p=q=1$. 
The underlying CFT is a $\Bbb{Z}/4\Bbb{Z}$ theory which
has four basic primary fields. The $\lambda =2$ field 
has dimension $1/2$ and corresponds
to the Dirac fermion. In this example, 
they coincide with Luttinger fermions and with 
edge fermions discussed in the previous section.
The $\lambda =\pm 1$ fields have dimensions $1/8$ and
are twists fields. The $\lambda=0$ field is the identity field.

\medskip

The partition function for the Dirac theory in the sector of charge $a$ is 
given by:
\begin{equation}
Z_{\mathrm{Fermi}}^{(a)}(\tau,\bar{\tau})=
|\chi _a(\tau)+\chi _{a+2}(\tau)|^2=|\varphi _a^{(+)}(\tau)|^2
.
\end{equation}
Here $\omega =1$ and the partition functions are factorized as expected
since these are squared modulus of fermionic determinants \cite{Ginsp}.  
Explicit computations give the following formulae:
\begin{eqnarray}
\varphi _0^{(+)} & = & \chi _0+\chi _2 = \vartheta_3/\eta \\
\varphi _1^{(+)} & = & \chi _1+\chi_3 = \vartheta_2/\eta \\
\varphi _0^{(-)} & = & \chi _0-\chi _2 = \vartheta_4/\eta \\
\varphi _1^{(-)} & = & \chi _1-\chi _3 = i\,\vartheta_1/\eta
,
\end{eqnarray}
which gives the correspondence between the $\Bbb{Z}/4\Bbb{Z}$ formulation of this
model and classical theta functions.
The $a=1$ partition function is nothing but $|\vartheta_2/\eta|^2$ which
is nothing but $Z_{PA}(\tau)$. This result is physically expected since adding
an extra electron to the free Dirac theory in the $(A,A)$ sector is equivalent
to considering it in the $(P,A)$ sector. Clearly, the charged excitations 
considered here are nothing
but free electrons.

\subsubsection{Laughlin general Hall fluid}

Here, we assume $\nu =1/q$ and therefore $\alpha =q$ is an odd integer. 
We can consider that $p=1$ and
therefore easily get $\omega \equiv -1\pmod{2q}$. The underlying
CFT is a $\Bbb{Z}/4q\Bbb{Z}$ field theory. Its partition function 
can be expressed quite easily in terms of generalized characters:
\begin{equation}
Z_{\mathrm{Lutt}}^{(a/q)}(\tau,\bar{\tau})=
\sum _{\lambda \in \Bbb{Z}/q\Bbb{Z}}
\varphi ^{(+)}_{2\lambda +a}(\tau)\,
\overline{\varphi ^{(+)}_{-2\lambda +a}(\tau)}
\end{equation}
All states corresponding to $\varphi ^{(+)}_{2\lambda +a}\,
\overline{\varphi ^{(+)}_{-2\lambda +a}}$ have left and right charges given,
up to integers, by:
$$\left({a\over 2q}+{\lambda \over q}, {a\over 2q}-{\lambda \over q}\right).$$
We see that the charge $a/q$ of Laughlin excitation is 
equally shared by the two edges. 
The partition function is periodic in $a$ with a period of $2q$.
This factor of two can be understood quite easily: only in this case, 
the contribution to the charge on each edge is
given by an integer. 

\medskip

Let us illustrate these considerations in the $\nu =1/3$ case~: 
the underlying CFT is a $\Bbb{Z}/12\Bbb{Z}$ CFT. It
has twelve primary fields with respect to ${\cal A}_+$.
In $a$ even sectors, 
the renormalized Luttinger fermion corresponds to $\chi _4\,\overline{\chi _2}$
and $\chi _{10}\,\overline{\chi _{8}}$
which has conformal dimensions 
$(h,\bar{h})= (2/3,1/6) $ (and therefore spin $1/2$). 
The spin $3/2$ conserved current corresponds to edge fermions.
The two point function exchange properties are governed by
$(z-\xi)^3$, exactly as for Laughlin wave function.

\section{Conclusion}
\label{secConclusion}

In this paper, we have described in an accurate and
comprehensive way the effective
conformal field theory of the spinless Luttinger liquid
in a gapless regime. Our
description is valid in a finite size and at
finite temperature.
The partition function and correlation functions of charge and current
densities have been
computed for any value of the coupling constant, within the framework
of a {\it non chiral} bosonization scheme.
The conserved currents generate 
the symmetry algebra of all Luttinger CFTs. Their correlation functions give access
to the response function of the system and to
permanent currents induced by an
external magnetic flux.
Primary fields with respect to this
$\widehat{U(1)}_R\times \widehat{U(1)}_L$ symmetry are vertex operators
whose correlation functions are exactly computed for the Luttinger CFT. 

Using this description of Luttinger CFTs, we have shown how the physics
of edge excitations of a fractional quantum Hall fluid on an annulus 
can be recovered. 
The filling fraction is related in a simple way to the Luttinger interaction 
parameter which plays the role of a moduli for the Luttinger CFT. The solution
of the Luttinger CFT has been
extended to twisted sectors in order to describe edge excitations
above a given bulk excitation. With
the $c=1$ Luttinger CFT, only Laughlin's fluids ($\nu ^{-1}\in 2\Bbb{N}+1$) can
be recovered but an extension of our analysis to several boson fields will
give access to other filling fractions\footnote{Work in progress}.
It turns out that the Luttinger CFTs that describe
edge excitations of a FQH fluid are rational: they possess extra conserved currents
which generate an extended symmetry algebra so huge that the Hilbert space
of states is decomposed in a finite direct sum of irreducible representations. 
For Laughlin's FQH fluids, this algebra is generated by creators and annihilators
of edge fermions. This provides a nice and simple physical picture of 
edge excitations in Laughlin's fluids: they are all obtained from a finite number of
them through the addition or substraction of fermions located on the left
or right edge, and through neutral deformations induced by $U(1)$ currents. 

\medskip

The formalism developed in the present
article can now be used for studying other
aspects of 1D strongly interacting systems such as impurity problems (for which a 
large amount of work has already been done), coupling
of a Luttinger circle or of a Hall cylinder to various external systems such as
phonons, quantized electromagnetic fields, 
tunnel junctions, other Luttinger systems etc. Performing 
computations in finite size and at finite temperature gives
a direct access to
the interplay between various effects such as discreteness of charges,  
interference effects and how the physics changes with temperature. 
Other strongly interacting systems may be described by the same techniques, although
the structure of the underlying Hilbert space may be different. The most natural
example of such systems is the XXZ spin~$1/2$ chain 
in its massless regime\footnote{Work in progress by P.~Degiovanni and S.~Dusuel.}. 
Introduction of backscattering disorder should lead to a different physics 
such as localization. However, this is a difficult problem for which conformal
methods may not be trivially applied. Maybe the study of the quantum group symmetries
of the massive phase or the use of the quantum inverse scattering method
could help to understand the transition from a gapless
regime to an insulating one and its interplay with finite size effects.

\paragraph{Acknowledgements}
We would like to thank B.~Dou\c{c}ot for many interesting and stimulating
remarks, questions and suggestions, 
during the elaboration of this work. We also thank M.~Bauer, M.~Blau,
and F.~Delduc for useful discussions about the role of Ward identities in 
bosonization. A.~Cappelli and 
D.~Carpentier are acknowledged for illuminating 
discussions and references about the
fractional quantum Hall effect. S.~Dusuel helped clarifying parts of
the text and questions related to the effects of holonomy 
shifts on partition functions.
Part of this work was
performed while P.~Degiovanni and 
R.~M\'elin were at the NEC Research Institute at Princeton. During this stay,
they benefited of useful discussions and references 
by B.~Altsh\"{u}ler, O.~Agam, A.~Andreev, I.~Smolyarenko, P.~Chandra, 
N.~Wingreen and C.~Tang.
Part of this work was also performed while R.~M\'elin was post-doctoral fellow
at the International School for Advanced Studies (SISSA) at Trieste. Finally, 
M.~Fabrizio and R.~Stora
pointed out to us several useful references.

\appendix

\section{Infrared limit Hamiltonians}
\label{secAppendRegis}

The aim of this appendix is to recall the derivation of the infrared
limit of the Hamiltonian of a one dimensional fermionic gas.
We begin with the free Hamiltonian and then treat the interacting
case.

\subsection{Free Hamitonian}
We have explained at the beginning of section
\ref{secCFTLuttinger} how
the dispersion relation can be linearized around the two
Fermi points. Let us derive the expression
(\ref{eqHamiltDirac})
of Dirac's Hamiltonian $H^{(0)}$ in terms of
right and left moving fields.
This is a well-known procedure which we recall here
for the sake of pedagogy.

\subsubsection{Lattice fermions}
Let us start with the second quantized lattice fermions Hamiltonian
\begin{equation}
H^{(0)} = \frac{t}{2} \sum_{n=1}^{N} \left(c_n^+ c_{n+1}
\label{Regis1}
+ c_{n+1}^+ c_n \right)
\end{equation}
with cyclic boundary conditions.
This is also the Hamiltonian
of Jordan-Wigner fermions of the antiferromagnetic XX0
spin chain, with $t=J$, the exchange constant.
The dispersion relation is
$\epsilon(k) = t \cos{(k)}$. 
The continuum limit is obtained by factoring out the
$k_F$ dependence of the fermion field as follows:
\begin{equation}
c_n = \sqrt{\frac{a}{2}} \left(
e^{i k_F \sigma} \psi_R(\sigma) + e^{-i k_F \sigma} \psi_L(\sigma) \right)
\label{Regis2}
,
\end{equation}
where $\psi_{R,L}(\sigma)$ are slowly varying fields of the variable
$\sigma=n a$. Notice that $c_n$ being dimensionless,
$\psi_R$ and $\psi_L$ have a scaling dimension $1/2$.
Moreover, once $\sigma$ becomes a continuous variable,
the canonical anticommutation relations $\{\psi_{R,L}^+(\sigma),
\psi_{R,L}(\sigma') \} = \delta(\sigma-\sigma')$ are compatible with the ones
of the lattice fermions $\{c_n^+,c_m\}=\delta_{n,m}$
provided the following convention for the continuous and
lattice delta functions: $\delta(\sigma-\sigma')=\delta_{n,n'}/a$.

Expanding (\ref{Regis1}) in terms of the right and left moving fields
in (\ref{Regis2}) leads to
\begin{equation}
2 H^{(0)} = v_F \int_0^{L} d\sigma \left(
\psi_R^+(\sigma) i \partial_{\sigma} \psi_R(\sigma)
- \psi_L^+(\sigma) i \partial_{\sigma} \psi_L(\sigma) \right)
\label{Regis3}
,
\end{equation}
where $v_F=\partial \epsilon(k_F)/\partial k = a t \sin(k_F a)$.
The right hand side of (\ref{Regis3}) is nothing but
the Hamiltonian of Dirac's theory.
In deriving (\ref{Regis3}), we have
(i) neglected the fast varying fields with prefactors
$\exp{(\pm 2 i k_F \sigma)}$ since
these processes have a negligible probability, $2 k_F$
not being a vector of the reciprocal lattice. Notice
that in the presence of a dimerization (i.e. $t_{2 i}=t(1-\delta)$
and $t_{2i + 1}=t(1+\delta)$), these processes are no longer
negligible since $2 k_F$ is then a reciprocal lattice vector.
In this dimerized case, the $2 k_F$ processes are responsible for the
opening of a gap.
(ii) considered $\sigma$
as a continuous variable (iii) expanded $\psi_{R,L}^+(\sigma+a)
= \psi_{R,L}^+(\sigma) + a \partial_{\sigma} \psi_{R,L}^+(\sigma)$,
and neglected the higher order derivatives which do not contribute to
the infrared limit
(iv) dropped out a constant ground state energy.

\subsubsection{Non relativistic fermions}
We notice that the same analysis can be carried out
starting from the microscopic Hamiltonian
\begin{equation}
H^{(0)} = \int_0^L \Psi^+(\sigma) \frac{\Delta^2}{2 m}
\Psi(\sigma) d \sigma
.
\end{equation}
The fermion field $\Psi(\sigma)$ is related to the chiral fields
by
\begin{equation}
\Psi(\sigma) = \frac{1}{\sqrt{2}} \left(
e^{i k_F \sigma} \psi_R(\sigma) + e^{-i k_F \sigma} \psi_L(\sigma) \right)
\end{equation}
and the low energy effective Hamiltonian is nothing but (\ref{Regis3}).

\subsection{A lattice model with only $g_2$ and $g_4$ interactions}
The aim of this section is to point out that, most of the time,
the Hamiltonian (\ref{eqIntHam}) containing only $g_2$
and $g_4$ interactions is not the most general low energy Hamiltonian
of one dimensional spinless fermions. Additional interactions may
arise due to the possibility of {\it Umklapp} processes when an underlying
lattice is present. These Umklapp processes correspond to 
scattering events where the
transferred momentum is a multiple of the reciprocal
lattice vector. These terms give in general birth
to Gross-Neveu type interactions 
that may drastically change the physics.
For instance, in the XXZ chain, these Umklapp processes are
responsible for the flow to the massive Ising fixed point
when $J_z > J_x=J_y$.
These terms can be neglected away from half-filling since, under
this condition, scattering processes across the Fermi sea
cannot transfer a momentum belonging to the
reciprocal lattice.
However, under some very special circumstances, these
terms may vanish even at half-filling, and the long range
physics is then exactly the one
described in the present article.

To be more concrete, we consider a spinless lattice one-dimensional
fermion system with a hopping term (\ref{Regis1}) plus an
interaction term involving nearest-neighbor
and next-nearest-neighbor interactions
\begin{equation}
H^{(1)}(U,V) = U \sum_{n=1}^N c_n^+ c_n c_{n+1}^+ c_{n+1}
+ V \sum_{n=1}^N c_n^+ c_n c_{n+2}^+ c_{n+2}.
\end{equation}
Using the decomposition (\ref{Regis2}) into fast and slowly
varying modes, we get
\begin{equation}
c_n^+ c_n = \frac{a}{2}
\left(
\psi_R^+(\sigma) \psi_R(\sigma) + \psi_L^+(\sigma)
\psi_L(\sigma)
+ e^{-2 i k_F \sigma} \psi_R^+(\sigma) \psi_L(\sigma)
+ e^{2 i k_F \sigma} \psi_L^+(\sigma) \psi_R(\sigma)
\right)
.
\end{equation}
Using this last equation, we derive the low energy
interaction Hamiltonian:
\begin{eqnarray}
\nonumber
2 H^{(1)}(U,V) &=&
\frac{(U+V) v_F}{2 t\sin{(k_F a)}}
 \int_0^L \left({\cal J}^2(\sigma)+
\overline{\cal J}^2(\sigma) \right) d \sigma + 
\frac{(U+V) v_F}{t \sin{(k_F a)}} \int_0^L {\cal J}(\sigma)
\overline{\cal J}(\sigma) d \sigma\\
\nonumber
&& + \left\{ \frac{v_F e^{-2 i k_F a} (U+V e^{-2 i k_F a})}
{2 t \sin{(k_F a)}} 
\int
e^{-4 i k_F \sigma} (\psi_R^+(\sigma)\psi_L(\sigma))^2 + h.c. \right\}
,
\end{eqnarray}
where the currents ${\cal J}$ and
$\overline{\cal J}$ are defined in section \ref{secFree}.
The last term of this Hamiltonian correspond to the Umklapp
processes with momentum transfer $4k_F$.
At half filling ($k_F=\pi/2a$), these terms
vanish when $U=V$.
Under these conditions, the remaining interaction Hamiltonian
only involves $g_2$ and $g_4$ interactions:
\begin{equation}
\label{H_U=V}
2 H^{(1)}(U,U) = \frac{U v_F}{t} \int_0^L \left({\cal J}^2(\sigma)+
\overline{\cal J}^2(\sigma) \right) d \sigma + 
\frac{2 U v_F}{t} \int_0^L {\cal J}(\sigma)
\overline{\cal J}(\sigma) d \sigma .
\end{equation}

\section{Explicit computations for the Dirac theory}
\label{secDirac}

\newcommand{\Z}[3]{\ensuremath{#3 \: \fbox{$#1$}\hspace{-2.5ex} \raisebox{-3ex}{$#2$}}\hspace{1.3ex}}  

This appendix details the computation of free fermionic partition functions, in the
presence of a magnetic flux and a chemical potential. The identification
of fermionic determinants with bosonic partition functions is discussed in
great details. Emphasis is put on modular properties of all partition
functions and on their behavior under various shifts of the magnetic flux
and of chemical potential.

\subsection{Anti-periodic fermions 
coupled to magnetic field and chemical potential}

\paragraph{Operator computations with fermions}
The computation of the partition function for Dirac's theory in the presence
of a magnetic flux and a chemical potential can be performed using a zeta
function 
renormalisation prescription. The contribution of left and right modes to the vacuum 
energy are given by:
\begin{equation}
E_R = -{\pi v_F\over L}\, \left({1\over 12}-(\lambda +a)^2\right)\quad
\mathrm{and}\quad
E_L = -{\pi v_F\over L}\, \left({1\over 12}-(\lambda -a)^2\right)
\end{equation}
where $\lambda = {-L\mu \over 2\pi v_F}$. The contribution of excitations
is easily computed using a particle-hole picture and, introducing
the fugacity $y=e^{\beta \mu}$, the final result is:
\begin{eqnarray}
Z^{(0)}_{AA}(a,\mu) & = & q^{{(\lambda +a)^2\over 2}}\,
\overline{q}^{{(\lambda -a)^2\over 2}}\, (q\overline{q})^{-{1\over 24}}\times 
\prod _{n=0}^{+\infty}
(1+y\,q^{n+a+1/2})(1+y^{-1}q^{n-a+1/2})\nonumber \\
& \times & 
\prod _{n=0}^{+\infty}
(1+y\,\overline{q}^{n-a+1/2})(1+y^{-1}\overline{q}^{n+a+1/2})\label{eqZAA1}
\end{eqnarray}
Let us remark immediately that $(q\overline{q})^{\lambda ^2/2}$ correspond
to an extensive contribution to the free energy.
All other contributions come from energy levels 
that scale as $L^{-1}$. Therefore, we
shall forget about the extensive contribution which is understood as
a zero energy. The $(q/\overline{q})^{\lambda a}$ contribution, which scales
as $L^0$ is also discarded. Therefore, 
the final expression for the partition function is:
\begin{equation}
\label{eqZAA2}
Z_{AA}(a,\mu) = (q\overline{q})^{{a^2\over 2}-{1\over 24}}
\prod _{n=0}^{+\infty}
(1+y\,q^{n+a+1/2})(1+y^{-1}q^{n-a+1/2})
\prod _{n=0}^{+\infty}
(1+y\,\overline{q}^{n-a+1/2})(1+y^{-1}\overline{q}^{n+a+1/2})
\end{equation}
Using Jacobi's triple product identity, we may express it
as follows:
\begin{equation}
\label{eqZAAbose}
Z_{AA}(a,\mu) = {1\over |\eta(\tau)|^2}\, \sum _{(m,\overline{m})\in \Bbb{Z}^2}
q^{{1\over 2}(m+a)^2}\, \overline{q}^{{1\over 2}(\overline{m}-a)^2}\; y^{m+\overline{m}}
\end{equation}
It is quite instructive to compare these expressions with the ones obtained
by Sachs and Wipf in \cite{Sachs:96-1}. Their computation is done using
a fermionic functional integral, and therefore involves the study of the zeta
function for the spectrum of the Dirac operator. They compare it to an
operator computation performed in an appendix of their paper and their results
agree with ours\footnote{They assume that $\tau $ is real and the extensive
contribution to the energy is also discarded.}.

\paragraph{Fermion/boson comparison in the $(A,A)$ sector}
We also need to compare $Z_{AA}(a,\mu)$ and $K_{AA}(a,\mu)$ defined 
as follows:
\begin{equation}
\label{eqKAA}
K_{AA}(a,\mu)= {1\over 2}\left(
Z_{[0,0]}(a,-{i\mu \beta \over 2\pi}) + 
Z_{[{1\over 2},0]}(a,-{i\mu \beta \over 2\pi}) 
+ Z_{[0,{1\over 2}]}(a,-{i\mu \beta \over 2\pi}) - 
Z_{[{1\over 2},{1\over 2}]}(a,-{i\mu \beta \over 2\pi})
\right)
\end{equation}
This expression is equal to (see section \ref{secBosonGauge}):
\begin{equation}
\label{eqKAAnew}
K_{AA}(a,\mu)=
{1\over |\eta(\tau)|^2}\sum _{{(l,k)\in \Bbb{Z}^2\atop k\equiv l\pmod{2}}}
y^l\, q^{{1\over 8}(l+k+2a)^2}\, \overline{q}^{{1\over 8}(l-k-2a)^2}
\end{equation}
Therefore, using $m=(l+k)/2$ and $\overline{m}=(l-k)/2$, the
right hand side of equation (\ref{eqZAAbose}) is easily recovered
since $m$ and $\overline{m}$
now belong to $\Bbb{Z}^2$ because of the parity condition $l\equiv k\pmod{2}$.
Therefore, the final relation between the fermionic zeta-renormalized 
partition function and the bosonic one has the form\footnote{We
have left extensive prefactors for convenience. They shall be discarded in the bulk 
of the present paper.}:
\begin{equation}
Z^{(0)}_{AA}(a,\mu)= (q\overline{q})^{{\lambda ^2\over 2}}\,
\left({q\over \overline{q}}\right)^{\lambda a}\; K_{AA}(a,\mu)
\end{equation}
This proves our bosonization formula for the $(A,A)$ sector, in the presence
of a magnetic flux and a uniform chemical potential. Let us now turn to the
$AP$ sector.

\paragraph{The $AP$ sector}
This sector is obtained from the $AA$ sector by changing $y$ into $-y$ in the
particle-hole contribution to $Z_{AA}$. The vacuum contribution is not modified. 
Therefore, we have: 
\begin{equation}
Z^{(0)}_{AP}(a,\mu)= (q\overline{q})^{{\lambda ^2\over 2}}\,
\left({q\over \overline{q}}\right)^{\lambda a}\; 
K_{AA}\left(a,\mu+{1\pi \over \beta}\right)
\end{equation}

\subsection{Shifting and modular invariance properties}

It is also useful to shift $a$ and the chemical potential (or equivalently $b$)
for the fermionic determinant. We shall first of all details the effects
of such shifts on bosonic partition functions, and then
on fermionic partition functions.

\paragraph{Effects of shifts on bosonic partition functions}
First of all, let us perform the shifts on the bosonic expressions. We introduce
three new bosonic expressions:
\begin{eqnarray}
K_{AP}(a,\mu) & = & K_{AA}\left(a,\mu +{i\pi \over \beta}\right)
\label{eqdefKAP}\\
& = & {1\over |\eta(\tau)|^2}\, \sum _{(m,\overline{m})\in \Bbb{Z}^2}
(-y)^{m+\overline{m}}q^{{1\over 2}(m+a)^2}\,
\overline{q}^{{1\over 2}(\overline{m}-a)^2}
\label{eqKAP}\\
K_{PA}(a,\mu) & = & K_{AA}\left(a+{1\over 2},\mu\right) \label{eqKPA}\\
 & = & {1\over |\eta(\tau)|^2}\, \sum _{(m,\overline{m})\in \Bbb{Z}^2}
y^{m+\overline{m}+1}
q^{{1\over 2}(m+a+1/2)^2}\, \overline{q}^{{1\over 2}(\overline{m}-a+1/2)^2}\\
K_{PP}(a,\mu) & = & K_{PP}\left(a+{1\over 2},\mu +{i\pi \over \beta}\right)\\
 & = & {1\over |\eta(\tau)|^2}\, \sum _{(m,\overline{m})\in \Bbb{Z}^2}
(-y)^{m+\overline{m}+1}
q^{{1\over 2}(m+a+1/2)^2}\, \overline{q}^{{1\over 2}(\overline{m}-a+1/2)^2}
\label{eqKPP}
\end{eqnarray}
Of course, these partition functions can also be expressed as:
\begin{eqnarray}
K_{AP}(a,\mu) & = & {1\over 2}\left( Z_{[0,0]} +Z_{[0,1/2]} -Z_{[1/2,0]}  
+Z_{[1/2,1/2]} 
\right) \\
K_{PA}(a,\mu) & = & {1\over 2}\left(Z_{[0,0]} -Z_{[0,1/2]} +Z_{[1/2,0]} 
+Z_{[1/2,1/2]} 
\right)\\
K_{PP}(a,\mu) & = & {1\over 2}\left( Z_{[0,0]} -Z_{[0,1/2]} -Z_{[1/2,0]}
-Z_{[1/2,1/2]} 
\right)
\end{eqnarray}
as easily follows from the obvious properties of $Z_{[\epsilon,\epsilon']}(a,b)$:
\begin{eqnarray}
Z_{[\epsilon,\epsilon']}\left(a+{1\over 2},b\right) & = & e^{2\pi i\epsilon'}\,
Z_{[\epsilon,\epsilon']}(a,b)\label{eqShiftB1}\\
Z_{[\epsilon,\epsilon']}\left(a,b+{1\over 2}\right) & = & e^{2\pi i\epsilon}\,
Z_{[\epsilon,\epsilon']}(a,b) \label{eqShiftB2}
\end{eqnarray}

\paragraph{Compatibility properties}
It is interesting to study the interplay between shifts on $a$ and $b$, modular
transformations properties of partition functions and bosonization formulae.
 
To be more precise, let us introduce a vector
notation, which gathers simultaneously the four fermionic sectors and
the four bosonic sectors. With this notation, the non-chiral bosonization 
has the following expression (valid for $(a,b)$ {\bf real}):
\begin{equation}
\label{eqBoz}
\begin{pmatrix}
\Z{F}{A}{A}\vspace{1ex}\\
\Z{F}{A}{P}\vspace{1ex}\\
\Z{F}{P}{A}\vspace{1ex}\\
\Z{F}{P}{P}
\end{pmatrix} = \underbrace{{1\over 2}\times
\begin{pmatrix}
1 & 1 & 1 & -1 \\
1 & 1 & -1 & 1 \\
1 & -1 & 1 & 1 \\
1 & -1 & -1 & -1 
\end{pmatrix}}_{\Lambda}\;
\begin{pmatrix}
Z_{[0,0]}\vspace{1ex}\\
Z_{[0,1/2]}\vspace{1ex}\\
Z_{[1/2,0]}\vspace{1ex}\\
Z_{[1/2,1/2]}
\end{pmatrix} 
\end{equation}
In this framework, shifts by $1/2$ on $a$ and $b$ are represented
by matrices, which we denote by $B_{(a,b)}$ for 
bosons\footnote{Diagonal with eigenvalues given by equation (\ref{eqShiftB1}) 
and (\ref{eqShiftB2}).} and $F_{(a,b)}$
for fermions. Then,  if $\Lambda $ denotes the matrix 
relating bosonic partition functions to the fermionic ones, we have:
\begin{equation}
\label{eqShiftComp}
F_{(a)}\times \Lambda = \Lambda \times B_{(a)}\quad \mathrm{and}\quad 
F_{(b)}\times \Lambda = \Lambda \times B_{(b)}
\end{equation}
It is an easy exercise to check that these relations are compatible
with the modular properties of bosonic partition functions 
$Z_{[\epsilon,\epsilon']}$. The $S$ and $T$ modular transformations can
be represented by $4\times 4$ matrices on bosonic and fermionic 
partition functions. More explicitly, these modular transformation matrices
are:
\begin{equation}
  \label{eq:actionS}
  S^{(B)}=
  \begin{pmatrix}
  1&0&0&0\\0&0&1&0\\0&1&0&0\\0&0&0&1
  \end{pmatrix}
  \;\;\mathrm{and}\;\;
  S^{(F)}=
  \begin{pmatrix}
  1&0&0&0\\0&0&1&0\\0&1&0&0\\0&0&0&1
  \end{pmatrix},
\end{equation}
 and for $T$:
\begin{equation}
  \label{eq:actionT}
  T^{(B)}=
  \begin{pmatrix}
  1&0&0&0\\0&1&0&0\\0&0&0&1\\0&0&1&0
  \end{pmatrix}
  \;\;\mathrm{and}\;\;
  T^{(F)}=
  \begin{pmatrix}
  0&1&0&0\\1&0&0&0\\0&0&1&0\\0&0&0&1
  \end{pmatrix}.
\end{equation}
Of course, we have 
\begin{equation}
\label{eqModComp}
T^{(B)}\times \Lambda = \Lambda \times T^{(F)}\quad \mathrm{and}\quad
S^{(B)}\times \Lambda = \Lambda \times S^{(F)}
\end{equation}
which expresses the compatibility of our bosonization matrix $\Lambda$ natural 
action of $SL(2,\Bbb{Z})$ on partition functions. 
Then, we also check that the shifting matrices are compatible with modular 
transformation matrices:
\begin{equation}
\label{eqModCompShifts}
\begin{cases}
S^{(B)}\ldotp B_{(a)} = B_{(b)}S^{(B)}\\
S\ldotp B_{(b)} = B_{(a)}S 
\end{cases}
\quad \mathrm{and}\quad 
\begin{cases}
T^{(B)} B_{(a)} = B_{(a)} T^{(B)}\\
T^{(B)} B_{(b)} = B_{(a)}B_{(b)} T^{(B)}
\end{cases}
\end{equation}
and similar equations for $F_{(a,b)}$. 

\medskip

Imposing compatibility between modular transformations and the bosonization
matrix only {\em partially} fixes it. To be precise, imposing (\ref{eqModComp}) 
fixes $\Lambda$'s form as follows:
\begin{equation}
\label{eq444}
\Lambda = \begin{pmatrix}
a&b&b&c\\a&b&c&b\\a&c&b&b\\d&e&e&e
\end{pmatrix}
\end{equation}
The bosonization formula for the $PP$ sector thus appears to be
independent from the one of the three other fermionic sectors.
Only imposing equations (\ref{eqShiftComp}) provides extra conditions on 
$\Lambda$ as was already noticed in \cite{Moore:87-1}.
With notations of (\ref{eq444}), we obtain $b=-c$, $e=c$ and $d=a$. 
Therefore $\Lambda$ now depends only on two coefficients, as could easily be
expected since $Z_{[0,0]}$ is modular invariant: 
\begin{equation}
\Lambda = \begin{pmatrix}
a&b&b&-b\\a&b&-b&b\\a&-b&b&b\\a&-b&-b&-b
\end{pmatrix}
\end{equation}
Henceforth, the  bosonization formula for the $AA$ fermionic sector fixes
bosonization formulae for all other sectors.

\medskip

However, this nice picture is disturbed when coupling fermions to a 
{\em chemical potential} or electric potential. The main point lies in the fact that
such a potential corresponds to an {\em imaginary twist} of boundary conditions $b$.
But, as noticed by Sachs and Wipf \cite[Section 3.1]{Sachs:96-1}, evaluating the
zeta renormalized partition function is not direct. The zeta function of Dirac's 
operator has a well-defined analytic continuation through a Poisson resummation
only for $a$ and $b$ real (relatively to the $AA$ sector). Therefore, introducing
a chemical potential forces us to use the analytic continuation of the
result for $b$ complex. But in this case, one should take care of
contributions arising from the part sensitive to these analytic continuations, that
is to say to vacuum energy contributions. 

It is now time to turn to the $PA$ and $PP$ sector which need special
care, as we shall see in the next section.

\subsection{The $PA$ and $PP$ sectors}

\paragraph{Shifting fermionic partition functions}
Fermionic partition functions can also be directly computed. The
computation in the anti-periodic fermionic sector have already been
performed and in this case, there 
Zeta renormalisation of the vacuum energy gives:
\begin{equation}
E_R = -{\pi v_F\over L}\, \left( (a+\lambda )^2+a+\lambda +{1\over 6} 
\right)\quad
\mathrm{and}\quad 
E_L= -{\pi v_F\over L}\, \left((a-\lambda)^2+\lambda -a +{1\over 6}  
\right)
\end{equation}
Using these vacuum energies, and taking the 
trace over the fermionic Fock's space
gives $Z_{PA}$ as a product of left and right contributions:
\begin{eqnarray}
Z^{(R)}_{PA}(a,\mu) & = & 
q^{{1\over 2}((\lambda +a)^2-a-\lambda)+{1\over 12}}
\prod_{n=0}^{+\infty}
(1+yq^{n+a})(1+y^{-1}q^{n+1-a})\\
Z^{(L)}_{PA}(a,\mu) & = & 
\overline{q}^{{1\over 2}((\lambda -a)^2+a-\lambda)+{1\over 12}}
\prod_{n=0}^{+\infty}
(1+y\overline{q}^{n-a})(1+y^{-1}\overline{q}^{n+1+a})
\end{eqnarray}
Jacobi's identity transforms the infinite products in power series:
\begin{equation}
Z^{(R)}_{PA}(a,\mu)Z^{(L)}_{PA}(a,\mu) = (q\overline{q})^{\lambda ^2/2}
\left({q\over \overline{q}}\right)^{\lambda a}\, (q\overline{q})^{-\lambda /2}\,
\sum_{(m,\overline{m})\in \Bbb{Z}^2}y^{m+\overline{m}}q^{{1\over 2}(m+a-1/2)^2}
\,\overline{q}^{{1\over 2}(\overline{m}-a-1/2)^2}
\end{equation}
Then, let us shift $m\mapsto m+1$ and $\overline{m}\mapsto \overline{m}+1$, and 
remember that $(q\overline{q})^{-\lambda /2}=y^{-1}$. We finally obtain:
\begin{equation}
Z^{(0)}_{PA}(a,\mu) = (q\overline{q})^{\lambda^2/2}\,
\left({q\overline{q}}\right)^{\lambda a}\, \times 
\sum _{(m,\overline{m})\in \Bbb{Z}^2}
y^{m+\overline{m}+1}q^{{1\over 2}(m+a+1/2)^2}\,\overline{q}^{{1\over 2}(m-a+1/2)^2}
\end{equation}
Therefore, comparison with bosonic expressions is described by
\begin{equation}
Z_{PA}(a,\mu) = (q\overline{q})^{\lambda^2/2}\,
\left({q\overline{q}}\right)^{\lambda a}\, \times 
K_{PA}(a,\mu)
\end{equation}
It is now instructing to compute the fermionic partition function in the $PP$ 
sector. In this case, the vacuum contribution remains the same than for 
the $PA$ partition function. The only change is a shift from $y$ to $-y$ in
the particle-hole contributions to the partition functions. Henceforth, we
have: 
\begin{eqnarray}
Z^{(R)}_{PP}(a,\mu) & = & 
q^{{1\over 2}((\lambda +a)^2-a-\lambda)+{1\over 12}}
\prod_{n=0}^{+\infty}
(1-yq^{n+a})(1-y^{-1}q^{n+1-a})\\
Z^{(L)}_{PP}(a,\mu) & = & 
\overline{q}^{{1\over 2}((\lambda -a)^2+a-\lambda)+{1\over 12}}
\prod_{n=0}^{+\infty}
(1-y\overline{q}^{n-a})(1-y^{-1}\overline{q}^{n+1+a})
\end{eqnarray}
Using Jacobi's triple product identity again, we obtain:
\begin{equation}
Z^{(R)}_{PP}(a,\mu)Z^{(L)}_{PP}(a,\mu) = 
(q\overline{q})^{{1\over 2}(\lambda ^2-\lambda)}
\left({q\over \overline{q}}\right)^{\lambda a}
\sum_{(m,\overline{m})\in \Bbb{Z}^2}(-y)^{m+\overline{m}}q^{{1\over 2}(m+a-1/2)^2}
\,\overline{q}^{{1\over 2}(\overline{m}-a-1/2)^2}
\end{equation}
And the same shifts as before give us:
\begin{equation}
Z^{(0)}_{PP}(a,\mu) = 
(q\overline{q})^{\lambda ^2/2}
\left({q\over \overline{q}}\right)^{\lambda a}
\sum_{(m,\overline{m})\in \Bbb{Z}^2}
(-1)^{m+\overline{m}}\,  y^{m+\overline{m}+1} 
q^{{1\over 2}(m+a+1/2)^2}
\,\overline{q}^{{1\over 2}(\overline{m}-a+1/2)^2}
\end{equation}
This can be compared with the corresponding bosonic partition function:
\begin{equation}
\label{eqExprPP}
Z_{PP}^{(0)}(a,\mu) = \boldmath{-}
(q\overline{q})^{\lambda ^2/2}\left({q\over \overline{q}}\right)^2
\times K_{PP}(a,\mu)
\end{equation}
This sign deserves some comments although we shall not 
use the $PP$ sector in this paper, which is mainly concerned with the $AA$ sector
of the massless Thirring model. First of all, let us stress that the signs
we have obtained ensure positivity of the partition functions $Z_{PP}$ and 
$Z_{PA}$. In zero magnetic field, the dominant terms at vanishing temperature are
$$\begin{cases}
Z_{PA} \simeq (q\overline{q})^{{1\over 12}}(y+y+2) 
= 4(q\overline{q})^{{1\over 12}}\cosh^2{\left({\beta\mu \over2}\right)}\\
Z_{PP} \simeq (q\overline{q})^{{1\over 12}}(y+y-2) 
= 4(q\overline{q})^{{1\over 12}}\sinh^2{\left({\beta\mu \over2}\right)}
\end{cases}
$$
whereas $K_{PP}$ behaves like $2-y-y^{-1}=-\sinh^2{(\beta\mu/2)}$. The 
sign in (\ref{eqExprPP}) technically originates from the extraction of a $|y|$
factor coming from vacuum contributions to build the $K_{P\star }$ functions. A more
formal way to express this invokes, as explained before, 
the analyticity properties of zeta regularized
fermionic determinants and their behavior near vanishing points. 

Last but not
least, let us point out that taking into account properly this sign is important 
for finding the correct genus one effective CFT describing the finite size XXZ model
at finite temperature\footnote{S.~Dusuel and P.~Degiovanni, work in progress.}.

\section{Normalization condition for bosonic functional integrals}
\label{secNormalisation}

\subsection{Normalizations}

For a non compactified bosonic field $\varphi$, the integration measure
${\cal D}_g[\varphi]$ is usually (see \cite{Ginsp}) normalized by:
\begin{equation}
\label{eqNorm1}
\int {\cal D}_g[\varphi] \,\exp{\left(-{g\over 2\pi}\int \varphi ^2\right)}=1
.
\end{equation}
Let us decompose $\varphi$ into a constant part $\varphi_0$ and a 
zero-average part
$\tilde{\varphi}$:
$\varphi = \varphi_0+\tilde{\varphi}$. We introduce ${\cal D}_{\perp,g}$
the restriction of our measure to the space
of zero-average functions.  Then, equation 
(\ref{eqNorm1}) implies
\begin{equation}
\label{eqNorm2}
\int  {\cal D}_{\perp,g}
[\tilde{\varphi}] \,\exp{\left(-{g\over 2\pi}\int \tilde{\varphi} ^2\right)}
=\sqrt{{g\,{\cal A}\over 2\pi^2}}
.
\end{equation}
On the space of zero-average functions, the measures ${\cal D}_{\perp,g}$
are related to each other by
\begin{equation}
\label{eqNorm3}
 {\cal D}_{\perp,g}=\sqrt{{g\over g'}}\; {\cal D}_{\perp,g'}
,
\end{equation}
as can be easily inferred from equation (\ref{eqNorm2}). 

\medskip

For compactified fields over a circle of radius $R$, we have to specify the
integration over the zero mode $\varphi_0={\cal A}^{-1}\int \varphi$
of the field. We choose the usual volume $d\varphi_0$ of $\Bbb{R}/2\pi R\Bbb{Z}$.
There is an instanton sum and a zero-average fluctuating part $\xi$ remains.
The integration measure is thus decomposed as:
\begin{equation}
\label{eqNorm4}
{\cal D}_{R,g}[\varphi]=
\sum _{\mathrm{instantons}}\,d\varphi_0\;{\cal D}_{\perp,g}[\xi]
.
\end{equation}

\subsection{Fluctuation contribution}

We briefly recall how to compute the fluctuation contribution
\begin{equation}
\label{eqNorm102}
Z_{\mathrm{fluct}}=\int  {\cal D}_{\perp,g}
[\varphi] \,\exp{\left(-{g\over 2\pi}\int (d\tilde{\varphi})^2\right)}
.
\end{equation}
Using equation (\ref{eqNorm2}), this partition function can be 
expressed in terms of a determinant of
the Laplacian on the space of zero-average functions as:
\begin{equation}
\label{eqNorm101}
Z_{\mathrm{fluct}}=\sqrt{g{\cal A}\over 2\pi^2}\times
\mathrm{Det}_{\perp}^{-1/2}\left(-\Delta\right)
.
\end{equation}
The simplest way to compute the determinant is to use the so-called zeta function
renormalization procedure. 
From the spectrum of the Laplacian on the torus we get
\begin{equation}
Z_{\mathrm{fluct}}={1\over \sqrt{2\pi^2 {\cal A}}}\;{1\over |\eta(\tau)|^2}
,
\end{equation}
the Dedekind $\eta$ function being
\begin{equation}
\label{eqDefEta}
\eta(\tau)=e^{i\pi \tau/12}\prod_{n=1}^{+\infty}(1-q^n)
.
\end{equation}
This final result is independent on $g$ as can be 
expected from the Jacobian (\ref{eqNorm3}).

\section{Explicit computations in the bosonic theory}
\label{secCalculs}

In this section, we recall how to compute some partition functions
of the theory of a free boson, compactified on a circle of
radius $R$. 
The functional integral we want to compute is defined 
by equation (\ref{eqZtarget}). In order to ``define'' it,
a precise normalization prescription for
functional integrals is needed. All details and definitions relative
to these matters have been gathered in appendix \ref{secNormalisation}.
The reader is referred to this section for details. The present appendix
focuses on the ``instanton'' aspects of the computation.

\subsection{Instantons}
\label{Instantons}

We look for instantons with the choice of boundary conditions
$[\epsilon,\epsilon']$. A given complex number $z$ can be
uniquely written as $z=x \omega_1 + t \omega_2$, where
$(x,t) \in \Bbb{R}^{2}$ are given by:
\begin{eqnarray}
x & = & \Re{\left(\frac{z}{\omega_1} \right)}
- \frac{\Re{(\tau)}}{\Im{(\tau)}} \Im{\left(
\frac{z}{\omega_1} \right)}\\
t & = & \frac{\Im{(z/\omega_1)}}{\Im{(\tau)}}.
\end{eqnarray}
Instantons are classical solutions, that is harmonic functions
of the torus with $[\epsilon,\epsilon']$ boundary conditions,
and therefore have the form
\begin{equation}
\varphi^{(I)}_{n+\epsilon,m+\epsilon'}(x,t)
= 2 \pi R \left( (n + \epsilon) x +
(m+\epsilon') t \right),
\label{instanton}
\end{equation}
with $(m,n) \in \ZZ^{2}$.
The solution (\ref{instanton}) is the only one with monodromy
$(2\pi (n+\epsilon)R, 2\pi (m+\epsilon ')R)$. 
For the sake of simplicity,
let us introduce $v=2 \pi (n+\epsilon)R$ and $w=2 \pi (m+\epsilon')R$,
so that $\varphi_{I}(x,t) = x v + t w$. The action of a
the configuration (\ref{instanton}) is thus
\begin{eqnarray}
S[\varphi^{(I)}_{n+\epsilon,m+\epsilon'}] & = &
\frac{g}{2 \pi} \int_{{\bf T}_{\Gamma}}
(\nabla \varphi^{(I)}_{n+\epsilon,m+\epsilon'})^{2}\nonumber \\
& = & \frac{ 2 g}{\pi}
\int_{D_{\Gamma}} (\partial_z \varphi^{(I)}_{n+\epsilon,
m+\epsilon'})(\partial_{\overline{z}}
\varphi^{(I)}_{n+\epsilon,m+\epsilon'})
\frac{d \overline{z} \wedge dz}{2 i},
\end{eqnarray}
where $D_{\Gamma}$ is an elementary cell of the
lattice $\Gamma$. We obtain easily
\begin{equation}
\begin{cases}
\partial_z \varphi^{(I)}_{n+\epsilon,m+\epsilon'}
= \frac{1}{2 i \omega_1} \left(
i v + \frac{w}{\Im{(\tau)}} - \frac{Re(\tau)}{\Im{(\tau)}} v \right)\\
\partial_{\overline{z}} \varphi^{(I)} =
\frac{1}{2 i \overline{\omega}_1} \left(
i v - \frac{w}{\Im{(\tau)}} + \frac{\Re{(\tau)}}{\Im{(\tau)}} v \right)
.
\end{cases}
\end{equation}
Finally, the action of a $(n+\epsilon,m+\epsilon')$ instanton
is
\begin{equation}
S[\varphi^{(I)}_{\epsilon+n,
\epsilon'+m}] = 2 \pi R^{2} g \Im{(\tau)} \left( (n+\epsilon)^{2} +
\left( \frac{m + \epsilon' - \Re{(\tau)}(n + \epsilon)}
{\Im{(\tau)}} \right)^{2} \right).
\end{equation}


\subsection{Calculation of the partition function}
\label{Calculation of ...}

If $\varphi$ and $\varphi^{(I)}_{\epsilon+n,
\epsilon'+m}$ both have the same monodromies, then, $\phi=\varphi-
\varphi^{(I)}_{\epsilon+n,\epsilon'+m}$ has a vanishing monodromy.
Moreover, since the action is quadratic and since $\varphi ^{(I)}$
extremalizes the action, we have
\begin{equation}
S[\varphi^{(I)}_{\epsilon+n,\epsilon'+m} + \phi] =
S[\phi] + S[\varphi^{(I)}_{\epsilon+n,\epsilon'+m}].
\end{equation}
Henceforth, the partition function factorizes:
\begin{equation}
Z_{[\epsilon,\epsilon']}(g,R) = Z_{f} \sum_{(m,n) \in \ZZ^{2}}
e^{-S[\varphi^{(I)}_{\epsilon+n,\epsilon'+m}]},
\label{Part}
\end{equation}
where $Z_f$ contains the contribution of the fluctuations
without monodromy, and we now compute the instanton contribution,
the fluctuation contribution having been already computed in
appendix \ref{secNormalisation}.

\subsection{Introduction of a gauge field and
instanton contribution}

The following property will be useful: let $A$ and $B$ be closed 
$1$-forms on the torus, then:
\begin{equation}
\label{eqFormuleIntegraleTopo}
\int_{\bf T} A \wedge B = \int_{(a)} A \int_{(b)} B
- \int_{(b)} A \int_{(a)} B,
\label{identity}
\end{equation}
where $(a)$ and $(b)$ are the cycles
\begin{equation}
(a):  t\in [0,1] \rightarrow t \mbox{ and }
(b):  t\in [0,1] \rightarrow \tau t.
\label{cycles}
\end{equation}
Using (\ref{identity}), taking the holonomies (\ref{eqHolon})
\begin{equation}
\int _{(a)}A=2\pi \,a\ \mathrm{and}\quad 
\int _{(b)}A=2\pi \,b
,
\end{equation} 
and the monodromies (\ref{eqBCexplicit2}) of $\varphi$ 
along the $(a)$ and $(b)$ cycles, we obtain
\begin{equation}
S[\varphi,A] = \frac{g}{2\pi} \int (\nabla \varphi)^{2}
+ 4\pi i((\epsilon + n)b-(\epsilon'+m)a)
.
\end{equation}
Clearly, the coupling to the gauge field adds a topological
term to the action.

\paragraph{Contribution of the instantons:}
If $f$ is a square integrable function on $\RR$, let the 
Fourier transform be defined as
\begin{equation}
({\cal F}\ldotp f)(y) = 
\int_{- \infty}^{+ \infty} dx\, e^{2 i \pi x y} f(x),
.
\end{equation}
With this convention, the Poisson resummation formula reads:
\begin{equation}
\sum_{n=-\infty}^{+ \infty} f(n) = \sum_{n=-\infty}^{+ \infty}
({\cal F}\ldotp f)(n).
\label{Poisson}
\end{equation}
Using (\ref{Poisson}), we deduce an expression for the summation
over $m$ in (\ref{Part}), and finally the contribution of
the instantons reads
\begin{equation}
\left( \frac{ \Im{(\tau)}}{2 R^{2} g} \right)^{1/2}
\sum_{(m,n) \in \ZZ^{2}} e^{-2 i \pi m \epsilon'}
q^{\frac{1}{2} \left( (n+\epsilon)R \sqrt{g} + \frac{m}{2 R \sqrt{g}}
\right)^{2}}
\overline{q}^{\frac{1}{2} \left( (n+\epsilon)R \sqrt{g} -
\frac{m}{2 R \sqrt{g}}
\right)^{2}}.
\label{Zinst}
\end{equation}
\section{Elliptic and theta functions}
\label{secElliptic}

The aim of this appendix is to recall a few definitions and
basic formulae about elliptic
and theta functions. Numerous references are available on this
vast subject, but \cite{Serre}, \cite{Lang2} and \cite{whittaker-watson}
cover the material needed in this paper.

\subsection{Basic definitions}
\label{secEllipticDefs}

Let us denote by $
\tau $ an element of the upper-half plane, and $q=e^{2\pi i \tau}$. 
Let
$\Gamma =\omega _1\Bbb{Z}\oplus \omega_2\Bbb{Z}$
($\Im{(\omega_2/\omega_1)}>0$) be a 
lattice in $\Bbb{C}$, and we define by $\tau =\omega_2/\omega_1$
the modular parameter.
An elliptic curve is associated with each lattice by
$\Bbb{T}_{\Gamma}=\Bbb{C}/(\omega_1\Bbb{Z}\oplus \omega_2\Bbb{Z})$. 
Its complex structure is parametrized by the
modular parameter $\tau $ modulo the action of $PSL(2,\Bbb{Z})$ through
homographies. 

\medskip

The Weierstrass function associated with the lattice $\Gamma$ is
defined by 
\begin{equation}\label{eqDefWF}
\wp_{\Gamma}(z) = {1\over z^2}+\sum _{\omega \in \Gamma \setminus\{0\}}
\left(
{1\over (z-\omega)^2}-{1\over \omega^2}
\right)
.
\end{equation}
It is doubly periodic, thus defining a meromorphic function on
the torus $\Bbb{T}_{\Gamma}$ . The limit $\tau \rightarrow +i\infty$ is
interesting:
\begin{equation}
\label{eqLimWP}
\lim_{\tau \rightarrow +i\infty}(\wp_{\Bbb{Z}\oplus \tau \Bbb{Z}}(z))
= {1\over z^2}+\sum _{n\in \Bbb{Z}\setminus\{0\}}
{1\over (z-n)^2}
={\pi ^2\over \sin{(\pi z)}^2}
.
\end{equation}

\medskip

Riemann theta functions with characteristics are defined
by:
\begin{equation}
\label{eqDefTheta}
\vartheta \left[
\begin{array}{c}
a \\ b
\end{array}\right] (z,\tau) = 
\sum _{n\in \Bbb{Z}+a}q^{n^2/2}\; e^{2\pi i n(z+b)}
.
\end{equation}
Specializing to $(a,b)\in \{0,1/2\}^2$ gives the famous Jacobi functions. Using
the conventions of \cite[Appendix 9.A]{ItzDrou}:
\begin{eqnarray}
\label{eqDefTheta1}
\vartheta _1 (z,\tau) & = & 
\vartheta \left[
\begin{array}{c}
1/2 \\ 1/2
\end{array}\right] (z,\tau)\\
\label{eqDefTheta2}
\vartheta _2 (z,\tau) & = & 
\vartheta \left[
\begin{array}{c}
1/2 \\ 0
\end{array}\right] (z,\tau)\\
\label{eqDefTheta3}
\vartheta _3 (z,\tau) & = & 
\vartheta \left[
\begin{array}{c}
0 \\ 0
\end{array}\right] (z,\tau)\\
\label{eqDefTheta4}
\vartheta _4 (z,\tau) & = & 
\vartheta \left[
\begin{array}{c}
0 \\ 1/2
\end{array}\right] (z,\tau)
.
\end{eqnarray}
Technically, these are holomorphic sections of spin bundles over the torus 
$\Bbb{T}_{(\Bbb{Z}\oplus \tau \Bbb{Z})}$ and they appear in the
theory of chiral fermions on a torus \cite{Moore:87-1}.

\subsection{Relations between elliptic and theta functions and 
the inverse of the Laplacian on the torus}
\label{secLaplacienInv}

The inverse of the Laplacian on the torus 
$\Bbb{T}_{\Gamma}$ can
be expressed in terms of $\vartheta_1$ \cite{ItzDrou}:
\begin{equation}
\label{eqLaplacienInverse}
\Delta^{-1}(z)= 
{1\over 2\pi}\log{\left(
\left|\omega_1e^{-i\pi {z\over \omega_1}}
{\vartheta_1(z/\omega_1,\tau)\over \vartheta_1'(0,\tau)}\right|
\right)}
-{1\over 2}\left(
\Im{({z\over \omega_1})}+{
\Im{({z\over \omega_1})}^2\over \Im{(\tau)}}
\right)
.
\end{equation}
It satisfies the following Green equation:
\begin{equation}
\label{eqGreen}
\Delta (\Delta ^{-1}) = \delta -{1\over {\cal A}}
,
\end{equation}
where ${\cal A}=\Im{(\omega_1\,\overline{\omega_2})}$ 
is the area of the torus. 
Connection with Weierstrass' $\wp$ function arises through
\begin{eqnarray}
\label{eqDerholom}
(\partial_z ^2\Delta^{-1})(z) & = & {1\over 4\omega_1^2\,\Im{(\tau)}}
-{1\over 4\pi}\,\wp_{\Gamma}(z)\\
\label{eqDerAntiHolom}
(\partial_{\bar{z}} ^2\Delta^{-1})(z) & = & {1\over 4\overline{\omega_1}^2\,
\Im{(\tau)}}
-{1\over 4\pi}\,\overline{\wp_{\Gamma}(z)}
.
\end{eqnarray}
Computing correlators in the operator formalism 
requires to disymmetrize the role of $\Re{(z)}$ and $\Im{(z)}$,
which are space and imaginary time components on the torus. 
Therefore, it proves useful to have a Fourier
expansion for Weierstrass' function. If we note
$x=\exp{(2\pi iz)}$, we have for $|q|<|x|<|q|^{-1}$ \cite{Lang2}:
\begin{equation}
\label{eqWPTF1}
-{\wp (z)\over 4\pi ^2} = 
{1\over 12} -2\sum _{n=1}^{+\infty}{q^n\over (q^n-1)^2}
+ {x\over (1-x)^2} +
\sum _{n=1}^{+\infty}{nq^n\over 1-q^n}\,(x^n+x^{-n})
.
\end{equation}
The function 
$${x\over (1-x)^2} = {-1\over 4\sin^2{(\pi z)}}$$
may be expanded in powers series in $x$ on each domain $|x|>1$ and
$|x|<1$. 

\medskip

The Weierstrass function admits a complex primitive, known as the $\xi $
function: $\xi '=-\wp$. Its expansion is given by
\begin{equation}
\label{eqXWP3}
\xi (z) = {1\over z}+\sum _{\omega \in \Gamma \setminus 
\{0\}} \left({1\over z-\omega}+{z+\omega \over \omega^2}\right)
.
\end{equation}
This function is multivalued on the torus. If
$\eta_{1,2}=\xi (z+\omega_{1,2})-\xi(z)$ denotes its monodromies,
\begin{equation}
\label{eqWPquasiP}
\omega_1\eta_2 - \omega_2\eta_1=2\pi i
.
\end{equation}
Using the Fourier expansion of $\wp$, we easily get
$\eta_1$ as a function of $\omega_1$ and $\omega_2$:
\begin{equation}
\label{eqEta1}
\eta_1 = {4\pi ^2\over \omega_1}\left(
{1\over 12}-2\sum _{n=1}^{+\infty}{q^n\over (1-q^n)^2}
\right)
.
\end{equation}
In the functional integral approach of the Luttinger CFT, 
one needs to remember that derivatives of 
$\Delta ^{-1}$ are to be understood as derivatives of distributions. Let us recall
that if $T$ be a distribution and $f$ a 
test function, $\partial _{\mu}T$ is defined as:
\begin{equation}
\label{eqDefDerDis}
(\partial_{\mu}T)\ldotp f = -T\ldotp (\partial _{\mu}f)
.
\end{equation}
For the sake of precision, we shall denote by $[\varphi]$ the regular distribution defined
by the integrable function $\varphi$: 
$$[\varphi]\ldotp f = \int \varphi\,f.$$
On the plane, the Laplacian Green's function is 
\begin{equation}
\Delta _{\Bbb{C}}^{-1}(z)={1\over 2\pi}\,\log{(|z|)}
.
\end{equation}
Its first derivative $\partial_z[\Delta _{\Bbb{C}}^{-1}]$, as a distribution, is 
$\left[{1\over 4\pi z}\right]$. 
Even if $1/z^2$, not being integrable, does {\em not} define a regular
distribution, 
the second derivative of $[\Delta _{\Bbb{C}}^{-1}]$ 
may however be related to  
$z\mapsto z^{-2}$ through an integration by
parts formula. Let $K$ be a 
bounded region of $\Bbb{C}$ with boundary $\partial K$,
then:
\begin{equation}
\label{eqIntPart}
\int _K(\partial_zf)g = -\int _K f\,(\partial_zg) +{i\over 2}\int _{\partial K}
(fg)(z,\bar{z})\,d\bar{z}
.
\end{equation}
Now, splitting the integral $\int (\partial_zf)/z$ 
into two contributions ($|z|>\varepsilon$ and $|z|<\varepsilon $) and
applying the above formula, 
one easily gets, for any test function $f$,
\begin{equation}
\partial_z\left[ {1\over z}\right]\ldotp f
= \lim_{\varepsilon\rightarrow 0^+}\left(
\int _{|z|>\varepsilon }{f(z,\bar{z})\over z^2}\, d^2z
\right)
.
\end{equation}
This ``principal value'' formula connects 
$\partial _z^2[\Delta ^{-1}_{\Bbb{C}}]$ to $z\mapsto z^{-2}$.
In the same way, formula (\ref{eqIntPart}) immediately leads to
the well-known formula
\begin{equation}
\partial_z\left[ {1\over \bar{z}}\right]= \pi \delta
,
\end{equation}
where the right hand side arises from $\partial\{z,\,|z|>\varepsilon\}$.
It is quite easy to generalize these formulae on the torus
$\Bbb{C}/(\Bbb{Z}\oplus \tau\Bbb{Z})$. Taking into account periodicity 
relative to the lattice, the same kind of 
analysis as before around singularity gives
\begin{equation}
\label{eqDerGreen}
\partial_z\left[ \Delta ^{-1}\right] = 
{1\over 4\pi }\,\left[ 
{\vartheta '_1(z,\tau)\over \vartheta_1(z,\tau)}+ 
2\pi i {\Im{(z)}\over \Im{(\tau)}}\right]=
{1\over 4\pi }\,\left[ \xi(z)-\eta_1z+2\pi i {\Im{(z)}\over \Im{(\tau)}}\right]
.
\end{equation}
Differentiating once again with respect to $z$ gives,
for any test function $f$ 
on the torus\footnote{Of course, the integral is restricted to a 
fundamental cell of the $\Bbb{Z}\oplus \tau\Bbb{Z}$ lattice, which 
for convenience, is supposed to be centered on $0$.}:
\begin{equation}
\label{eqDerGreenSecond}
\partial_z^2\left[ \Delta ^{-1}\right] \ldotp f =
\lim _{\varepsilon \rightarrow 0^+}
\left( \int _{|z|>\varepsilon} d^2 z\, (
{1\over 4\Im{(\tau)}}-{1\over 4\pi}\,\wp(z) )\,f(z,\bar{z})\right)
,
\end{equation}
where the integral is performed over a unit cell of the lattice
$\Bbb{Z}\oplus \tau 
\Bbb{Z}$ centered on zero, with the restriction $|z|>\varepsilon$. 
In the same way, we recover
that
\begin{equation}
\label{eqEllipticDelta}
\partial_z\partial _{\bar{z}}\left[ \Delta ^{-1}\right] =
{1\over 4}\,\left(\delta -{1\over {\cal A}}\right)
.
\end{equation}

\subsection{Modular properties of theta functions}
\label{secThetaModule}

Modular properties of Riemann theta functions are easily 
found. Let us recall that the modular group $SL(2,\Bbb{Z})$ is
generated by two matrices \cite[Chapter 7]{Serre}:
$$t=\left(\begin{array}{cc}
1 & 1 \\
0 & 1 
\end{array}\right)
\quad \mathrm{and}\ s=\left(
\begin{array}{cc}
0 & -1 \\
1 & 0 
\end{array}\right).$$
We then have
\begin{eqnarray}
\label{eqThetaT}
\vartheta \left[
\begin{array}{c}
a \\ b
\end{array}\right] (z+,\tau+1) & = & e^{i\pi  a(1-a)}\,\vartheta \left[
\begin{array}{c}
a \\ b+a+1/2
\end{array}\right]
(z,\tau)\\
\label{eqThetaS}
\vartheta\left[
\begin{array}{c}
a \\ b
\end{array}
\right]({-z\over \tau},-{1\over \tau}) & = & 
(-i\tau)^{1/2}\,e^{-2\pi izb}\,e^{i\pi z^2/\tau}\,
\vartheta \left[
\begin{array}{c}
b \\ -a
\end{array}
\right](z,\tau)
.
\end{eqnarray}
The Dedekind $\eta$ function, defined in (\ref{eqDefEta})
transforms as
\begin{eqnarray}
\label{eqEtaT}
\eta(\tau+1) & = & e^{i\pi /12}\eta(\tau)\\
\label{eqEtaS}
\eta(-1/\tau) & = & (-i\tau)^{1/2} \eta(\tau)
.
\end{eqnarray}
Combining Dedekind's function with Riemann's theta functions leads to
finite dimensional unitary representations of the modular group. For 
$N\geq 1$ and $n\in \Bbb{Z}/N\Bbb{Z}$, let us define
\begin{equation}
\label{eqDefChiN}
\chi _n(\tau)= {\vartheta 
\left[
\begin{array}{c}
n/N\\ 0
\end{array}\right]
(0,N\tau)\over \eta(\tau)}
.
\end{equation}
Then, one has 
\begin{eqnarray}
\label{eqChiNT}
\chi _n(\tau +1) & = & e^{2\pi i ({n^2\over 2N}-{1\over 24})}\, \chi _n(\tau)\\
\label{eqChiS}
\chi _n(-1/\tau) & = & {1\over \sqrt{N}}\,\sum _{m\in \Bbb{Z}/N\Bbb{Z}}
e^{-2\pi i nm/N}\,\chi _m(\tau)
.
\end{eqnarray}
This unitary linear
representation of $SL(2,\Bbb{Z})$
arises in many conformal field theories, such as the rational 
Gaussian model \cite{DijkTh} 
which is heavily used in the present paper, but also in the
$SU(N)$ Wess-Zumino-Witten models at level one \cite{ITZ}. 
The kernel of this representation contains one of the congruence 
subgroups $\Gamma (12N)$ of $SL(2,\Bbb{Z})$. Remember that
\begin{equation}
\label{eqDefGN}
\Gamma (N) = \left\{M\in SL(2,\Bbb{Z}),\quad M\equiv \left(
\begin{array}{cc}
1 & 0 \\ 0 & 1
\end{array}\right)\pmod{N}\right\}
\end{equation}
and therefore, it appears as a representation of a quotient of 
$SL(2,\Bbb{Z})/\Gamma (12N)\simeq SL(2,\Bbb{Z}/12N\Bbb{Z})$. 

\section{Non relativistic fermions}
\label{secFermiSea}

In this section, we compute density-density correlations of
a gas 
of non relativistic fermions. For the sake of simplicity, we work
at 
zero temperature, thus avoiding to use the detailed dynamics of fermions. 
Our goal is to shed light on the Luttinger CFT results obtained
in section \ref{secCFTQJCorr}.

\subsection{Response to an external potential}
\label{secChargeSusceptibility}

Let us compute the linear response of the charge density to an external 
potential for non-relativistic free fermions. The external potential
may conveniently be treated as a stationary perturbation. Standard 
stationary perturbation theory gives us the first order 
correction to the eigenstate $|\psi _n\rangle$:
\begin{equation}
\label{eqPert1}
|\psi _n^{(1)}\rangle = |\psi _n\rangle + 
\sum _{m\neq n} {V_{mn}\over E_n-E_m}\,|\psi _m\rangle ,
\end{equation}
with $V$ the perturbation. This gives 
gives the first order correction to the matrix element $A_{mn}=
\langle \psi _n^{(1)}|A |\psi _m^{(1)}\rangle $:
\begin{equation}
\label{eqPert2}
\sum _{l\neq m} {V_{lm}A_{nl}\over E_m-E_l}
+\sum _{l\neq n} {V_{nl}A_{lm}\over E_n-E_l}.
\end{equation}
The Fourier transform of the density response $\rho(q)$ to an
external potential $V(q)$ modulated at a wave vector $q$ can
therefore be
expressed at the linear order in terms of the density of
states of particle-hole excitations by $\rho(q) = \nu(q) V(q)$,
with 
\begin{equation}
\label{eqPert31}
\nu(q)  =  \int _0^{+\infty}{d\omega \over \omega}
\, \nu (q,\omega).
\end{equation}
The density $\nu(q,\omega)$ of particle-hole excitations is defined by:
\begin{equation}
\label{eqPet32}
\nu(q,\omega)  =  \sum _q n(q)(1-n(q))\delta (\omega -
\epsilon(k)-\epsilon(q+k))
.
\end{equation}
Evaluating this expression leads to the explicit expression
of the density response to an external potential: 
\begin{equation}
\label{eqPert4}
\nu(q)= -{L \over \pi v_F}\, {1\over 2x}\log{\left(
\left|{1+x\over 1-x}\right|
\right)},\quad x={q\over 2k_F}
.
\end{equation}
This expression deserves two comments: first, there is a logarithmic
singularity (Kohn singularity) at $q=2k_F$ which 
is the sign of the so-called Peierls instability
\cite{Peierls:Quantum}. This instability is
responsible for the opening of a gap and a transition to
a charge density wave if, for instance, the system is coupled
to lattice distortions. Electronic interactions by themselves
can also be responsible for such a transition.
Next, at small momenta, we have $\nu(q)\simeq -L/\pi v_F$, which is,
as expected, nothing but the CFT result (\ref{eqChargeResp}).

\subsection{Density-density correlations}
\label{secNRCorr}

We consider here fermionic modes of momenta $2\pi n/L$. The charge
density operator is given in terms of creation and annihilation operators
by the operator
\begin{equation}
\widehat{\rho}(\sigma)=
{1\over L}
\sum _{k,q}e^{i\,q\sigma}\,\widehat{c}^{\dagger}_{k+q}\,
\widehat{c}_k
\end{equation}
At zero temperature, the system is in the Fermi vacuum 
$|F\rangle$ where all fermionic
states of momenta $k$ such that $|k|<k_F$ are occupied, 
and all other states
are empty. We have $N=2n_F+1$ particles in the system and the Fermi 
wave vector is 
equal to
$k_F=2\pi (n_F+1/2)/L$. A ultra violet cut-off is introduced
in the system: momenta are
bounded in absolute value by $\pi /a$, where $a$ is 
the lattice spacing.
We want to compute the equal time two point function:
\begin{equation}
\label{eqNR2pointsfct}
\langle F|\,\widehat{\rho}(\sigma)\widehat{\rho}(0)\,|F\rangle
= \frac{1}{L^2} \sum_{k,k,q,q'} e^{i q' \sigma}
\langle F| c_{c'+q'}^+ c_{k'} c_{k+q}^+ c_k |F \rangle
,
\end{equation}
and we separate between two cases: (i) $q=0$ and (ii) $q \ne 0$.

\paragraph{Case $q=0$}

Non vanishing contributions arise from $|k'|<k_F$ and
$q'=0$. The contribution to (\ref{eqNR2pointsfct}) is
$(N/L)^2 = (k_F/\pi)^2$, which is nothing but the
the square of the average of $\rho (\sigma)$.
Taking normal ordered products cancels this contribution,
which is therefore not relevant in our discussion.

\paragraph{Case $q\neq 0$}

Then $|k+q|>k_F$, and we should have $k'=k+q$ and
$q'=-q$ otherwise the matrix element in (\ref{eqNR2pointsfct}) vanishes. 
After a straightforward calculation of the different
sums, we obtain an explicit expression for the connected
density-density correlation:
\begin{eqnarray}
\nonumber
\langle F|\,\widehat{\rho}(\sigma)\widehat{\rho}(0)\,|F\rangle_c
 &=& \frac{2 k_F}{\pi L} \frac{
\cos{\left(\left(\frac{\pi}{2 a}+ k_F - \frac{\pi}{L} \right) \sigma \right)}
\sin{\left(\left(\frac{\pi}{2 a} - k_F \right) \sigma \right)}}
{ \sin{\left(\frac{\pi \sigma}{L} \right)}}\\
&+&
\label{dens-dens-connected}
\frac{k_F}{\pi L} \frac{\sin{\left( \left( 2 k_F - \frac{\pi}{L} \right)
\sigma \right)}}{\sin{\left(\frac{\pi \sigma}{L}\right)}}
-
\frac{1}{L^2} \frac{\sin^2{(k_F \sigma)}}{\sin^2{\left(\frac{\pi \sigma}{L}
\right)}}.
\end{eqnarray}
The first term arises from the contribution $|q| \ge 2 k_F$
(all of the $k$-matrix elements are non-zero,
with $k$ inside the Fermi sea),
and the last two terms arise from the contribution $|q| < 2 k_F$
(only some of these matrix elements are non zero).

\medskip

In order to recover the CFT results, one should take
the thermodynamic limit and consider only long distance
correlations:
$1/k_F \ll \sigma \ll L$ and
$1/a \ll \sigma \ll L$, $k_F$ and $a$ being kept fixed.
The connected density-density correlation in this limit is then
\begin{equation}
\label{dens-dens-result}
\langle F|\,\widehat{\rho}(\sigma)\widehat{\rho}(0)\,|F\rangle_c
\sim \frac{2 k_F}{\pi} \delta(\sigma) - \frac{1}{2 L^2}
\sin^{-2}{\left(\frac{\pi \sigma}{L}\right)}.
\end{equation}
In order to derive the $\delta(\sigma)$ contribution, 
the following identity is needed:
\begin{equation}
{1\over L}\,{\sin{\left({\pi \lambda \sigma \over L}\right)}\over 
\sin{\left({\pi \sigma \over L}\right)}}\longrightarrow \delta ({\sigma})
\end{equation}
in the limit $\lambda \rightarrow + \infty$, $\lambda $ being an {\em odd integer}. 
The other contribution
in (\ref{dens-dens-result})
has been obtained by performing an average over length scales $l$
such that $1/k_F \ll l \ll \sigma$:
$\sin^2{(k_F \sigma)}$ has been replaced by $1/2$.

\medskip

The CFT prediction for this correlation function are given by 
equations (\ref{eqCorrQB}) and (\ref{eqCorrQC}). 
Without any external potential, the CFT result reads
\begin{equation}
\langle \rho (\sigma,t)\rho(0,0)\rangle_c
={-1\over 2\pi^2\alpha}\,\Re{(\wp(\sigma+i\,v_St))}
+{1\over 2\pi\alpha}\delta (\sigma)\delta(t).
\end{equation}
Since we are interested here in the zero temperature limit,
we make use of the limiting behavior (\ref{eqLimWP})
of the Weierstrass $\wp$
function in the limit $\tau \rightarrow
+ i \infty$, from which
the equal time and zero temperature correlator can be deduced:
\begin{equation}
\label{d-d-CFT-result}
- {1\over 2\alpha L^2}\sin^{-2}{\left(\pi{\sigma\over L}\right)}+
{1\over 2\pi \alpha }\delta (\sigma) \delta(0).
\end{equation}

The second term is rather ill-defined and deserves some comments:
the $\delta (0)$ contribution arises from the equal time condition 
$t=0$. In fact, as
we have just seen, this delta distribution in the complex plane must
be regularized. The detailed computation we have just performed
from the non relativistic fermions problem shows 
that the divergence is smeared on a length scale of order $k_F^{-1}$.
The first term in (\ref{d-d-CFT-result}) is nothing but the
average of the non relativistic correlator over length scales large compared
to the microscopic scale $k_F^{-1}$, as we have already seen.
This completes the identification of the CFT result in the
non interacting limit with the non relativistic free fermions result.

\section{Operator computations and comparison with the CFT results}
\label{secOperators}

This Appendix is devoted to recover some of the CFT results in
terms of {\it chiral} mode bosonization. Contrary to appendix
\ref{secFermiSea},
we work here from the beginning with relativistic fermions,
but computations in the interacting theory can be carried
out by making use of the Bogoliubov transformation
(\ref{Bogo-trans-I}) and (\ref{Bogo-trans-II}).
As an example,
we reconsider the density-density correlations
within this chiral framework. More precisely,
we compute the generating functional for equal time 
density correlators. The functional result
will of course be recovered.

\medskip

Introducing an external potential coupled linearly to the
density, or equivalently a source field for density correlations
amounts to solve a bosonic Hamiltonian of the generic form
$h = \omega a^+ a + \lambda (a^+ + a)$, the second term arising
from the coupling to the potential. This problem is then
diagonalized by a unitary transformation.
The unitary operator $U[V]$ to be used is
\begin{equation}
\label{Unitary-transfo}
U[V(\sigma)] = \exp{\left(
{i\over 2v_S\sqrt{\alpha}}
\int _0^Ld\sigma \eta(\sigma)\,(J(\sigma)+\overline{J}(\sigma))
\right)},
\end{equation}
where $\eta'(\sigma)= {\cal V}(\sigma)$. Its action on
the $U(1)$ currents is given by:
\begin{eqnarray}
\label{eqJtransfo}
J'(\sigma) &=& U[V(\sigma)]\, J(\sigma)\,U[V(\sigma)]^{-1}=J(\sigma ) + 
{{\cal V}(\sigma)\over 2\pi v_S\sqrt{\alpha}}\\
\label{eqJbarTransfo}
\overline{J}'(\sigma) &=& U[V(\sigma)]\,\overline{J}(\sigma)
\,U[V(\sigma)]^{-1} =\overline{J}(\sigma ) +
{{\cal V}(\sigma)\over 2\pi v_S\sqrt{\alpha}}
.
\end{eqnarray}
These transformed operators
satisfy $\widehat{U(1)}$ commutation relations.
The total Hamiltonian (involving the kinetic term plus the
fermion-fermion interactions plus the coupling of the
density to the external potential)
transforms as follows
under the unitary transformation (\ref{Unitary-transfo}):
\begin{equation}
\label{eqHtransfo}
H[V(\sigma),J,\overline{J}] =
H[0,J',\overline{J}'] + {1\over 2\pi \alpha v_S}
\int _0^L{\cal V}(\sigma)^2\,d\sigma,
\end{equation}
where we have used the form (\ref{Hamil-J-inter}) of
the free plus fermion-fermion interaction terms.
We also have:
\begin{equation}
\int _0^Lb(\sigma)\rho_{[J,\overline{J}]}(\sigma) =
\int _0^Lb(\sigma)\rho_{[J',\overline{J}']}(\sigma) - {1\over \pi \alpha 
v_S} \int _0^L{\cal V}(\sigma)b(\sigma)\, d\sigma
\label{eqFtransfo}
.
\end{equation}
Therefore, using equations (\ref{eqJtransfo}) to (\ref{eqFtransfo}) 
and the cyclicity of the trace, the generating functional
(\ref{d-gene-func}) reads
\begin{equation}
\label{eqLinearExtract}
W^{(0)}_{[V(\sigma),\chi,q]}[b(\sigma)]= 
W^{(0)}_{[0,\chi,q]}[b(\sigma)]\,
e^{-{1\over \pi \alpha}\int _0^Lb(\sigma){\cal V}(\sigma)\,d\sigma}
.
\end{equation}
The  average $b_0$ of $b$ couples to the total charge and contributes to the 
linear term (\ref{eqCorrQA}) through
$$\int_0^L\rho(\sigma)b_0\,d\sigma = {q\over L}\int _0^Lb(\sigma)\,d\sigma .$$
Therefore, we have recovered the linear part corresponding to (\ref{eqCorrQA}).
Notice that the ${1\over 2\pi \alpha v_S}\int {\cal V}(\sigma)^2$ term
that appears in the right hand side of equation (\ref{eqHtransfo}) 
corresponds to the chiral gauge transformation anomaly, up to a factor
of $v_S$ that has been put to unity in (\ref{eqWID1}).

\medskip

The quadratic contribution to the generating functional
may easily be computed by noticing that 
$e^{\int b(\sigma) \rho(\sigma)d\sigma}$ is of
the form $e^{A+A^{\dagger}}$
where $[A,A^{\dagger}]$ is proportional to the identity. More
precisely\footnote{In this paragraph, $b$ has a
vanishing average over the circle.}
\begin{equation}
\label{eqDefA}
A= {1\over\sqrt{\alpha}}\sum _{n=1}^{+\infty}
(b_{-n}\,J_n+b_n\,\overline{J}_n)
.
\end{equation}
and the commutator of interest reads
$$ [A,A^{\dagger}]= {2\over \alpha }\sum _{n=1}^{+\infty}\;n\,|b_n|^2$$
Glauber formula leads to computing the finite temperature 
average of $<e^{A^{\dagger}}\;e^A>$.
But for a single harmonic oscillator, the following identity between finite 
temperature averages holds:
\begin{equation}
\label{eqOscillateur}
\left\langle e^{\lambda a^{\dagger}}\;e^{\mu a}\right\rangle
= e^{\lambda \mu\,<a^{\dagger}\;a>}
.
\end{equation}
Applied to the present context, this leads to
\begin{equation}
\label{eqW0Result}
W^{(0)}_{[0,\chi,q]}[b(\sigma)]= \exp{\left(
{1\over \alpha}\sum _{n=1}^{+\infty} n b_n b_{-n}\, {1+q^n\over 1-q^n}
\right)}
,
\end{equation}
where $q=\exp{(-2 \pi v_S/L)}$.
Then, one recognizes the Fourier expansion of the Weierstrass function. More
precisely, one formally has
\begin{equation}
\lim_{\varepsilon \rightarrow 0^+}
\left({\wp (\sigma +i\varepsilon )+\wp (\sigma -i\varepsilon) \over 2}
\right)= -\eta_1 +2\pi^2
\sum _{n=1}^{+\infty}n{1+q^n\over 1-q^n}\,(x^n+x^{-n})
.
\end{equation}
where $\eta_1$ denotes $\xi$'s monodromy as in appendix \ref{secElliptic}.
However, since $b$'s constant term is not taken in account in
equation (\ref{eqW0Result}), equations
(\ref{eqCorrQB}) and (\ref{eqCorrQC}) are recovered.

\section{Duality for the free boson revisited}
\label{secDuality}

The explicit 
expression of the partition function of the free boson on the torus
shows that theories at $\alpha =g R^2$ and $\alpha ^{-1}/4$ have the
same partition functions. This symmetry is called {\em duality} by
conformal field theorists \cite{DVVc1}. 
We analyze here the interplay 
between this duality and the boundary conditions that are imposed
on the bosonic field. Of course we focus on the generating functional
corresponding to the massless Thirring model coupled to a gauge field
as described in section \ref{secGauge}.
Our method will rely on direct functional integrals manipulation. The 
normalization prescription is important for these computations.

\subsection{Introduction of auxiliary fields}

The starting point has been known for a long time and consists in
introducing a vector field so that:
\begin{equation}
\exp{\left( -{g\over 2\pi}\int (\partial _{\mu}\varphi)^2\right)}
=\int {\cal D}[b_{\mu}]\, \exp{\left(
-{\pi \over 2g}\int b_{\mu}b^{\mu}+i\int b_{\mu}
\epsilon ^{\mu\nu}\partial _{\nu}\varphi\right)}
.
\end{equation}
The next step consists in integrating over $\varphi$. An effective action
for the field $b_{\mu}$ comes out. More precisely, let us consider
\begin{equation}
Z_{{\cal C}}[A]=\int_{{\cal C}} {\cal D}_{R,g}[\varphi]
\int {\cal D}[b_{\mu}]\,
\exp{\left(
-{\pi \over 2g}\int (b-{A\over \pi R})^2+i\int b\wedge d\varphi \right)}
.
\end{equation}
Let us now decompose $\varphi $ in a classical solution 
$\varphi _c$ having the
required monodromy, the constant part (zero mode), 
and a fluctuating part $\xi$~: $\varphi =\varphi _c+\varphi _0+\xi$. 
Integrating by parts shows 
that 
$$\int b\wedge d\varphi = \int 
b\wedge d\varphi _c +
\int \xi \, db.
$$
Roughly speaking, the integration over $\xi $
shows that $b$ is a flat connection.
We can therefore decompose it as $b=h+d\alpha $ where $h$ is a constant
$1$-form on the torus and $\alpha $ a zero form. Part of the dual bosonic
field will arise from this zero form. In the next section of this appendix,
this idea will be expressed more precisely.

\subsection{Compactification of the dual field}

Formula (\ref{eqFormuleIntegraleTopo}) implies
the vanishing of the integral of $d\alpha \wedge d\varphi_c$ since $d\alpha $
is exact. Therefore, if $b$ is closed,
$$ \int b\wedge d\varphi _c$$
can be expressed in terms of the
$\varphi_c$'s monodromies and the $b$'s
holonomies, or equivalently the $h$'s ones. Let us denote by $h_{(a,b)}$
these holonomies along the $(a)$ and $(b)$ cycles respectively. Let us also
introduce $n_{(a,b)}$ such that the $\varphi_c$'s monodromies are
$2\pi R\, n_{(a,b)}$. We then have: 
$$ \int b\wedge d\varphi _c=2\pi R\, (h_{(a)}n_{(b)}-h_{(b)}n_{(a)}).$$
The summation over the discrete set of values for $n_{(a,b)}$ discretizes
the $h_{(b,a)}$:
\begin{equation}
\sum _{n\in \Bbb{Z}}\exp{(2\pi i R\, h\,n)}= {1\over R}
\sum _{m\in \Bbb{Z}}\delta (h-{m\over R})
.
\end{equation}
This is the reason why a compactified discretized field also appears
in the dual theory. 

\medskip

At this stage, a careful treatment of the integration measures is
necessary. Definitions are given in appendix \ref{secNormalisation}. We
assume that $A$ is a {\em constant} gauge field. The case of a general
gauge field will be treated in section \ref{secWardDual}

\medskip

Let us now perform all manipulations on the modular invariant 
partition function
\begin{eqnarray}
\label{eqNorm5}
Z[A] & = & 2\pi R\,\sum _{(n,m)\in \Bbb{Z}^2}\int {\cal D}[b]e^{-{\pi \over 2g}\int b^2}
\int {\cal D}_{\perp,g}[\xi]
\exp{\left(i\int (b+{A\over \pi R})\wedge d(\varphi_{n,m}+\xi)\right)}\nonumber \\
 & = &  2\pi R\,\int {\cal D}[b]\;
e^{-{\pi \over 2g}\int (b-{A\over \pi R})^2}W_{\mathrm{inst}}[b]
\int {\cal D}_{\perp,g}[\xi]\, e^{i\int b \wedge d\xi}
\label{eqNorm6}
.
\end{eqnarray}
The functional $W_{\mathrm{inst}}[b]$ arises from the sum over instantons. Let us
decompose $b=h+\tilde{b}$, where the constant $1$-form $h$ is 
$b$'s average over the torus. 
The integration measure for the auxiliary fields factorizes as 
${\cal D}[b]=d^2h\, {\cal D}_{\perp}[\tilde{b}]$.
We then have
\begin{equation}
\label{eqNorm7}
W_{\mathrm{inst}}[h] = {1\over R^2} \sum _{(m_a,m_b)\in \Bbb{Z}^2}
\delta \left(\int _{(a)}h-{m_a\over R}\right)
\delta \left(\int _{(b)}h-{m_b\over R}\right)
.
\end{equation}
Therefore, denoting by $h_{(l_a,l_b)}$ the constant $1$-form of holonomies 
$(l_a,l_b)$, we obtain:
\begin{equation}
\label{eqNorm8}
\int d^2h\,W_{\mathrm{inst}}[h]\, \Phi[h]={1\over {\cal A}\,R^2}\;
\sum _{(m_a,m_b)\in \Bbb{Z}^2}\Phi[h_{R^{-1}\,(m_a,m_b)}]
.
\end{equation}
We shall now compute the integrals over $\tilde{b}$ and $\xi$. Let 
$$F[A,h]=
\int {\cal D}_{\perp}[\tilde{b}]
\int {\cal D}_{\perp,g}[\xi]\,
e^{i\int \tilde{b}\wedge d\xi}\,
\exp{\left(-{\pi \over 2g}\int (h-{A\over \pi R}+\tilde{b})^2\right)}
.
$$
Integrating over $\tilde{b}$, taking into account the normalization condition
over $b$ and using that $A$ is a constant gives:
\begin{equation}
\label{eqNorm9}
F[A,h]={{\cal A}\over 2g}\;\int  {\cal D}_{\perp,g}[\xi]\,
\exp{\left(-{g\over 2\pi}\int (d\xi)^2-{\pi \over 2g}\int (h-{A\over \pi R})^2\right)}
.
\end{equation}
Changing variables for $\tilde{\chi}=g\xi$ and using the Jacobian (\ref{eqNorm3})
leads to:
\begin{equation}
\label{eqNorm10}
F[A,h]={{\cal A}\over 2}\;\int  {\cal D}_{\perp,g^{-1}}[\tilde{\chi}]
\exp{\left(-{1\over 2\pi g}\int (d\tilde{\chi}+\pi h-{A\over R})^2\right)}
.
\end{equation}
Putting all parts together, we finally get
\begin{equation}
\label{eqNormalDual}
Z[A]=\int {\cal D}_{{1\over 2R},{1\over g}}[\chi]\,
\exp{\left(
-{1\over 2\pi g}\int (d\chi-{A\over R})^2
\right)}
,
\end{equation}
and $\chi $ is compactified over a circle of radius $R'=1/2R$.
The resulting parameter is $\alpha '=R^{'2}g^{-1}=1/4\alpha$ as
expected. 

\subsection{Duality for the Luttinger Conformal Field Theory}

We shall now go along the computation for the Luttinger case. Here, 
the functional integral boundary conditions are defined by 
equations (\ref{eqBoundaryTwisted}).
The summation over monodromies at fixed $(\varepsilon ,\varepsilon ')$
are performed first. The $h_{(a,b)}$ holonomies should therefore
be integer multiples of $1/R$. We shall then perform the summation over
$(\varepsilon ,\varepsilon ')$ and obtain the following result:
\begin{equation}
\label{eqDualite}
Z^{(r)}_{\mathrm{Lutt}}[\alpha,A]=\sum _{(u_a,u_b)\in \{0,1\}^2}
(-1)^{u_au_b}\, e^{i\pi ru_b}\int _{(\pi R^{-1}\, u_a,\pi R^{-1}\, u_b)}
{\cal D}_{{1\over 2R},{1\over g}}[\chi]\,
e^{-{1\over 2\pi g}\int (\partial _{\mu}\chi - {A\over R})^2}
.
\end{equation}
Then, by shifting to a field compactified on a circle of radius $1/R$,
we restore a $1/2$ factor due to the zero mode measure.
Developing the exponential, we obtain:
\begin{equation}
\label{eqLuttingerDual}
Z^{(r)}_{\mathrm{Lutt}}[\alpha,A] = 
Z^{(0)}_{\mathrm{Lutt}}[\alpha ^{-1},{iA^*\over \alpha}+b_r]\times 
\exp{\left({1\over 2\pi \alpha }\int A^2\right)}
,
\end{equation}
where $b_r$ corresponds to a magnetic flux of $r\,\Phi _0/2$. This arises
from 
\begin{equation}
e^{i\pi ru_b}=e^{{i\over \pi R^{-1}}\int b_r\wedge d\chi}\ 
\mathrm{where}\quad b_r=\pi {r\over L}\,d\sigma
.
\end{equation}
It is interesting to notice that duality still holds but
$\alpha '=\alpha ^{-1}$ instead of $4/\alpha$. 
Contrarily to the modular invariant case,
the free Dirac theory is the self dual point\footnote{This little difference
with the usual modular invariant case
had indeed been noticed by Klassen {\it et al.} \cite{Klassen:92-1} some time ago.}. 
We also notice that the roles 
of electric and magnetic fields are exchanged. The presence of a 
given charge in the system corresponds to a magnetic flux in the 
dual theory. Therefore, although the partition function in the $r=0$
sector, without any external field, is invariant under 
$\alpha \mapsto 1/\alpha $, the physics is not the same.

\subsection{Ward identities in the dual theory}
\label{secWardDual}

Finally, we have obtained duality formulae for the
compactified boson and the Luttinger theory in its bosonic form coupled
to a constant gauge field. Let us now extend these formulae to 
any gauge field. 
Let us first recall
that the original theory is gauge invariant under the transformation
\begin{equation}
\begin{cases}
A \mapsto A+d\beta\\
\varphi \mapsto \varphi
.
\end{cases}
\end{equation}
But the dual theory is also gauge invariant under
\begin{equation}
\begin{cases}
A \mapsto A+d\beta\\
\chi \mapsto \chi +\pi \beta 
.
\end{cases}
\end{equation}
In the dual theory, chiral gauge transformations correspond to:
\begin{equation}
\begin{cases}
A \mapsto A+d^*\beta \\
\chi \mapsto \chi
,
\end{cases}
\end{equation}
whereas, in the dual theory $\varphi $ is changed into 
$\varphi +i\beta /gR$. It is obvious that, 
under chiral gauge transformations, both the original and
the dual theory have the same transformation properties (\ref{eqWID1}). 
They are also invariant under normal gauge transformations.
Therefore, using Hodge's
theorem, we can extend formulae (\ref{eqNormalDual}) and 
(\ref{eqLuttingerDual}) to
any vector potential $A$.

\providecommand{\bysame}{\leavevmode\hbox to3em{\hrulefill}\thinspace}

\end{document}